\documentclass[preprint,superscriptaddress,floatfix]{revtex4-1}  

\usepackage{preamble}

\usepackage{mathrsfs}
\usepackage{amsmath}
\usepackage{amssymb}
\usepackage{graphicx}

\usepackage{epstopdf}
\usepackage{grffile}
\usepackage{esint}
\usepackage{textcomp}
\usepackage{url}
\usepackage[english]{babel}
\usepackage{todonotes}

\usepackage[bookmarksnumbered=true,unicode=true]{hyperref}

\graphicspath{{graphics/}}

\begin{document}

\title{Theory of voltammetry in charged porous media}

\date{\today}

\author{Edwin Khoo}
\affiliation{Department of Chemical Engineering, Massachusetts Institute of Technology, Cambridge, MA 02139, USA}

\author{Martin Z. Bazant}
\email[Corresponding author: ]{bazant@mit.edu}
\affiliation{Department of Chemical Engineering, Massachusetts Institute of Technology, Cambridge, MA 02139, USA}
\affiliation{Department of Mathematics, Massachusetts Institute of Technology, Cambridge, MA 02139, USA}

\begin{abstract}
  We couple the Leaky Membrane Model, which describes the diffusion and electromigration of ions in a homogenized porous medium of fixed background charge, with Butler-Volmer reaction kinetics for flat electrodes separated by such a medium in a simple mathematical theory of voltammetry. The model is illustrated for the prototypical case of copper electro-deposition/dissolution in aqueous charged porous media. We first consider the steady state with three different experimentally relevant boundary conditions and derive analytical or semi-analytical expressions for concentration profiles, electric potential profiles, current-voltage relations and overlimiting conductances. Next, we perform nonlinear least squares fitting on experimental data, consider the transient response for linear sweep voltammetry and demonstrate good agreement of the model predictions with experimental data. The experimental datasets are for copper electrodeposition from copper(II) sulfate solutions in a variety of nanoporous media, such as anodic aluminum oxide, cellulose nitrate and polyethylene battery separators, whose internal surfaces are functionalized with positively and negatively charged polyelectrolyte polymers.  
\end{abstract}

\maketitle

\section{Introduction}

In recent years, there is a growing need to extend electrochemical methods and devices to include charged porous media, which are macroscopically neutral, but contain charged internal surfaces or sites that provide a significant total charge per volume, comparable to the additional neutral salt concentration.  Transport in a neutral confined channel or porous medium is described by the classical Nernst-Planck equations for diffusion and electromigration (also collectively known as electrodiffusion), which predict a diffusion-limited current that the current in the system cannot exceed~\cite{newman_electrochemical_2004,bard_electrochemical_2000}. Under potentiostatic conditions, an infinite voltage is required for the current to reach its diffusion-limited value. Under galvanostatic conditions, applying a current that is larger than its diffusion-limited value results in negative concentrations and singularities at Sand's time~\cite{sand_concentration_1899}. However, experiments for electrodialysis in ion-exchange membranes~\cite{rubinstein_voltage_1979,rosler_ion_1992,krol_concentration_1999,krol_chronopotentiometry_1999,rubinshtein_experimental_2002,rubinstein_direct_2008,deng_overlimiting_2013,schlumpberger_scalable_2015,nikonenko_intensive_2010,nikonenko_desalination_2014,strathmann_electrodialysis_2010} and for microchannels and nanochannels~\cite{kim_concentration_2007,yossifon_selection_2008,zangle_propagation_2009,zangle_theory_2010,zangle_effects_2010,nam_experimental_2015,schiffbauer_probing_2015} have demonstrated that it is possible for an electrochemical system to exceed the diffusion-limited current and achieve overlimiting current (OLC) beyond bulk electrodiffusion.

In a confined channel or a porous medium, there are three physical mechanisms for OLC~\cite{dydek_overlimiting_2011}: surface conduction (SC)~\cite{mani_propagation_2009,zangle_propagation_2009,zangle_theory_2010,zangle_effects_2010,mani_deionization_2011,dydek_nonlinear_2013}, electroosmotic flow (EOF)~\cite{yaroshchuk_coupled_2011,rubinstein_convective_2013} and electroosmotic instability (EOI)~\cite{rubinstein_electro-osmotically_2000,zaltzman_electro-osmotic_2007}. These mechanisms are a strong function of the pore size and for pore sizes in the nanometer scale, surface conduction is expected to be the dominant OLC mechanism~\cite{dydek_overlimiting_2011}. When surface charges are present on the pore walls in a charged nanoporous medium and a sufficiently large current or voltage is applied to deplete the coions at an ion-selective interface such as an electrode or ion-exchange membrane, a large electric field develops in the depletion region that drives electromigration of the counterions in the electric double layers, i.e., surface conduction. In the depletion region, because the concentration gradients of the coions and counterions are very small, surface conduction is responsible for carrying most of the current. Surface conduction therefore sustains the OLC beyond bulk electrodiffusion and causes the formation and propagation of a deionization shock where ions are depleted behind the shock in porous media~\cite{mani_deionization_2011,yaroshchuk_over-limiting_2012,dydek_nonlinear_2013} and in microchannels and nanochannels~\cite{dydek_overlimiting_2011,mani_propagation_2009,zangle_propagation_2009,zangle_theory_2010,zangle_effects_2010,nielsen_concentration_2014}. In addition, there are also chemical mechanisms for OLC such as water splitting~\cite{nikonenko_intensive_2010,nikonenko_desalination_2014} and current-induced membrane discharge caused by membrane deprotonation and water self-ionization~\cite{andersen_current-induced_2012}.

The key mathematical concept in the leaky membrane model for describing OLC due to surface conduction is the addition of a volume-averaged background charge density term to the macroscopic electroneutrality equation for an electrolyte containing two or more mobile charge carriers so that one of the charge carriers can be depleted. This concept also appears in closely related fields such as electrodialysis in ion-exchange membranes and semiconductor physics. For describing ion transport in ion-exchange membranes, a spatially averaged background charge density is commonly added to macroscopic electroneutrality in order to account for the fixed ions present in the membranes~\cite{hawkins_cwirko_transport_1989,verbrugge_ion_1990,verbrugge_ion_1990-1,bernardi_mathematical_1992}; this simplification, as opposed to using Poisson's equation for electrostatics to describe space charge, is also known as the Teorell-Meyer-Sievers (TMS) theory~\cite{wang_electrolyte_1995,peters_analysis_2016}. In doped semiconductors, the dopant concentration is analogous to the volume-averaged background charge density while the electrons and holes are analogous to the anions and cations of a binary electrolyte respectively~\cite{pierret_semiconductor_1996,rubinstein_electro-diffusion_1990,roosbroeck_theory_1950,sze_physics_2006}.

We first derive the governing equations for describing transport and electrochemical reaction kinetics in a charged nanoporous medium. To predict OLC due to surface conduction, we use the leaky membrane model to describe transport~\cite{dydek_overlimiting_2011,mani_deionization_2011,dydek_nonlinear_2013}. For electrochemical reaction kinetics, we use Butler-Volmer reaction kinetics~\cite{bazant_theory_2013,ferguson_nonequilibrium_2012,newman_electrochemical_2004,bard_electrochemical_2000} and focus on copper electrodeposition and electrodissolution as a classic example whose reaction mechanism and parameters are well studied~\cite{newman_electrochemical_2004,mattsson_galvanostatic_1959,bockris_mechanism_1962,brown_rate-determining_1965}. There are more sophisticated reaction models for copper electrodeposition and electrodissolution that, for example, take into account the adsorption of copper(I) ions on the electrode surface and do not assume any rate-determining step~\cite{huerta_garrido_eis_2006,lasia_remarks_2007,huerta_garrido_reply_2008}. In the interest of being able to derive analytical or semi-analytical expressions for quantities of interest, we do not account for these additional complications in the reaction model. We first study the model at steady state under three different experimentally relevant boundary conditions, including the Butler-Volmer boundary conditions. We then use the model to study copper electrodeposition and electrodissolution under linear sweep voltammetry (LSV) in charged nanoporous anodic aluminum oxide (AAO) membranes, which are ordered membranes, and cellulose nitrate (CN) and polyethylene (PE) membranes, which are random membranes, to demonstrate that the model can fit published experimental results~\cite{han_over-limiting_2014,han_dendrite_2016} for a variety of membranes with sufficient accuracy. In these porous membranes, the surface charge density on the pore walls is tuned by using the layer-by-layer technique of depositing multiple layers of negatively or positively charged polyelectrolytes~\cite{hammond_form_2004}.  

Table~\ref{tab:Symbols for variables, parameters and constants} in Appendix~\ref{sec:Symbols for variables, parameters and constants} provides the symbols for variables, parameters and constants that are used throughout the paper. Superscripted ``$\textn{a}$'' and ``$\textn{c}$'' refer to quantities evaluated at the anode and cathode respectively, superscripted $\Theta$ denotes standard state, and superscripted ``\textn{eq}'' denotes equilibrium.

\section{Model}\label{sec:Model}

\subsection{Transport in leaky membrane model}\label{sec:Transport in leaky membrane model}

\begin{figure}
  \centering
  \includegraphics[scale=0.6]{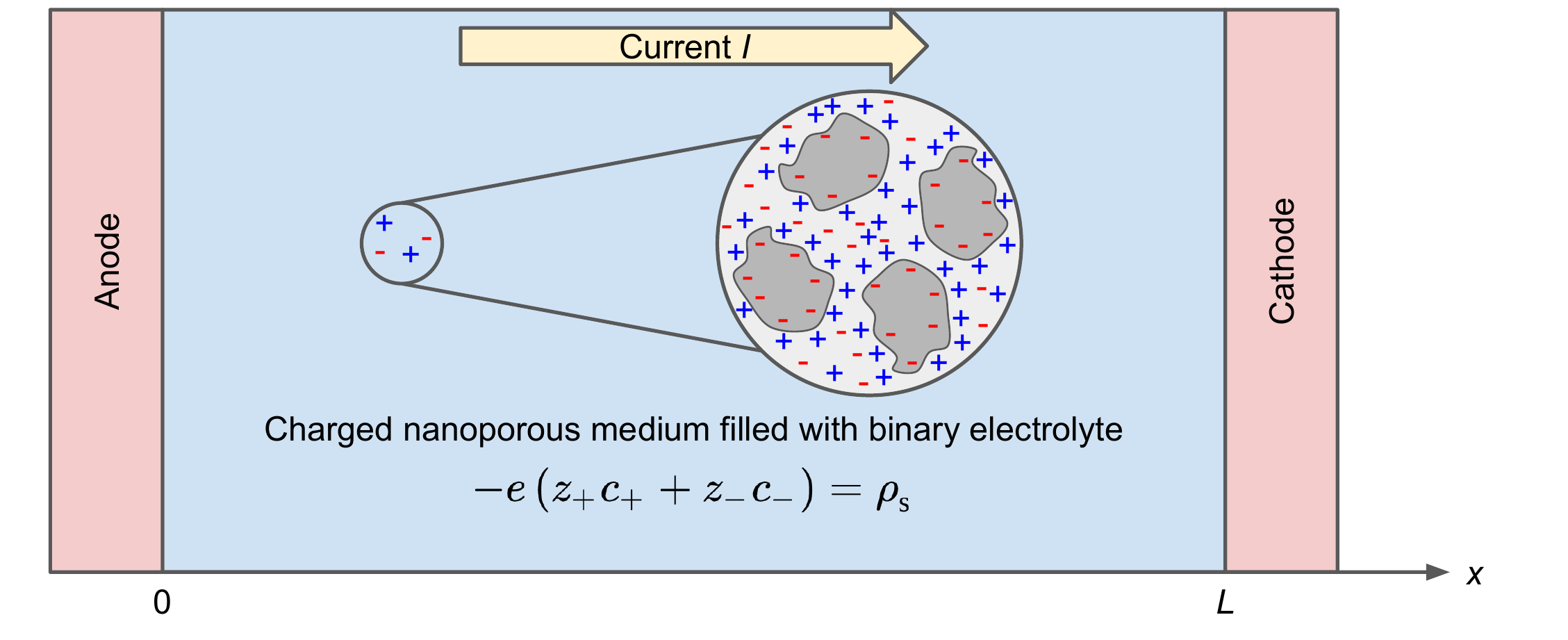}
  \caption{Schematic of system considered: charged nanoporous medium filled with binary electrolyte flanked on the left by anode and right by cathode. Current $I$ in system flows from left to right. The equation shown describes macroscopic electroneutrality given by Equation~\ref{eq:Macroscopic electroneutrality} where $\rho_\textn{s}$ is the volume-averaged background charge density.}\label{fig:Schematic}
\end{figure}

As illustrated in Figure~\ref{fig:Schematic}, we consider a charged nanoporous medium with a porosity $\epsilon_\textn{p}\left(r,t\right)$, an internal pore surface area/volume ratio $a_\textn{p}\left(r,t\right)$ and a pore surface charge/area ratio $\sigma_\textn{s}\left(r,t\right)$ where $r$ denotes the position vector. The porous medium is filled with a binary asymmetric electrolyte with unequal diffusivities. The chemical formula of the neutral salt is written as $\textn{c}_{\nu_+}^{z_+}\textn{a}_{\nu_-}^{z_-}$ where $\textn{c}^{z_+}$ and $\textn{a}^{z_-}$ represent the cations and anions respectively and $\nu_+$ and $\nu_-$ are the numbers of cations and anions produced respectively by the complete dissociation of 1 molecule of neutral salt. The anode and cathode are located on the left and right ends of the system respectively, therefore the current $I$ in the system flows from left to right.

Based on linear irreversible thermodynamics~\cite{bazant_theory_2013,ferguson_nonequilibrium_2012}, the diffusional molar flux $F_i$ of species $i \in \{+,-\}$ is given by
\begin{equation}
  F_i = -\frac{\epsilon_\textn{p}D_i^\textn{m}c_i}{\tau k_\textn{B}T}\nabla\mu_i, \quad \mu_i = k_\textn{B}T\ln a_i + z_i e\phi + \mu_i^\Theta,
\end{equation}
where $a_i = \gamma_i\hat{c}_i$ is the activity of species $i$ and $\hat{c}_i \equiv \frac{c_i}{c_i^\Theta}$ is the concentration of species $i$, $c_i$, normalized by its standard concentration $c_i^\Theta$, and the  $\Theta$ superscript denotes standard state. $T$ and $\phi$ are the temperature and electric potential of the electrolyte respectively and $k_\textn{B}$ is the Boltzmann constant. A natural scale for electric potentials is the thermal voltage given by $\frac{k_\textn{B}T}{e} = \frac{RT}{F} \approx 26\,\textn{mV}$ at $T = 298\,\textn{K}$ (room temperature) where $F = N_\textn{A}e$ and $R = N_\textn{A}k_\textn{B}$. $D_i^\textn{m}$, $\mu_i$, $z_i$, $\mu_i^\Theta$ and $\gamma_i$ are the molecular (free solution) tracer diffusivity, electrochemical potential, charge number, standard electrochemical potential and activity coefficient of species $i$ respectively. We account for corrections due to the porosity $\epsilon_\textn{p}$ and tortuosity $\tau$ of the charged nanoporous medium in $F_i$ and we ignore dispersion effects. We assume isothermal conditions, i.e., $T$ is constant, and that the material properties $\epsilon_\textn{p}$ and $\tau$ are uniform and constant.

$\gamma_\pm$ is generally a function of $c_\pm$. Modeling diffusion as an activated process, $D_i^\textn{m} = D_{i0}^\textn{m}\frac{\gamma_i}{\gamma_{\ddagger,i}^\textn{d}}$ where $D_{i0}^\textn{m}$ and $\gamma_{\ddagger,i}^\textn{d}$ are the molecular (free solution) tracer diffusivity in the dilute limit and the activity coefficient of the transition state for activated diffusion of species $i \in \{+,-\}$ respectively~\cite{bazant_theory_2013}. Throughout this paper, we set all activity coefficients to $1$ and ignore non-ideal effects because we are primarily interested in studying the effects of coupling Butler-Volmer reaction kinetics with the leaky membrane model, therefore we set $D_\pm^\textn{m} = D_{\pm0}^\textn{m}$.

The macroscopic diffusivities need to account for corrections due to the tortuosity of the charged nanoporous medium $\tau$. Following~\cite{ferguson_nonequilibrium_2012}, we define the macroscopic tracer diffusivity in the dilute limit of species $i \in \{+,-\}$, $D_{i0}$, as $D_{i0} \equiv \frac{D_{i0}^\textn{m}}{\tau}$. Thus, $F_\pm$ becomes
\begin{equation}
  F_\pm = -\epsilon_\textn{p}D_{\pm0}\left(\nabla c_\pm + \frac{z_\pm ec_\pm}{k_\textn{B}T}\nabla\phi\right).
\end{equation}
For AAO membranes that have parallel straight cylindrical pores with a constant pore radius, $\tau = 1$ while for random porous membranes such as CN and PE membranes, we can use the Bruggeman relation given by $\tau = \epsilon_\textn{p}^{-\frac{1}{2}}$ to estimate their tortuosities as a function of porosity.

The leaky membrane model consists of the Nernst-Planck equations that are coupled with the algebraic constraint given by macroscopic electroneutrality. Assuming no convection, the Nernst-Planck equations are given by
\begin{equation}
  \epsilon_\textn{p}\frac{\partial c_\pm}{\partial t} + \nabla\cdot F_\pm = 0, \label{eq:Nernst-Planck equations}
\end{equation}
where we account for corrections due to the porosity of the charged nanoporous medium $\epsilon_\textn{p}$ and assume that there are no homogeneous reactions. Macroscopic electroneutrality implies that
\begin{equation}
  \rho_\textn{s} \equiv \frac{\sigma_\textn{s}}{h_\textn{p}} = \frac{a_\textn{p}\sigma_\textn{s}}{\epsilon_\textn{p}} = -e\left(z_+c_+ + z_-c_-\right) \label{eq:Macroscopic electroneutrality}
\end{equation}
where we define the effective pore size $h_\textn{p} \equiv \frac{\epsilon_\textn{p}}{a_\textn{p}}$ and $\rho_\textn{s}$ is the volume-averaged background charge density. We assume that the material properties $a_\textn{p}$, $h_\textn{p}$, $\sigma_\textn{s}$, $\rho_\textn{s}$ are uniform and constant. Since we are invoking macroscopic electroneutrality, the electric double layers are assumed to be at equilibrium and their structures are not explicitly considered. For the electroneutrality of 1 molecule of neutral salt, we require $z_+\nu_+ + z_-\nu_- = 0$. The current density $J$ is given by the linear combination of the diffusional molar fluxes of all species weighted by their charges, i.e.,
\begin{equation}
  J = e\left(z_+F_+ + z_-F_-\right).
\end{equation}
Multiplying $z_i e$ to the Nernst-Planck equation of each species $i \in \{+,-\}$ and summing all such equations gives the charge conservation equation
\begin{equation}
  \nabla\cdot J = 0. \label{eq:Charge conservation equation}
\end{equation}
We denote the positions of the anode/electrolyte and cathode/electrolyte interfaces as $r_\textn{m}^\textn{a}\left(t\right)$ and $r_\textn{m}^\textn{c}\left(t\right)$ respectively and the ``$\textn{a}$'' and ``$\textn{c}$'' superscripts denote the anode and cathode respectively. The current $I$ is given by
\begin{equation}
  I = \int\left.\hat{n}\cdot J\right\rvert_{r=r_\textn{m}^\textn{c}}\ud S^\textn{c} = \int-\left.\hat{n}\cdot J\right\rvert_{r=r_\textn{m}^\textn{a}}\ud S^\textn{a} \label{eq:I}
\end{equation}
where we define $\hat{n}$ as the unit normal that points outwards from the electrolyte and the surface integral is performed over the total surface area of the anode or cathode, i.e., including both the electrolyte and matrix phases. Because of charge conservation, the current entering the cathode must be equal to the current leaving the anode, which is enforced in Equation~\ref{eq:I}. We will use Butler-Volmer reaction kinetics to describe electrochemical reactions at the electrodes and we denote the Faradaic current densities at the anode and cathode as $J_\textn{F}^\textn{a}$ and $J_\textn{F}^\textn{c}$ respectively. Because of the conservation of charges across the anode/electrolyte and cathode/electrolyte interfaces and because the volumetric porosity of a porous medium is equal to its areal porosity~\cite{bear_dynamics_1988}, we require
\begin{equation}
  \left.\hat{n}\cdot J\right\rvert_{r=r_\textn{m}^\textn{a,c}} = \left.\epsilon_\textn{p}J_\textn{F}^\textn{a,c}\right\rvert_{r=r_\textn{m}^\textn{a,c}}. \label{eq:Conservation of charges across electrode/electrolyte interfaces}
\end{equation}
The boundary condition given by Equation~\ref{eq:Conservation of charges across electrode/electrolyte interfaces} describes the coupling between transport in the charged nanoporous medium and electrochemical reaction kinetics at the electrode/electrolyte interfaces.

For a binary electrolyte, we define the neutral salt bulk concentration, which is denoted by $c$, that can be depleted, i.e., reach $0$. Regardless of the sign of $\rho_\textn{s}$, the concentration of the ions whose charge has the same sign as $\rho_\textn{s}$, i.e., the coions, can be depleted. For $\rho_\textn{s} \leq 0$, $c_-$ can reach $0$ while for $\rho_\textn{s} \geq 0$, $c_+$ can reach $0$. Therefore, we define
\begin{align}
  c &\equiv
  \begin{cases}
    \frac{c_-}{\nu_-}, & \rho_\textn{s} \leq 0 \\
    \frac{c_+}{\nu_+}, & \rho_\textn{s} \geq 0.
  \end{cases} \label{eq:Neutral salt bulk concentration}
\end{align}
Hence, rearranging Equation~\ref{eq:Neutral salt bulk concentration},
\begin{equation}
  c_- = \nu_-c - \frac{\rho_\textn{s} + \left\lvert\rho_\textn{s}\right\rvert}{2z_-e}. \label{eq:c_- expressed as a function of c}
\end{equation}
The initial neutral salt bulk concentration $c\left(t=0\right)$ is specified as an initial condition and we determine $c_-\left(t=0\right)$ from Equation~\ref{eq:c_- expressed as a function of c}.

\subsection{Electrochemical reaction kinetics}\label{sec:Electrochemical reaction kinetics}

Generally, for an electron transfer reaction involving $n$ electrons, the Faradaic current density $J_\textn{F}$ can be written in terms of the exchange current density $J_0$ and overpotential $\eta$ as
\begin{equation}
  J_\textn{F} = J_0\left[\exp\left(-\frac{\alpha_\textn{c}ne\eta}{k_\textn{B}T}\right) - \exp\left(\frac{\alpha_\textn{a}ne\eta}{k_\textn{B}T}\right)\right], \quad \alpha_\textn{c} + \alpha_\textn{a} = 1, \label{eq:General J_F}
\end{equation}
where $\alpha_\textn{c}$ and $\alpha_\textn{a}$ are the cathodic and anodic charge transfer coefficients respectively~\cite{bazant_theory_2013,ferguson_nonequilibrium_2012} and $J_0$ is generally a function of the activities of the oxidized and reduced species and electrons. We define $\eta = \Delta\phi - \Delta\phi^\textn{eq}$ where $\Delta\phi = \phi_\textn{e} - \phi$ is the interfacial electric potential difference, $\phi_\textn{e}$ is the electric potential of the electrode and $\Delta\phi^\textn{eq}$ is the Nernst potential, which is generally a function of the activities of the oxidized and reduced species and electrons. The overpotential provides the driving force for a Faradaic reaction to go out of equilibrium and results in a nonzero Faradaic current density.

As a prototypical example of electrochemical reaction kinetics, we consider copper electrodeposition and electrodissolution, which have been studied extensively in literature~\cite{newman_electrochemical_2004,mattsson_galvanostatic_1959,bockris_mechanism_1962,brown_rate-determining_1965}. A more general theoretical treatment of electrochemical reaction kinetics based on nonequilibrium thermodynamics can be found at~\cite{bazant_theory_2013,ferguson_nonequilibrium_2012}. All variables here are evaluated at the electrode/electrolyte interface ($r=r_\textn{m}^\textn{a,c}$). We assume that only ions exist in the electrolyte while only electrons and neutral atoms exist in the solid electrode, i.e., we do not consider mixed ion-electron conductors that are used in applications such as solid oxide fuel cells.

The reaction mechanism for copper electrodeposition and electrodissolution~\cite{newman_electrochemical_2004,mattsson_galvanostatic_1959,bockris_mechanism_1962,brown_rate-determining_1965} can be written as
\begin{align}
  \textn{Cu}^{2+}\textn{(aq)}+\textn{e}^- &\rightleftharpoons \textn{Cu}^+\textn{(ads)}, \\
  \textn{Cu}^+\textn{(ads)}+\textn{e}^- &\rightleftharpoons \textn{Cu}\textn{(s)},
\end{align}
where (aq) indicates aqueous, (ads) indicates adsorbed on the electrode surface, (s) indicates solid, and the mechanism involves the overall transfer of $n = 2$ electrons. We assume that the first step is the rate-determining step (RDS) while the second step is at equilibrium and that Butler-Volmer reaction kinetics~\cite{bazant_theory_2013,ferguson_nonequilibrium_2012,newman_electrochemical_2004,bard_electrochemical_2000} applies to both steps. We also assume that the activity of the electrons is $1$, i.e., we ignore non-ideal effects associated with the electrons. Denoting $J_{\textn{F},1}$ and $J_{\textn{F},2}$ as the Faradaic current densities for the first and second steps respectively, in terms of $\Delta\phi$, we obtain
\begin{align}
  J_{\textn{F},1} &= \frac{e}{\gamma_{\ddagger,1}^\textn{r}}\left \{k_{\textn{c},1}a_{\textn{Cu}^{2+}}\exp\left(-\frac{\alpha_1e\Delta\phi}{k_\textn{B}T}\right) - k_{\textn{a},1}a_{\textn{Cu}^+}\exp\left[\frac{\left(1-\alpha_1\right)e\Delta\phi}{k_\textn{B}T}\right]\right \}, \label{eq:J_F_1} \\
  J_{\textn{F},2} &= \frac{e}{\gamma_{\ddagger,2}^\textn{r}}\left \{k_{\textn{c},2}a_{\textn{Cu}^+}\exp\left(-\frac{\alpha_2e\Delta\phi}{k_\textn{B}T}\right) - k_{\textn{a},2}\exp\left[\frac{\left(1-\alpha_2\right)e\Delta\phi}{k_\textn{B}T}\right]\right \},
\end{align}
where $\gamma_{\ddagger,i}^\textn{r}$, $k_{\textn{c},i}$, $k_{\textn{a},i}$ and $\alpha_i$ are the activity coefficient of the transition state for the Faradaic reaction, cathodic rate constant, anodic rate constant and charge transfer coefficient of step $i \in \{1,2\}$ respectively. We assume that $a_\textn{Cu}$ remains constant at $1$ in $J_{\textn{F},2}$ because Cu\textn{(s)} is a solid metal at room temperature. Because we assume that the first step is the RDS while the second step is at equilibrium, $J_\textn{F} = 2J_{\textn{F},1}$, where the factor of $2$ accounts for the overall transfer of $2$ electrons, and $J_{\textn{F},2} \approx 0$. Like in Section~\ref{sec:Transport in leaky membrane model}, we assume that all activity coefficients are equal to 1 and ignore non-ideal effects. Therefore,
\begin{equation}
  J_\textn{F} = 2e\left \{k_{\textn{c},1}\hat{c}_{\textn{Cu}^{2+}}\exp\left(-\frac{\alpha_1e\Delta\phi}{k_\textn{B}T}\right) - \frac{k_{\textn{a},1}k_{\textn{a},2}}{k_{\textn{c},2}}\exp\left[\frac{\left(2-\alpha_1\right)e\Delta\phi}{k_\textn{B}T}\right]\right \}.
\end{equation}
At equilibrium, $J_\textn{F}=0$ and $\Delta\phi=\Delta\phi^\textn{eq}$ and we recover the Nernst equation given by
\begin{equation}
  \Delta\phi^\textn{eq} = \frac{k_\textn{B}T}{2e}\ln\left(\frac{k_{\textn{c},1}k_{\textn{c},2}\hat{c}_{\textn{Cu}^{2+}}}{k_{\textn{a},1}k_{\textn{a},2}}\right).
\end{equation}
where the ``eq'' superscript denotes equilibrium. At standard conditions, $\hat{c}_{\textn{Cu}^{2+}} = 1$ and we obtain
\begin{equation}
  E^\Theta \equiv \Delta\phi^{\textn{eq},\Theta} = \frac{k_\textn{B}T}{2e}\ln\left(\frac{k_{\textn{c},1}k_{\textn{c},2}}{k_{\textn{a},1}k_{\textn{a},2}}\right) \label{eq:E_Theta}
\end{equation}
where $E^\Theta$ is the standard electrode potential for $\textn{Cu}^{2+}$ ions. We express $J_\textn{F}$ in terms of $J_0$ and $\eta$ as
\begin{equation}
  J_\textn{F} = J_0\left \{\exp\left(-\frac{\alpha_1e\eta}{k_\textn{B}T}\right) - \exp\left[\frac{\left(2-\alpha_1\right)e\eta}{k_\textn{B}T}\right]\right \}, \quad J_0 = 2e\left(k_{\textn{c},1}\hat{c}_{\textn{Cu}^{2+}}\right)^{1-\frac{\alpha_1}{2}}\left(\frac{k_{\textn{a},1}k_{\textn{a},2}}{k_{\textn{c},2}}\right)^{\frac{\alpha_1}{2}}. \label{eq:J_F}
\end{equation}
Comparing Equation~\ref{eq:J_F} with Equation~\ref{eq:General J_F}, we identify
\begin{equation}
  \alpha_\textn{c} = \frac{\alpha_1}{2}, \quad \alpha_\textn{a} = 1 - \frac{\alpha_1}{2}.
\end{equation}
Given the value of $J_0$ at a given reference value $\hat{c}_{\textn{Cu}^{2+}}^\textn{ref}$ and denoting this value of $J_0$ as $J_0^\textn{ref}$, we can rewrite $J_0$ as
\begin{equation}
  J_0 = J_0^\textn{ref}\left(\frac{\hat{c}_{\textn{Cu}^{2+}}}{\hat{c}_{\textn{Cu}^{2+}}^\textn{ref}}\right)^{1-\frac{\alpha_1}{2}} = ek_0\hat{c}_{\textn{Cu}^{2+}}^{1-\frac{\alpha_1}{2}}, \quad k_0 = \frac{J_0^\textn{ref}}{e\left(\hat{c}_{\textn{Cu}^{2+}}^\textn{ref}\right)^{1-\frac{\alpha_1}{2}}},
\end{equation}
where $k_0$ is the overall reaction rate constant.

To compare the reaction rate with the diffusion rate, we define the Damkohler number $\textn{Da}$ as the ratio of these two rates. Taking Equation~\ref{eq:Conservation of charges across electrode/electrolyte interfaces} into consideration, the scale for the Faradaic current density can be estimated as $e\epsilon_\textn{p}k_0$ while the scale for the current density in the electrolyte due to diffusion and electromigration is set by the limiting current density $J_\limn$, which is given in Equation~\ref{eq:Limiting current density}. Therefore, the Damkohler number $\textn{Da}$ is given by the ratio of these two scales:
\begin{equation}
  \textn{Da} = \frac{e\epsilon_\textn{p}k_0}{J_\limn}.
\end{equation}
A large $\textn{Da}$, i.e., $\textn{Da} \gg 1$, means that the system is diffusion-limited while a small $\textn{Da}$, i.e., $\textn{Da} \ll 1$, means that the system is reaction-limited.

\subsection{Boundary conditions, constraints and initial conditions}\label{sec:Boundary conditions, constraints and initial conditions}

For this paper, because we are interested in a 1D model where the electrodes are located at the endpoints of the 1D domain, we prescribe boundary conditions only at these endpoints. For 2D and 3D models, we would need to prescribe appropriate boundary conditions at boundaries that are not electrode/electrolyte interfaces.

\subsubsection{Boundary conditions}\label{sec:Boundary conditions}

We denote the anode and cathode electric potentials as $\phi_\textn{e}^\textn{a,c}$. We arbitrarily choose the anode to be on the left end of the system and the cathode to be on the right end of the system and ground the anode at all times, i.e., $\phi_\textn{e}^\textn{a} = 0$.

We assume that the electrode/electrolyte interfaces are stationary. In reality, these interfaces move because of copper electrodeposition and electrodissolution, therefore we can relate their normal velocities to the normal current densities using mass conservation and the mass-average velocity of the liquid electrolyte is nonzero~\cite{sundstrom_morphological_1995,elezgaray_linear_1998,deen_analysis_2011}. Nonetheless, these velocities are usually negligible and will be ignored in this paper. Mass conservation of the inert anions implies that $\hat{n} \cdot F_-\left(r=r_\textn{m}^\textn{a,c}\right) = 0$. Conservation of charges across the electrode/electrolyte interfaces requires $\hat{n} \cdot J\left(r=r_\textn{m}^\textn{a,c}\right) = \epsilon_\textn{p}J_\textn{F}^\textn{a,c}$ as discussed in Section~\ref{sec:Transport in leaky membrane model}.

\subsubsection{Constraints from galvanostatic and potentiostatic conditions and linear sweep voltammetry}\label{sec:Constraints from galvanostatic and potentiostatic conditions and linear sweep voltammetry}

For galvanostatic conditions where we impose a current $I_\textn{applied}$ on the system, we require $\int \hat{n} \cdot J\left(r=r_\textn{m}^\textn{c}\right) \ud S^\textn{c} = \int -\hat{n} \cdot J\left(r=r_\textn{m}^\textn{a}\right) \ud S^\textn{a} = I_\textn{applied}$. For potentiostatic conditions where we impose an electric potential $V$ on the cathode, we set $\phi_\textn{e}^\textn{c} = V$. For linear sweep voltammetry (LSV) where we impose a linearly time-varying electric potential on the cathode, $\phi_\textn{e}^\textn{c} = \beta_\textn{LSV}t$ where $\beta_\textn{LSV}$ is the sweep rate.

\subsubsection{Initial conditions}\label{sec:Initial conditions for c_-, r_m^a and r_m^c}

Based on the discussion in Section~\ref{sec:Transport in leaky membrane model} about the neutral salt bulk concentration, we specify the initial condition for $c$ as $c\left(t=0\right) = c_0$, therefore $c_-\left(t=0\right) = \nu_-c_0 - \frac{\rho_\textn{s}+\left\lvert\rho_\textn{s}\right\rvert}{2z_-e} \equiv \beta_1$.

\section{Model implementation}

For all results, we specialize the model to one spatial dimension $x$. To numerically solve the steady state equations in Section~\ref{sec:Steady state current-voltage relations and overlimiting conductances}, we use MATLAB's $\mathtt{bvp4c}$ boundary value problem solver. The form of equations that is appropriate for use with the $\mathtt{bvp4c}$ solver is given in Section I of the Supplementary Material. We also provide all the necessary Jacobians to the $\mathtt{bvp4c}$ function to increase convergence rate; they are especially useful for the highly nonlinear Butler-Volmer boundary conditions. The expressions for these Jacobians are given in Section II of the Supplementary Material. For computing the semi-analytical steady state current-voltage relation for Butler-Volmer boundary conditions in Section~\ref{sec:Butler-Volmer boundary conditions at anode and cathode}, we use MATLAB's $\mathtt{fsolve}$ and $\mathtt{fzero}$ functions with default relative and absolute tolerances to invert the nonlinear algebraic Butler-Volmer equations.

In Section~\ref{sec:Copper electrodeposition and electrodissolution in AAO, CN and PE membranes}, to fit the experimental datasets with the steady state current-voltage relation for Butler-Volmer boundary conditions, we use MATLAB's $\mathtt{lsqnonlin}$ function to perform nonlinear least squares fitting. For the time-dependent linear sweep voltammetry (LSV) numerical simulations, we implement the model in COMSOL Multiphysics 5.3a, which uses the finite element method, by using the General Form PDE interface. After the numerical data are generated, we use MATLAB R2017b to postprocess and plot them. We also use MATLAB's $\mathtt{polyfit}$ function to estimate the experimental overlimiting conductances.

\section{Results}

\subsection{Limiting current density and limiting current}\label{sec:Limiting current density and limiting current}

We derive the limiting current density, which is denoted by $J_\limn$, and the limiting current, which is denoted by $I_\limn$. To do so, we assume the following: 1) $\rho_\textn{s} = 0$, 2) a 1D system at steady state $\left(\frac{\partial c_\pm}{\partial t} = 0\right)$ where the domain of the system is $x\in\left[0,L\right]$ and we arbitrarily choose the anode to be at $x=0$ and the cathode to be at $x=L$, and 3) ignore Faradaic reactions at both the anode and cathode. We assume that the anions are inert and cannot leave the system, therefore the boundary conditions for the system are $F_-\left(x=0\right) = F_-\left(x=L\right) = 0$. That the anions cannot leave the system also implies that the number of anions in the system is conserved, which is expressed by the integral constraint $\int_0^{L}c_-\ud x = \nu_-c_0L$ where $c_0$ is the neutral salt bulk concentration. Limiting current is attained when the concentrations of both the cations and anions vanish at the cathode. Therefore, using the boundary conditions and integral constraint and setting $c_-=0$ at $x=L$, we obtain  
\begin{equation}
  J_\limn = \frac{2z_+e\epsilon_\textn{p}D_{+0}\left(1 - \frac{z_-}{z_+}\right)\nu_-c_0}{L} = \frac{2\left(\nu_+ + \nu_-\right)z_+e\epsilon_\textn{p}D_{+0}c_0}{L}. \label{eq:Limiting current density}
\end{equation}
We note that $\left(\nu_+ +\nu_-\right)c_0$ is the sum of the concentrations of the cations and anions. Hence,
\begin{equation}
  I_\limn \equiv J_\limn A = \frac{2z_+e\epsilon_\textn{p}D_{+0}\left(1 - \frac{z_-}{z_+}\right)\nu_-c_0A}{L} = \frac{2\left(\nu_+ + \nu_-\right)z_+e\epsilon_\textn{p}D_{+0}c_0A}{L}
\end{equation}
where $A$ is the total surface area of the anode or cathode.

\subsection{Steady state current-voltage relations and overlimiting conductances}\label{sec:Steady state current-voltage relations and overlimiting conductances}

It is convenient to simplify the model at steady state in order to derive analytical or semi-analytical expressions for $c_-$, $\phi$, steady state current-voltage relation and overlimiting conductance that can be easily used for fitting experimental data. Overlimiting conductance is only defined for $\rho_\textn{s} < 0$ and not for $\rho_\textn{s} \geq 0$ because the system can exceed the limiting current, i.e., become overlimiting, only when $\rho_\textn{s} < 0$. We consider three types of boundary conditions that are commonly realized in experiments: 1) reservoir boundary condition at the anode, 2) no-anion-flux boundary condition at the anode, and 3) Butler-Volmer boundary conditions at the anode and cathode. To verify these analytical or semi-analytical expressions, we compare them with numerical solutions obtained from solving the equations using MATLAB's $\mathtt{bvp4c}$ boundary value problem solver.  

All parameters used in this section are given in Table~\ref{tab:Parameters for copper electrodeposition and electrodissolution for AAO membranes}. In this section, copper electrodeposition and electrodissolution occurs in AAO membranes containing copper(II) sulfate ($\textn{CuSO}_4$) as the electrolyte. For AAO membranes that have parallel straight cylindrical pores with the same length and a constant pore radius, the assumptions that $\epsilon_\textn{p}$, $\tau$, $a_\textn{p}$ and $h_\textn{p}$ are uniform and constant are reasonable. Denoting the pore radius as $r_\textn{p}$, we obtain $h_\textn{p} = \frac{\epsilon_\textn{p}}{a_\textn{p}} = \frac{r_\textn{p}}{2}$. The electrodes are circular with a radius $r_\textn{e}$, therefore $A=\pi r_\textn{e}^2$ where $A$ is the total surface area of the anode or cathode.  

\begingroup
\squeezetable  
\begin{table} \caption{Parameters for copper electrodeposition and electrodissolution for AAO membranes at $T = 298\,\textn{K}$ and copper(II) sulfate ($\textn{CuSO}_4$) electrolyte ($\nu_+ = 1, \nu_- = 1, z_+ = 2, z_- = -2$).}\label{tab:Parameters for copper electrodeposition and electrodissolution for AAO membranes}  
  \centering
  \begin{ruledtabular}
  \begin{tabular}{ccc}
    Parameter & Value & Notes and references \\ \hline  
    $E^\Theta$ & $0.3419\,\textn{V}$ & Ref.~\cite{haynes_crc_2016} \\
    $D_{+0}^\textn{m}$ & $7.14 \times 10^{-10}\,\textn{m}^2/\textn{s}$ & Ref.~\cite{haynes_crc_2016} \\
    $D_{-0}^\textn{m}$ & $1.065 \times 10^{-9}\,\textn{m}^2/\textn{s}$ & Ref.~\cite{haynes_crc_2016} \\
    $J_0^\textn{ref}$ & $ 2.9\,\textn{mA}/\textn{cm}^2$ & Mean of exchange current densities for E electrodes in Table 2 of~\cite{mattsson_galvanostatic_1959} \\
    $c_{\textn{Cu}^{2+}}^\textn{ref}$ & $75\,\textn{mM}$ & Ref.~\cite{mattsson_galvanostatic_1959} \\
    $\alpha_1$ & 0.75 & Compromise between $0.5$ in~\cite{mattsson_galvanostatic_1959} and $1.16$ in~\cite{newman_electrochemical_2004} \\
    $M_\textn{m}$ & $63.546\,\textn{g}/\textn{mol}$ & Ref.~\cite{haynes_crc_2016} \\
    $\rho_\textn{m}$ & $8.96\,\textn{g}/\textn{cm}^3$ & Ref.~\cite{haynes_crc_2016} \\
    $r_\textn{p}$ & $175\,\textn{nm}$ & Mean of product specification of $150\,\textn{nm}-200\,\textn{nm}$ \\
    $L$ & $60\,\mu\textn{m}$ & Product specification \\
    $\epsilon_\textn{p}$ & 0.375 & Mean of product specification of $0.25-0.50$ \\
    $r_\textn{e}$ & $6\,\textn{mm}$ & Product specification \\
    $\tau$ & 1 & Straight pores \\
    $c_{\textn{Cu}^{2+}}^\Theta$ & $1\,\textn{M} = 10^3\,\textn{mol}\,\textn{m}^{-3}$ & Standard concentration \\
  \end{tabular}
  \end{ruledtabular}
\end{table}
\endgroup

\subsubsection{Case 1: reservoir boundary condition at anode}\label{sec:Reservoir boundary condition at anode}

We make the same assumptions used to derive $J_\limn$ and $I_\limn$ in Section~\ref{sec:Limiting current density and limiting current} except that we assume $\rho_\textn{s}\neq0$. For the boundary conditions, we assume that there is a reservoir at $x=0$ and an ideal cation-selective and anion-blocking surface at $x=L$. We also set $\phi$ at the anode at $x=0$ to $0$ and $\phi$ at the cathode at $x=L$ to $-V$ where $V\geq0$ so that the current $I$ flows from $x=0$ to $x=L$. In summary, the boundary conditions are given by
\begin{align}
  c_-\left(x=0\right) &= \nu_-c_0 - \frac{\rho_\textn{s}+\left\lvert\rho_\textn{s}\right\rvert}{2z_-e} \equiv \beta_1, \label{eq:Reservoir boundary condition for anion concentration at anode} \\
  \phi\left(x=0\right) &= 0, \label{eq:Reservoir boundary condition for electrolyte electric potential at anode} \\
  F_-\left(x=L\right) &= 0, \label{eq:No anion flux at cathode} \\
  \phi\left(x=L\right) &= -V, \quad V \geq 0. \label{eq:Electrolyte electric potential at cathode}
\end{align}
Using these boundary conditions, we obtain
\begin{equation}
  I = JA = \frac{1}{2}I_\limn\frac{\beta_1}{\nu_-c_0}\left[1-\exp\left(\frac{z_-eV}{k_\textn{B}T}\right)\right] - \frac{z_+e\epsilon_\textn{p}D_{+0}\rho_\textn{s}A}{Lk_\textn{B}T}V. \label{eq:Steady state I-V relation for reservoir boundary condition at anode}
\end{equation}
Setting $\rho_\textn{s}=0$, we define the limiting current $I_\limn^\textn{reservoir}$ as
\begin{equation}
  I_\limn^\textn{reservoir} \equiv \lim_{V\rightarrow\infty}\lim_{\rho_\textn{s}\rightarrow0}I = \frac{1}{2}I_\limn. \label{eq:Limiting current for reservoir boundary condition at anode}
\end{equation}
For $\rho_\textn{s}<0$, Equation~\ref{eq:Steady state I-V relation for reservoir boundary condition at anode} shows that $I>I_\limn^\textn{reservoir}$ for sufficiently large values of $V$, i.e., the current $I$ becomes overlimiting. Therefore, for $\rho_\textn{s}<0$,
\begin{equation}
  \lim_{V\rightarrow\infty}I\left(\rho_\textn{s}<0\right) = I_\limn^\textn{reservoir} + \sigma_\textn{OLC}V \label{eq:I at large V for reservoir boundary condition at anode}
\end{equation}
where we define the overlimiting conductance $\sigma_\textn{OLC}$ as
\begin{equation}
  \sigma_\textn{OLC} \equiv -\frac{z_+e\epsilon_\textn{p}D_{+0}\rho_\textn{s}A}{Lk_\textn{B}T} = -\frac{z_+e\epsilon_\textn{p}D_{+0}^\textn{m}\sigma_\textn{s}A}{\tau Lk_\textn{B}Th_\textn{p}}, \quad \rho_\textn{s}<0. \label{eq:Overlimiting conductance for reservoir boundary condition at anode}
\end{equation}
Equation~\ref{eq:I at large V for reservoir boundary condition at anode} predicts that $I$ varies linearly with $V$ for a sufficiently large $V$ and the overlimiting conductance $\sigma_\textn{OLC}$ is the gradient of this linear relationship. For $\rho_\textn{s} \leq 0$, because $c_-$ can reach $0$, there are no restrictions on how large $V$ can be. For $\rho_\textn{s} > 0$, there is a finite maximum value of $V$, which is denoted by $V_\maxn^\textn{reservoir}$, for which the steady state $I$-$V$ relation is valid. The current that corresponds to $V_\maxn^\textn{reservoir}$ is denoted as $I_\maxn^\textn{reservoir}$. $V_\maxn^\textn{reservoir}$ is determined by setting $c_+ = 0$, or equivalently, $c_- = -\frac{\rho_\textn{s}}{z_-e}$, and $\phi = -V_\maxn^\textn{reservoir}$ at $x = L$:
\begin{equation}
  V_\maxn^\textn{reservoir} = \frac{k_\textn{B}T}{z_-e}\ln\left(\frac{\rho_\textn{s}}{-z_-e\nu_-c_0+\rho_\textn{s}}\right), \quad \rho_\textn{s} > 0.
\end{equation}
Because $I_\maxn^\textn{reservoir} < I_\limn^\textn{reservoir}$, the presence of a positive background charge effectively reduces the diffusion-limited current, which is defined in Equation~\ref{eq:Limiting current for reservoir boundary condition at anode} for $\rho_\textn{s} = 0$.

We define the following nondimensionalization to make the equations more compact: $\tilde{x} \equiv \frac{x}{L}$, $\tilde{c}_\pm \equiv \frac{c_\pm}{\nu_\pm c_0}$, $\tilde{\beta}_1 \equiv \frac{\beta_1}{\nu_-c_0} = 1 + \frac{\tilde{\rho}_\textn{s}+\left\lvert\tilde{\rho}_\textn{s}\right\rvert}{2}$, $\tilde{\phi} \equiv \frac{e\phi}{k_\textn{B}T}$, $\tilde{D}_{\pm 0} \equiv \frac{D_{\pm 0}}{D_{\textn{amb}0}}$, $\tilde{I} \equiv \frac{I}{I_\limn}$ and $\tilde{\rho}_\textn{s} \equiv \frac{\rho_\textn{s}}{z_+\nu_+ec_0} = -\frac{\rho_\textn{s}}{z_-\nu_-ec_0}$ where $D_{\textn{amb}0}$ is the ambipolar diffusivity of the neutral salt in the dilute limit and is given by $D_{\textn{amb}0} = \frac{\left(z_+ - z_-\right)D_{+0}D_{-0}}{z_+D_{+0}-z_-D_{-0}}$~\cite{newman_electrochemical_2004}. Therefore,
\begin{align}
  \tilde{I}_\limn^\textn{reservoir} &= \frac{1}{2}, \\
  \tilde{c}_- &= \tilde{\beta}_1\exp\left(-z_-\tilde{\phi}\right), \\
  \tilde{I}\tilde{x} &= \frac{1}{2}\left[\tilde{\beta}_1 - \tilde{c}_- + \frac{z_+\tilde{\rho}_\textn{s}}{z_+-z_-}\ln\left(\frac{\tilde{c}_-}{\tilde{\beta}_1}\right)\right] = \frac{1}{2}\left \{\tilde{\beta}_1\left[1-\exp\left(-z_-\tilde{\phi}\right)\right] - \frac{z_+z_-\tilde{\rho}_\textn{s}}{z_+-z_-}\tilde{\phi}\right \}, \\
  \tilde{I} &= \frac{1}{2}\left \{\tilde{\beta}_1\left[1-\exp\left(z_-\tilde{V}\right)\right] + \frac{z_+z_-\tilde{\rho}_\textn{s}}{z_+-z_-}\tilde{V}\right \}. \label{eq:Dimensionless steady state I-V relation for reservoir boundary condition at anode}
\end{align}
It is possible to express $\tilde{c}_-$ and $\tilde{\phi}$ as explicit functions of $\tilde{x}$. We first define $\tilde{\alpha}_1 \equiv \frac{z_+\tilde{\rho}_\textn{s}}{z_+-z_-}$ and $\tilde{\alpha}_2 \equiv z_-\tilde{\alpha}_1$. If $\tilde{\rho}_\textn{s} = 0$, then
\begin{align}
  \tilde{c}_- &= 1 - 2\tilde{I}\tilde{x}, \\
  \tilde{\phi} &= -\frac{1}{z_-}\ln\tilde{c}_- = -\frac{1}{z_-}\ln\left(1 - 2\tilde{I}\tilde{x}\right). \label{eq:Electrolyte electric potential as a function of x for rho_s = 0 for reservoir boundary condition at anode}
\end{align}
If $\tilde{\rho}_\textn{s} \neq 0$, we use the Lambert W function~\cite{corless_lambertw_1996}, which is denoted as $W\left(\cdot\right)$. Given the form of Equation~\ref{eq:Dimensionless steady state I-V relation for reservoir boundary condition at anode}, for $\tilde{\rho}_\textn{s} < 0$, we can interpret the bulk electrolyte as a diode that is connected in parallel to the electric double layers on the pore surfaces that act as a shunt resistor to conduct OLC via surface conduction in regions where the anions are depleted~\cite{dydek_overlimiting_2011,dydek_nonlinear_2013}. It is therefore not surprising that the Lambert W function is applicable here as it is invoked in describing current flow through a diode with series resistance under an applied voltage~\cite{banwell_exact_2000}, in describing current in solar cells with series and shunt resistances under an applied voltage~\cite{jain_exact_2004}, and in ion transport problems in ion-exchange membranes~\cite{a.moya_theory_2015,sistat_steady-state_1999,oliveira_steady_2016} and electrolysis cells~\cite{pohjoranta_copper_2010}. For physically valid concentration and electric potential profiles, we restrict the Lambert W function and its argument to be real-valued. In this case, the function consists of two branches that are denoted by $W_0$ and $W_{-1}$~\cite{corless_lambertw_1996}. Because we require $\tilde{c}_-$ to be nonnegative, we must use the $W_0$ branch for $\tilde{\rho}_\textn{s} < 0$ and the $W_{-1}$ branch for $\tilde{\rho}_\textn{s} > 0$. We will leave out the subscript in $W\left(\cdot\right)$ and the appropriate branch to be used is implied by the sign of $\tilde{\rho}_\textn{s}$. Therefore,
\begin{align}
  \tilde{c}_- &= -\tilde{\alpha}_1 W\left[-\frac{\tilde{\beta}_1}{\tilde{\alpha}_1}\exp\left(\frac{2\tilde{I}\tilde{x}-\tilde{\beta}_1}{\tilde{\alpha}_1}\right)\right], \\
  \tilde{\phi} &= -\frac{\tilde{\alpha}_1}{\tilde{\alpha}_2}\ln\left(\frac{\tilde{c}_-}{\tilde{\beta}_1}\right) = -\frac{\tilde{c}_- + 2\tilde{I}\tilde{x} - \tilde{\beta}_1}{\tilde{\alpha}_2}. \label{eq:Electrolyte electric potential as a function of x for rho_s < 0 for reservoir boundary condition at anode}
\end{align}
Noting that $\tilde{\phi}\left(\tilde{x}=1\right) = -\tilde{V}$, we can evaluate Equations~\ref{eq:Electrolyte electric potential as a function of x for rho_s = 0 for reservoir boundary condition at anode} and~\ref{eq:Electrolyte electric potential as a function of x for rho_s < 0 for reservoir boundary condition at anode} at $\tilde{x} = 1$ to express $\tilde{V}$ as a function of $\tilde{I}$:
\begin{align}
  \tilde{V} &=
  \begin{cases}
    \frac{1}{z_-}\ln\left(1-2\tilde{I}\right), & \tilde{\rho}_\textn{s} = 0 \\
    \frac{-\tilde{\alpha}_1W\left[-\frac{\tilde{\beta}_1}{\tilde{\alpha}_1}\exp\left(\frac{2\tilde{I}-\tilde{\beta}_1}{\tilde{\alpha}_1}\right)\right] + 2\tilde{I} - \tilde{\beta}_1}{\tilde{\alpha}_2}, & \tilde{\rho}_\textn{s} \neq 0.
  \end{cases}
\end{align}

\begin{figure}
  \centering
  \includegraphics[scale=0.6]{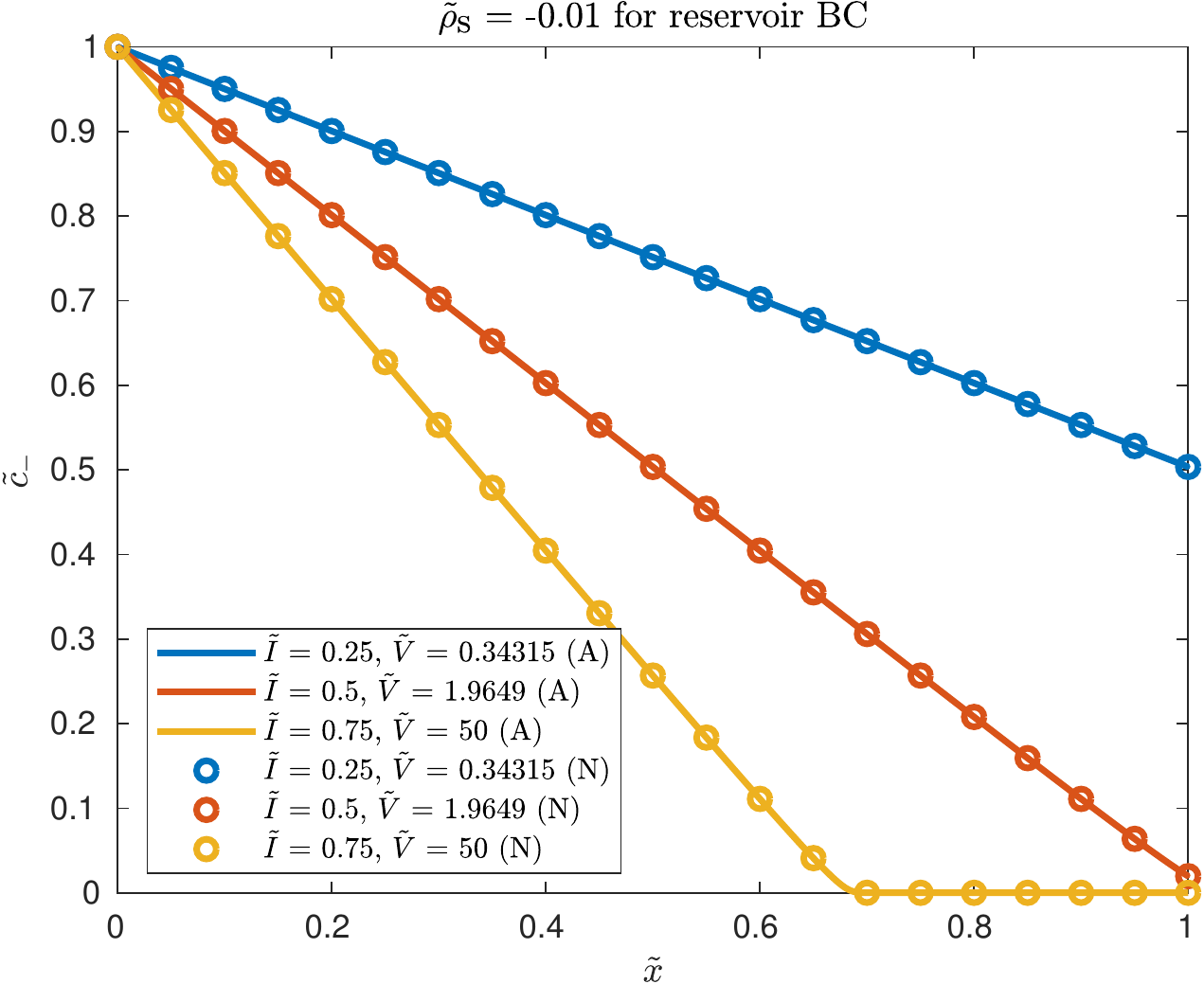}
  \includegraphics[scale=0.6]{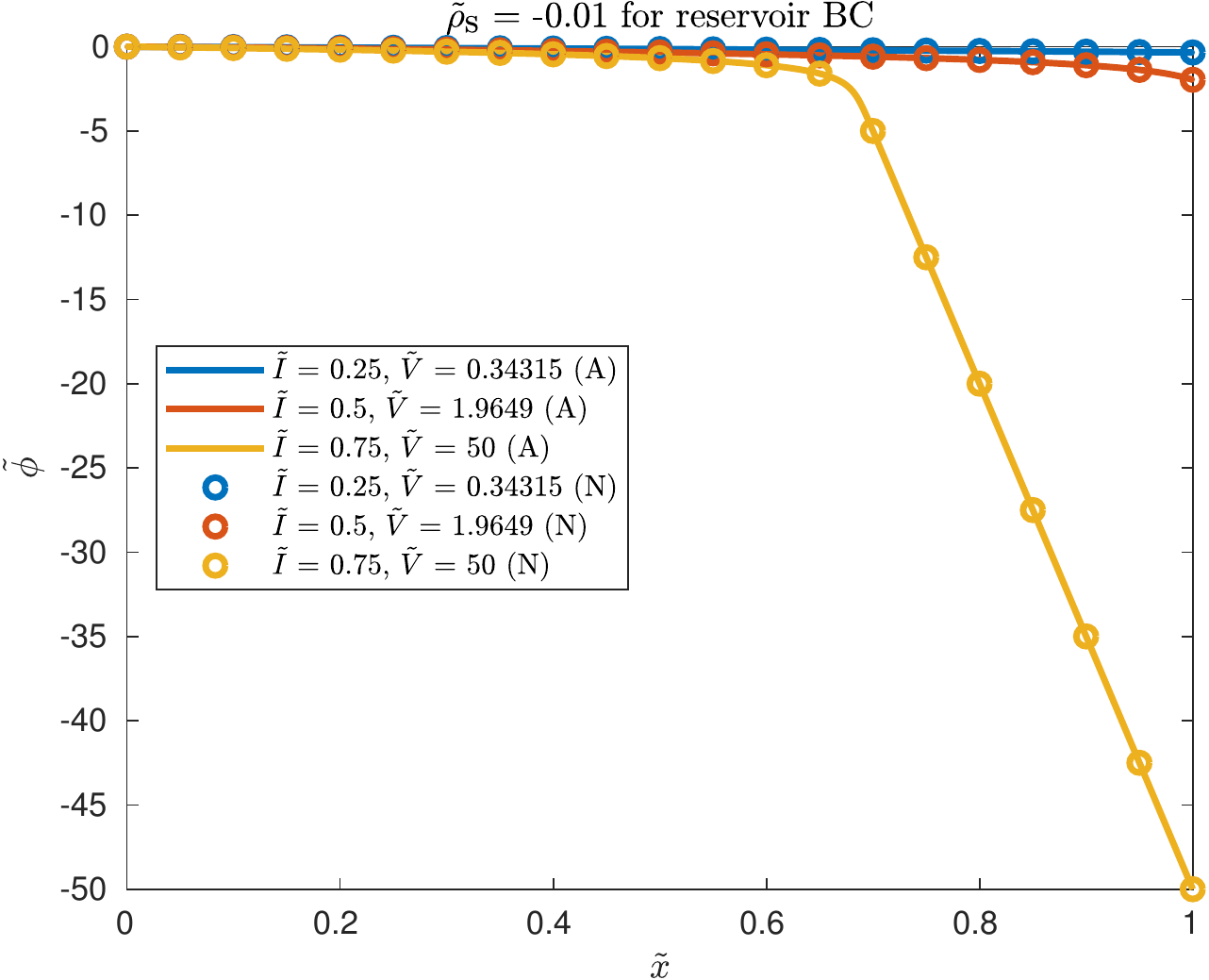}
  \includegraphics[scale=0.6]{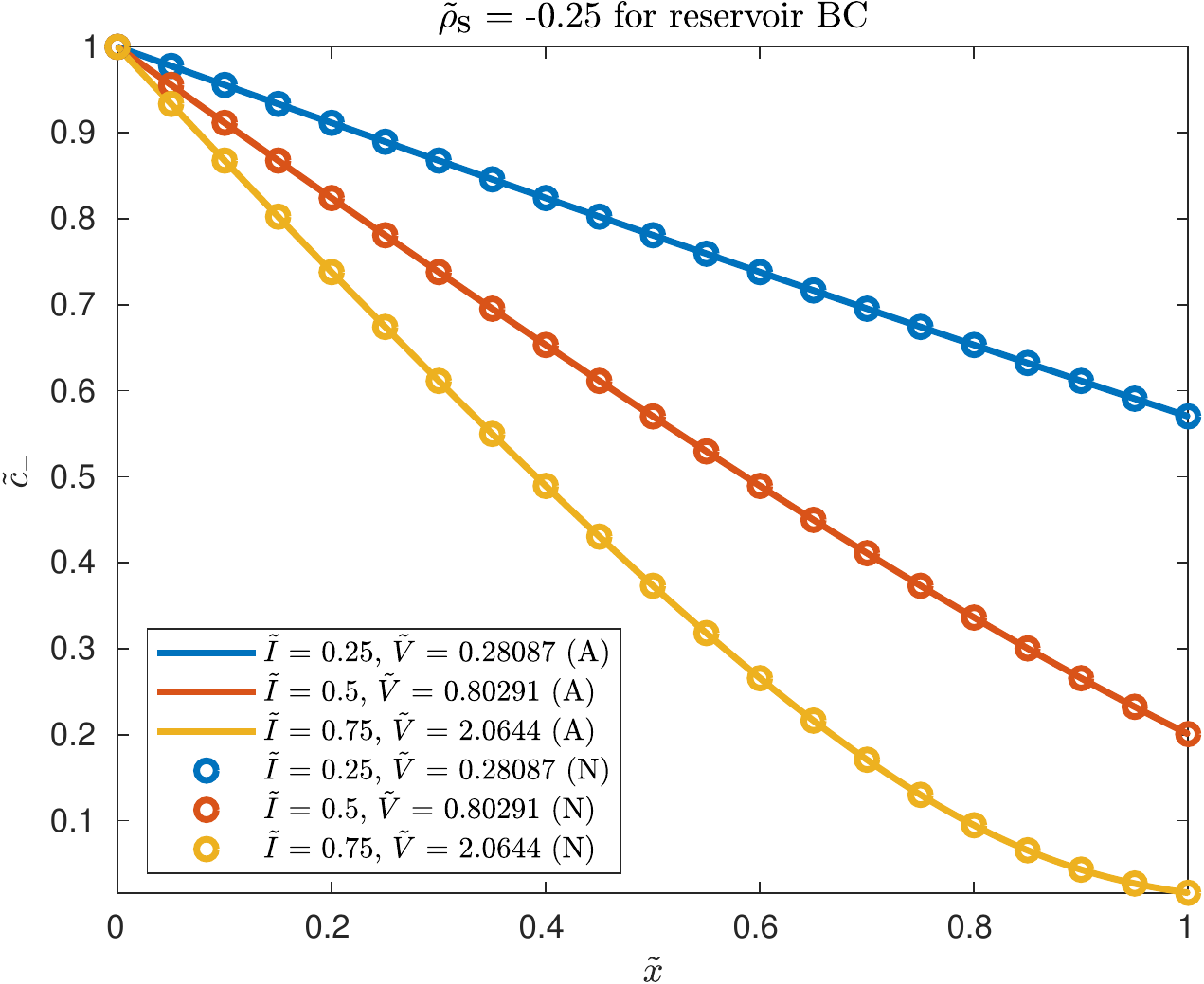}
  \includegraphics[scale=0.6]{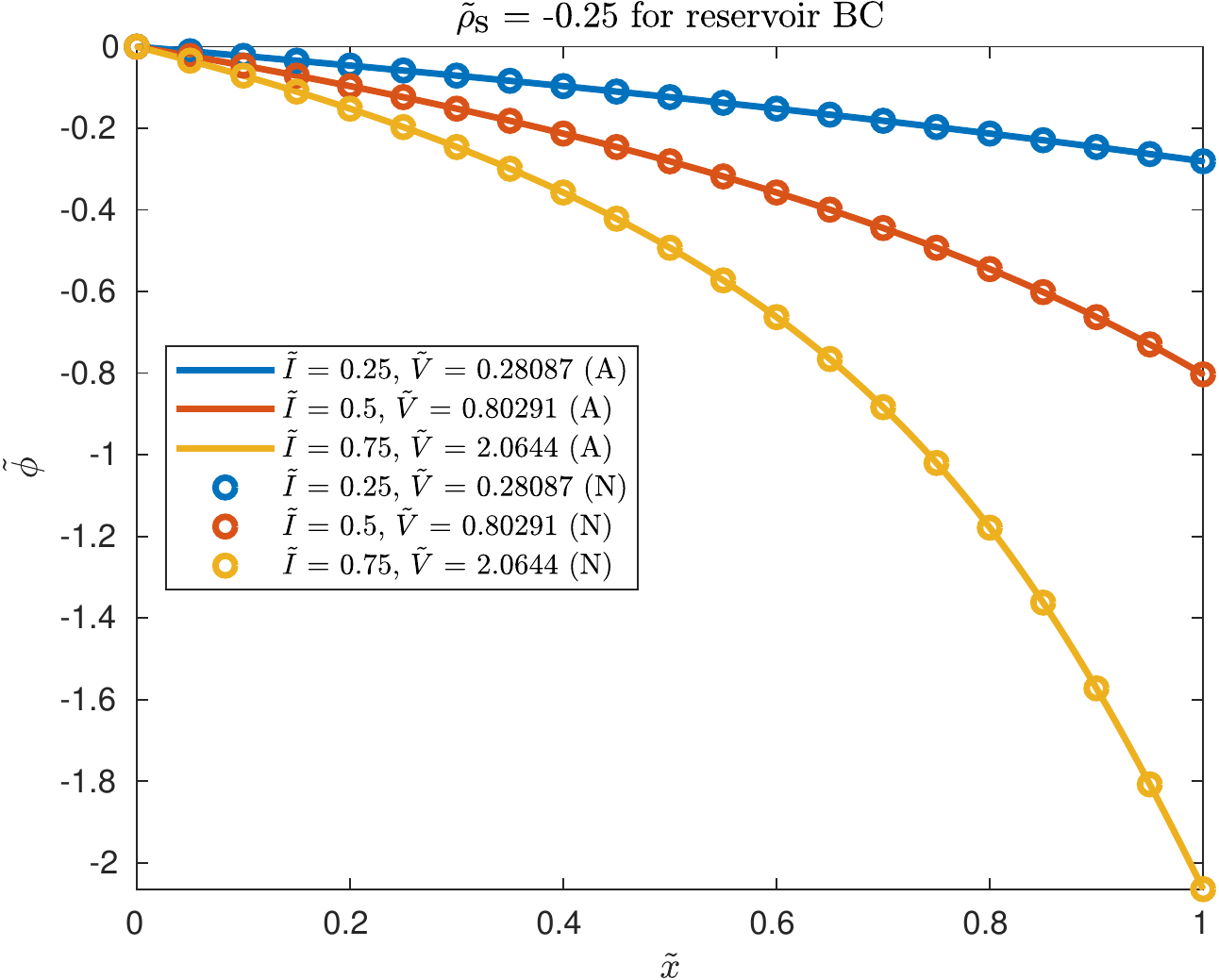}
  \caption{Plots of $\tilde{c}_-$ and $\tilde{\phi}$ against $\tilde{x}$ for $\tilde{\rho}_\textn{s} = -0.01, -0.25$ and $\tilde{I} = 0.25, 0.5, 0.75$ for reservoir boundary condition at anode. (A) refers to analytical solutions and (N) refers to numerical solutions.}\label{fig:c_- and phi against x for rho_s < 0 for reservoir boundary condition at anode}
\end{figure}

\begin{figure}
  \centering
  \includegraphics[scale=0.6]{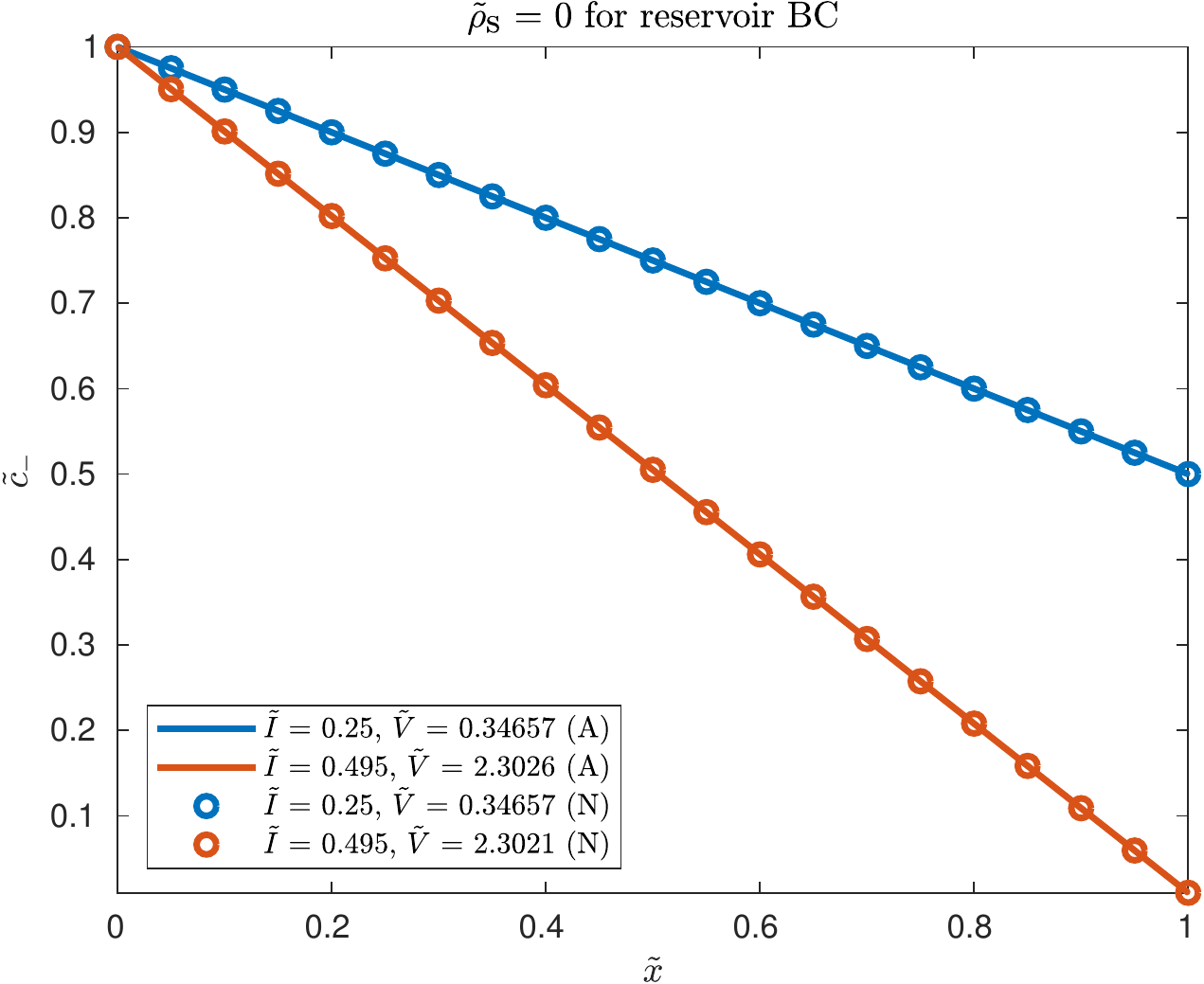}
  \includegraphics[scale=0.6]{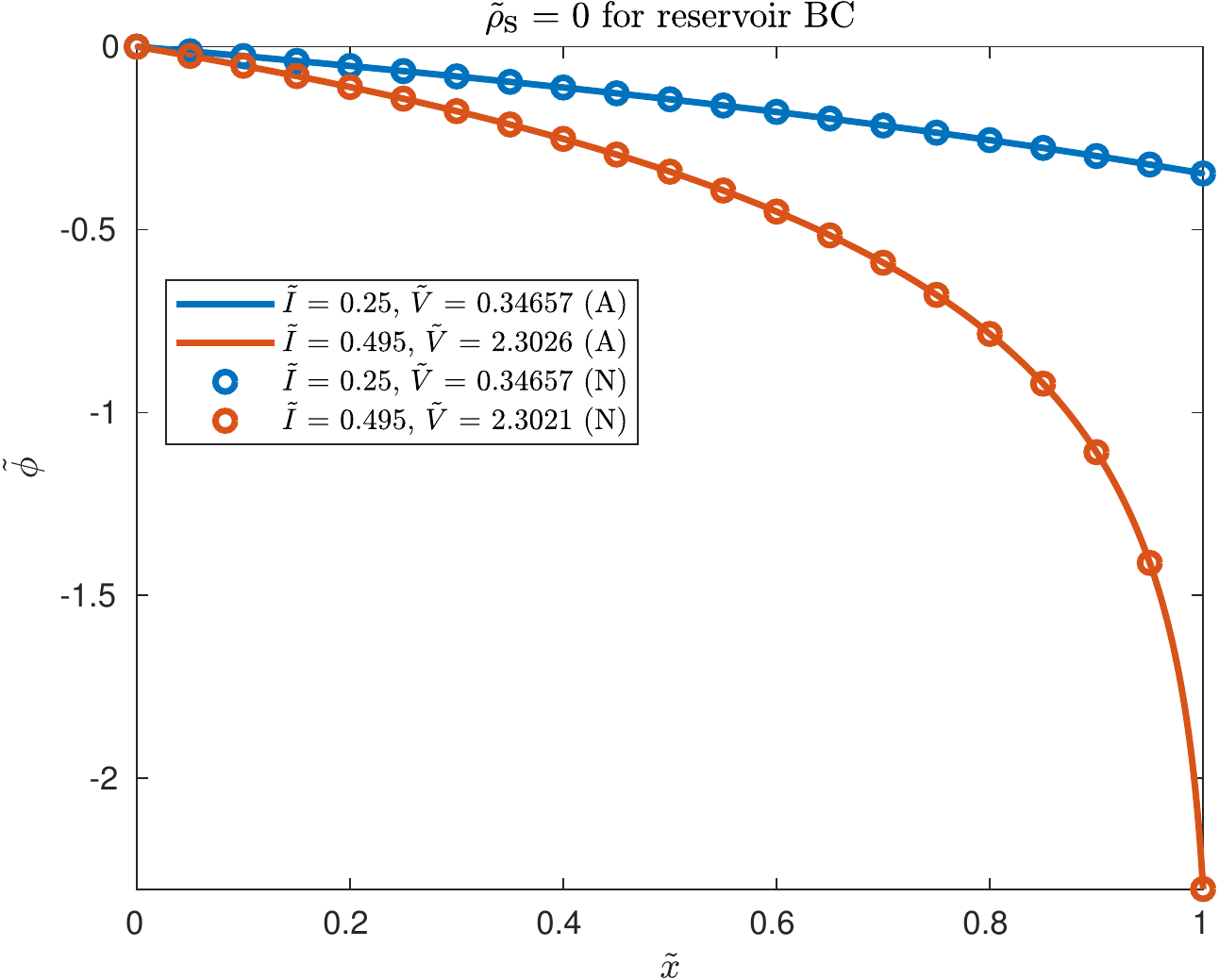}
  \includegraphics[scale=0.6]{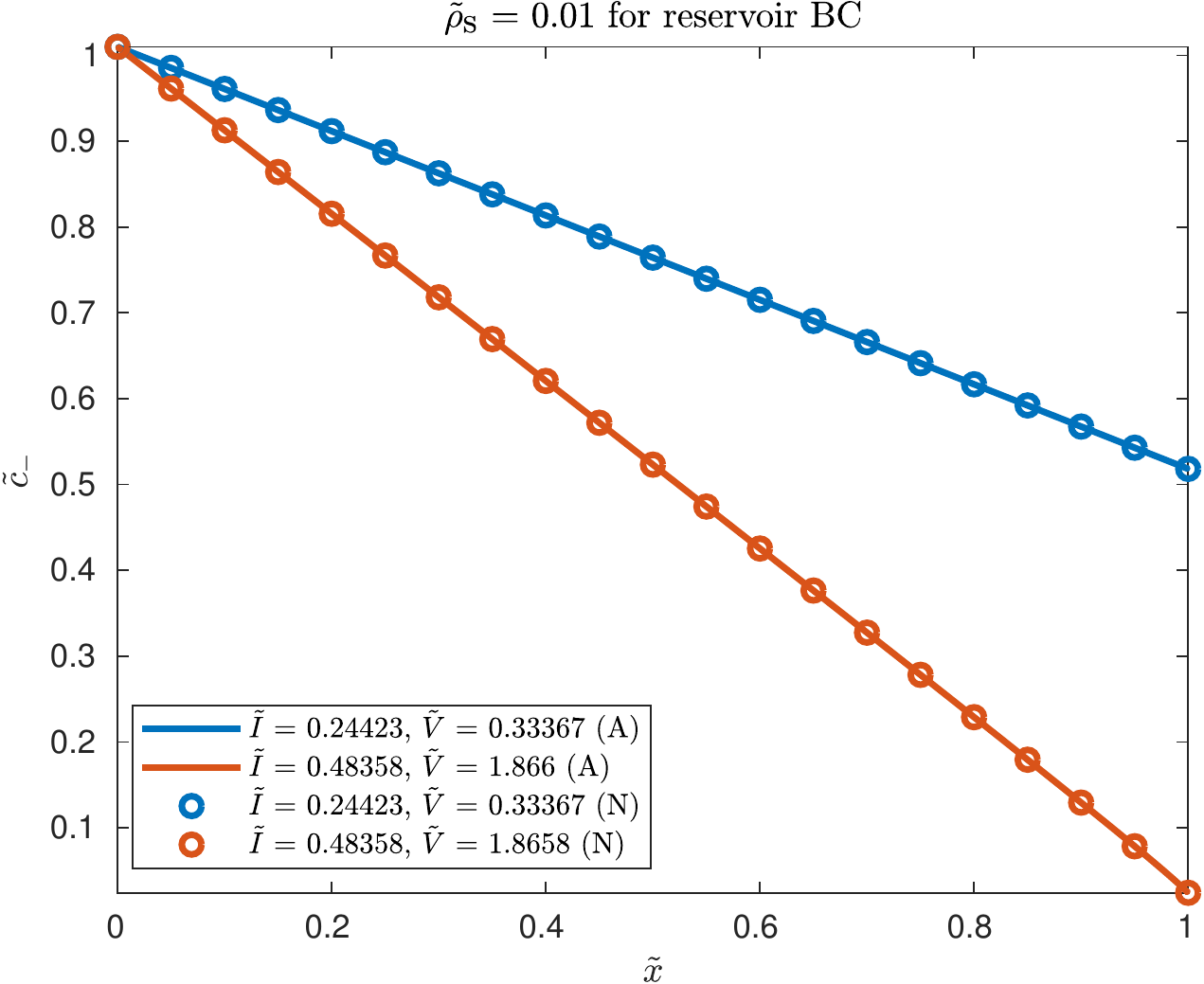}
  \includegraphics[scale=0.6]{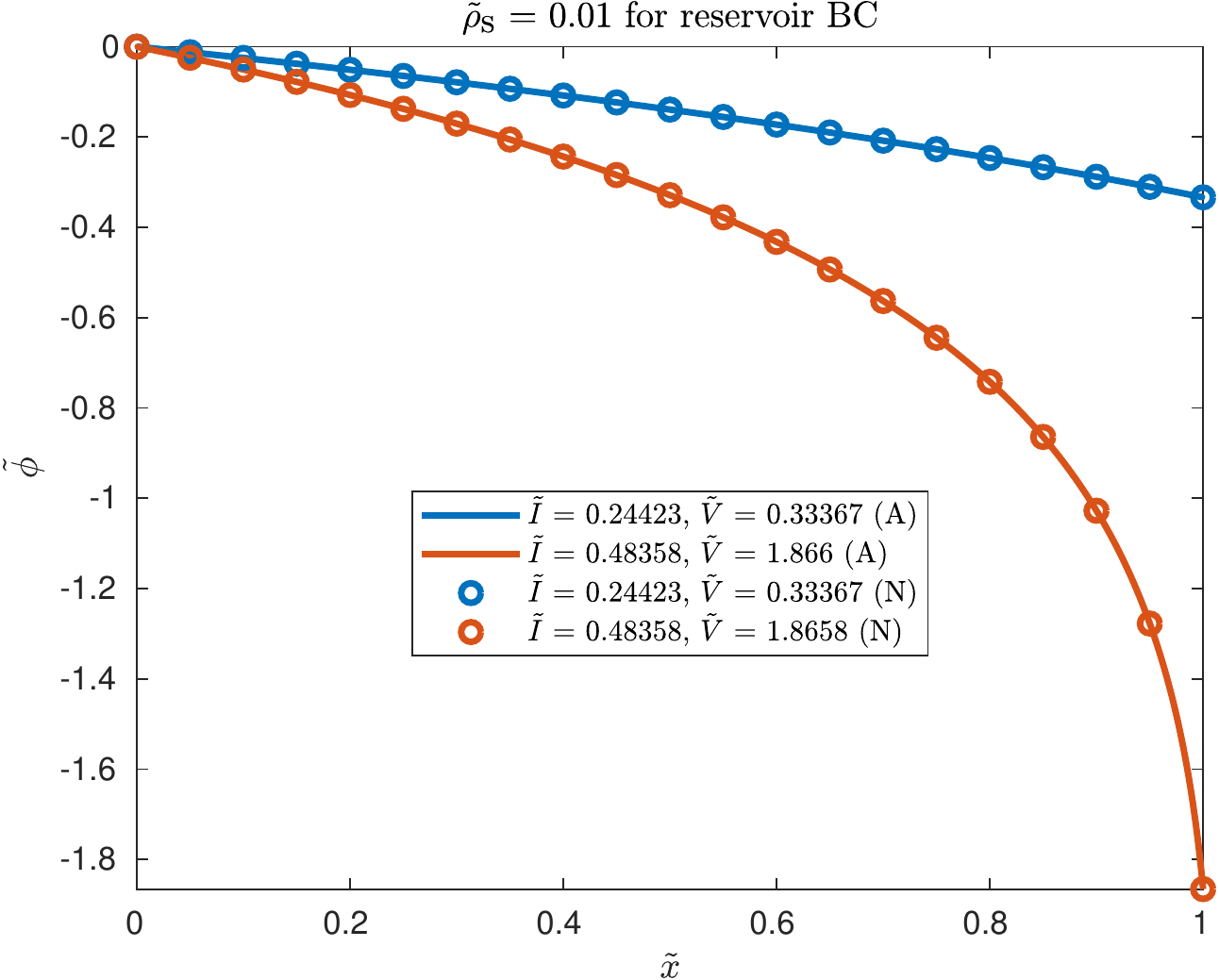}
  \includegraphics[scale=0.6]{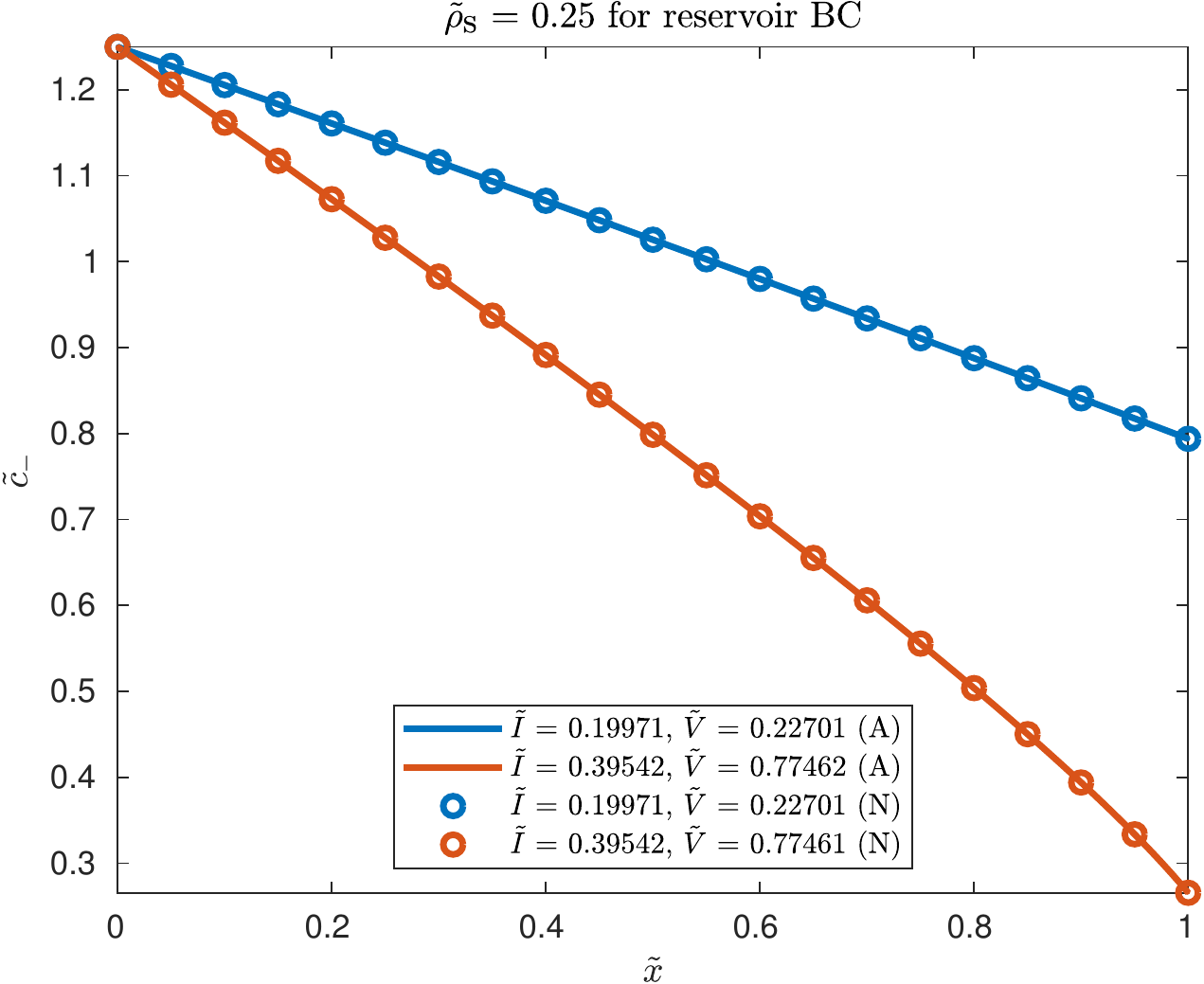}
  \includegraphics[scale=0.6]{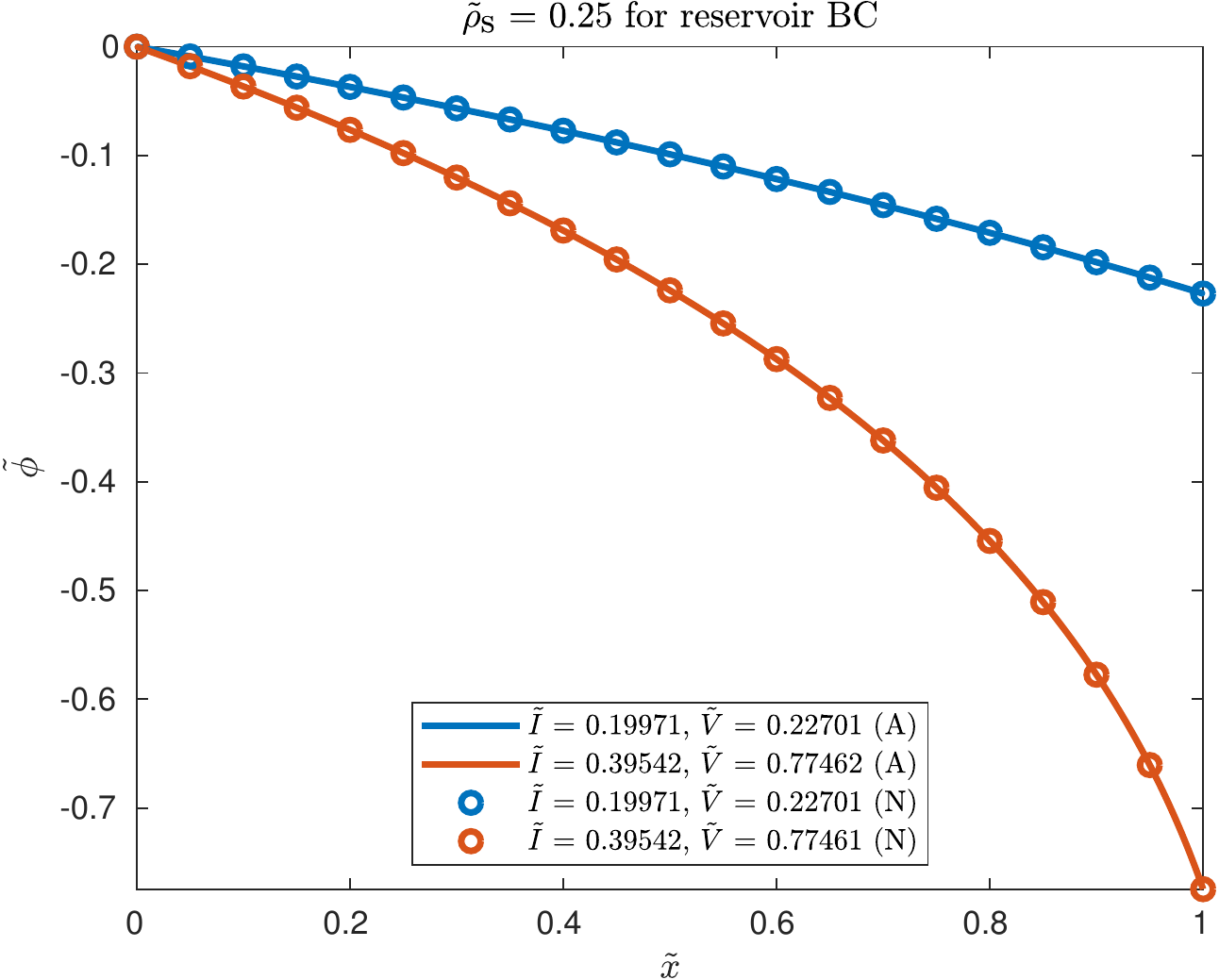}
  \caption{Plots of $\tilde{c}_-$ and $\tilde{\phi}$ against $\tilde{x}$ for 1) $\tilde{\rho}_\textn{s} = 0$ and $\tilde{I} = 0.25, 0.495$ (top row) and 2) $\tilde{\rho}_\textn{s} = 0.01, 0.25$ and $\tilde{I} = 0.5\tilde{I}_\maxn^\textn{reservoir}, 0.99\tilde{I}_\maxn^\textn{reservoir}$ (second and third rows) for reservoir boundary condition at anode. (A) refers to analytical solutions and (N) refers to numerical solutions.}\label{fig:c_- and phi against x for rho_s >= 0 for reservoir boundary condition at anode}  
\end{figure}

\begin{figure}
  \centering
  \includegraphics[scale=0.6]{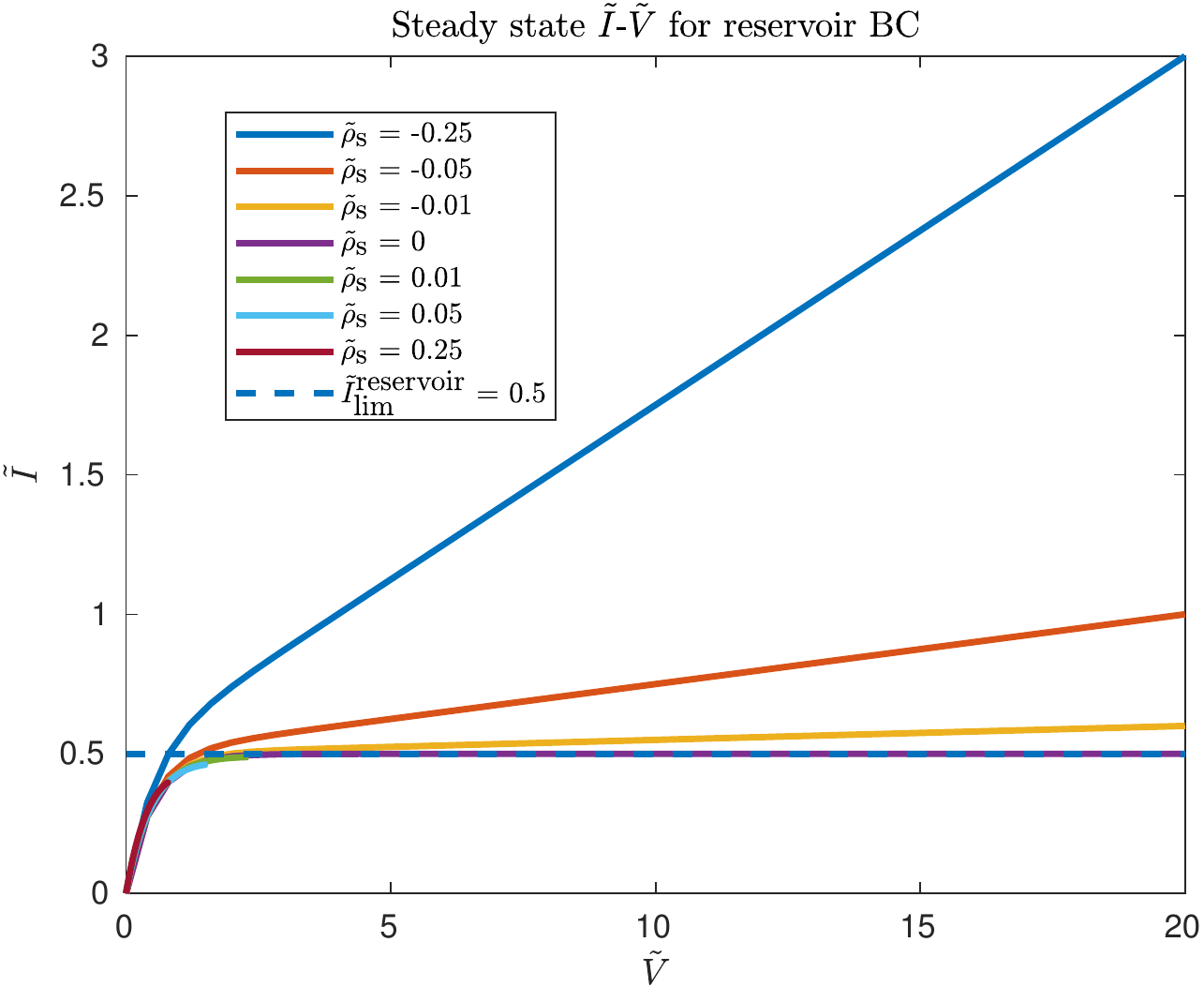}
  \includegraphics[scale=0.6]{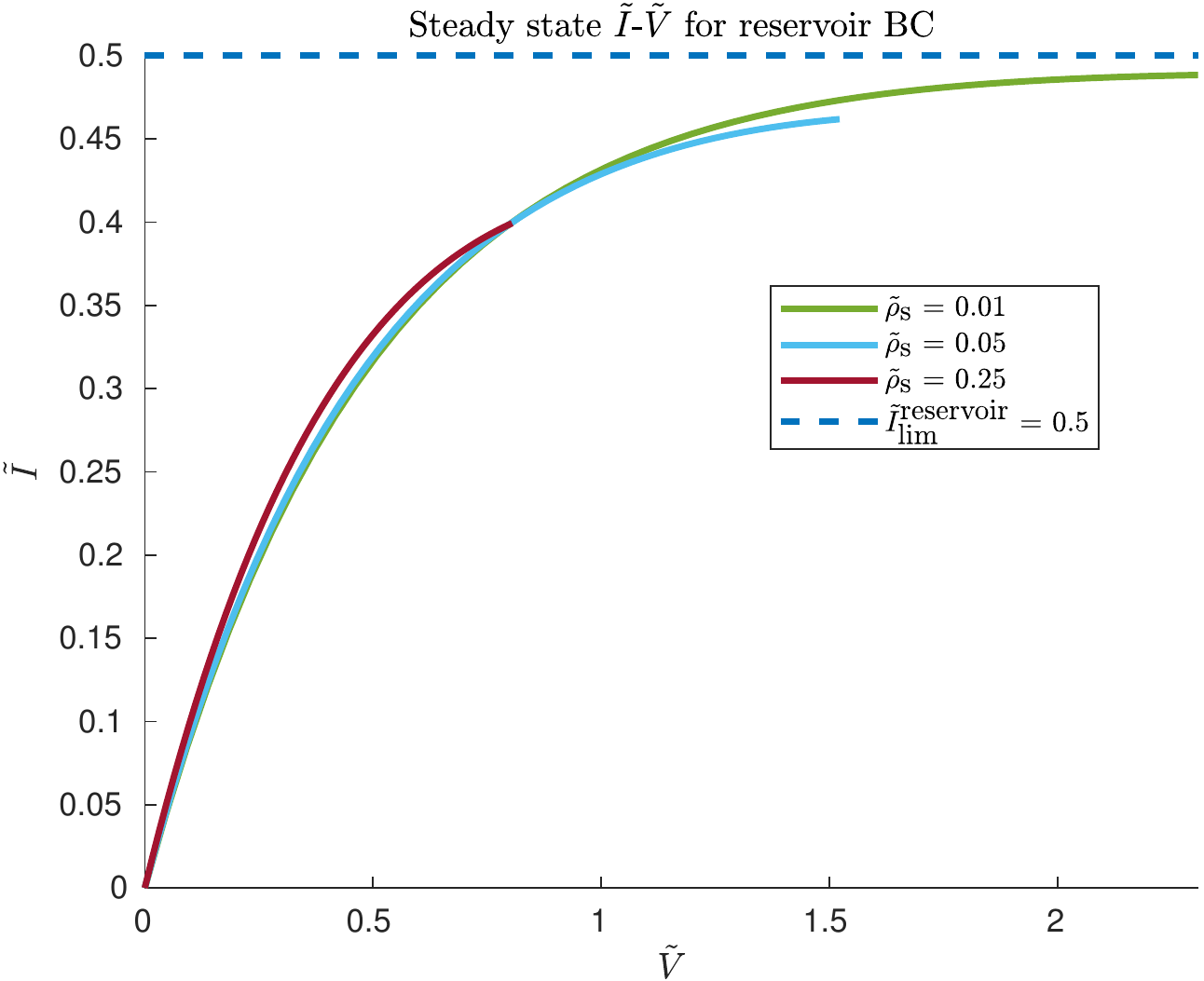}
  \caption{Left: steady state $\tilde{I}$-$\tilde{V}$ relations for $\tilde{\rho}_\textn{s} = 0, \pm 0.01, \pm 0.05, \pm 0.25$ for reservoir boundary condition at anode. Right: zoom-in view of left plot for only $\tilde{\rho}_\textn{s} = 0.01, 0.05, 0.25$. The dashed line denotes $\tilde{I}_\limn^\textn{reservoir} = 0.5$, which is the maximum $\tilde{I}$ that the system can reach when $\tilde{\rho}_\textn{s} = 0$.}\label{fig:Steady state I-V relations for reservoir boundary condition at anode}
\end{figure}

For both analytical expressions and numerical solutions, we plot $\tilde{c}_-$ and $\tilde{\phi}$ as functions of $\tilde{x}$ for $\tilde{\rho}_\textn{s} = -0.01, -0.25$ in Figure~\ref{fig:c_- and phi against x for rho_s < 0 for reservoir boundary condition at anode} and $\tilde{\rho}_\textn{s} = 0, 0.01, 0.25$ in Figure~\ref{fig:c_- and phi against x for rho_s >= 0 for reservoir boundary condition at anode}. For $\tilde{\rho}_\textn{s} = 0$, we choose $\tilde{I} = 0.25, 0.495$ and avoid $\tilde{I} = \tilde{I}_\limn^\textn{reservoir} = 0.5$. This is because $\tilde{I} = \tilde{I}_\limn^\textn{reservoir} = 0.5$ implies $\tilde{V} \rightarrow \infty$, which cannot be displayed exactly in $\tilde{\phi}$-$\tilde{x}$ plots and also cannot be attained in numerical simulations. For $\tilde{\rho}_\textn{s} = 0.01, 0.25$, we choose $\tilde{I} = 0.5\tilde{I}_\maxn^\textn{reservoir}, 0.99\tilde{I}_\maxn^\textn{reservoir}$ and for $\tilde{\rho}_\textn{s} = -0.01, -0.25$, we choose $\tilde{I} = 0.25, 0.5, 0.75$. We first observe that the analytical expressions agree very well with the numerical solutions, thus verifying that the analytical expressions are correct. Regardless of $\tilde{\rho}_\textn{s}$, when current is either underlimiting ($\tilde{I} = 0.25, 0.495, 0.5\tilde{I}_\maxn^\textn{reservoir}, 0.99\tilde{I}_\maxn^\textn{reservoir}$) or limiting ($\tilde{I} = \tilde{I}_\limn^\textn{reservoir} = 0.5$), $\tilde{c}_-$ is approximately linear in $\tilde{x}$, which is expected because the dominant physics at work is ambipolar diffusion. When current is overlimiting ($\tilde{I} = 0.75$), for small $\left\lvert\tilde{\rho}_\textn{s}\right\rvert$ values such as $\tilde{\rho}_\textn{s} = -0.01$, anions are depleted near and beyond the cathode and the depletion region extends for a finite distance from the cathode into the electrolyte. In the depletion region, $\tilde{\phi}$ is linear in $\tilde{x}$, which implies that electromigration under a constant electric field, i.e., surface conduction, is responsible for carrying current in this region. In contrast, because the concentration gradient is almost zero, diffusion only carries a negligible portion of the current. We also plot $\tilde{I}$ against $\tilde{V}$ for $\tilde{\rho}_\textn{s} = 0, \pm 0.01, \pm 0.05, \pm 0.25$ and $\tilde{V} \in \left[0,20\right]$ in Figure~\ref{fig:Steady state I-V relations for reservoir boundary condition at anode}. For $\tilde{\rho}_\textn{s} = 0$, $\tilde{I}$ asymptotically approaches $\tilde{I}_\limn^\textn{reservoir}$ as expected. For $\tilde{\rho}_\textn{s} = -0.01, -0.05, -0.25$, $\tilde{I}$ eventually becomes larger than $\tilde{I}_\limn^\textn{reservoir}$ at a sufficiently large $\tilde{V}$ and $\tilde{I}$ becomes a linear function of $\tilde{V}$ whose gradient gives the overlimiting conductance. On the other hand, for $\tilde{\rho}_\textn{s} = 0.01, 0.05, 0.25$, the right plot in Figure~\ref{fig:Steady state I-V relations for reservoir boundary condition at anode} illustrates that having a positive background charge imposes a finite maximum voltage, which corresponds to a finite maximum current that is smaller than the limiting current $\tilde{I}_\limn^\textn{reservoir} = 0.5$.

\subsubsection{Case 2: no-anion-flux boundary condition at anode}\label{sec:No-anion-flux boundary condition at anode}

We repeat the analysis done in Section~\ref{sec:Reservoir boundary condition at anode} except that we replace the boundary condition for $c_-$ given by Equation~\ref{eq:Reservoir boundary condition for anion concentration at anode} with $F_-\left(x=0\right) = 0$ such that the anode is also an ideal cation-selective and anion-blocking surface. Because the anions cannot leave the system, the number of anions in the system is conserved, which is expressed by the integral constraint $\int_0^{L}c_-\ud x = \beta_1L$. Using the boundary conditions and integral constraint, we obtain
\begin{align}
  I &= JA = \frac{1}{2}I_\limn\tilde{\alpha}_3\left[1-\exp\left(z_-\tilde{V}\right)\right]- \frac{z_+e\epsilon_\textn{p}D_{+0}\rho_\textn{s}A}{Lk_\textn{B}T}V, \label{eq:Steady state I-V relation for no-anion-flux boundary condition at anode} \\
  \tilde{\alpha}_3 &= \frac{-\tilde{A_2} + \sqrt{\tilde{A}_2^2-4\tilde{A_1}\tilde{A_3}}}{2\tilde{A_1}}, \label{eq:Dimensionless alpha_3} \\
  \tilde{A_1} &\equiv \frac{1}{2}\left(1-\frac{z_-}{z_+}\right)\left[1 - \exp\left(2z_-\tilde{V}\right)\right], \\
  \tilde{A_2} &\equiv -\left[\tilde{\rho}_\textn{s}+\left(1-\frac{z_-}{z_+}\right)\tilde{\beta}_1\right]\left[1-\exp\left(z_-\tilde{V}\right)\right], \\
  \tilde{A_3} &\equiv -z_-\tilde{\beta}_1\tilde{\rho}_\textn{s}\tilde{V}.
\end{align}
$\tilde{\alpha}_3$ is obtained by solving the quadratic equation $\tilde{A}_1\tilde{\alpha}_3^2 + \tilde{A}_2\tilde{\alpha}_3 + \tilde{A}_3 = 0$ and keeping only the positive root because we require physically valid concentration and electric potential profiles. Setting $\rho_\textn{s} = 0$, we define the limiting current $I_\limn^\textn{NAF}$ as
\begin{equation}
  I_\limn^\textn{NAF} \equiv \lim_{V\rightarrow\infty}\lim_{\rho_\textn{s}\rightarrow0}I = I_\limn \label{eq:Limiting current for no-anion-flux boundary condition at anode}
\end{equation}
where the ``NAF'' superscript denotes no anion flux. For $\rho_\textn{s} < 0$, Equation~\ref{eq:Steady state I-V relation for no-anion-flux boundary condition at anode} shows that $I > I_\limn^\textn{NAF}$ for sufficiently large values of $V$, i.e., the current $I$ becomes overlimiting. Therefore, for $\rho_\textn{s} < 0$,
\begin{equation}
  \lim_{V\rightarrow\infty}I\left(\rho_\textn{s}<0\right) = \frac{1}{2}I_\limn^\textn{NAF}\frac{\left(\tilde{\rho}_\textn{s}+1-\frac{z_-}{z_+}\right) + \sqrt{\left(\tilde{\rho}_\textn{s}+1-\frac{z_-}{z_+}\right)^2+2\left(1-\frac{z_-}{z_+}\right)z_-\tilde{\rho}_\textn{s}\tilde{V}}}{1-\frac{z_-}{z_+}} + \sigma_\textn{OLC}V \label{eq:I at large V for no-anion-flux boundary condition at anode}
\end{equation}
where we define the overlimiting conductance $\sigma_\textn{OLC}$ as
\begin{equation}
  \sigma_\textn{OLC} \equiv -\frac{z_+e\epsilon_\textn{p}D_{+0}\rho_\textn{s}A}{Lk_\textn{B}T} = -\frac{z_+e\epsilon_\textn{p}D_{+0}^\textn{m}\sigma_\textn{s}A}{\tau Lk_\textn{B}Th_\textn{p}}, \quad \rho_\textn{s}<0. \label{eq:Overlimiting conductance for no-anion-flux boundary condition at anode}
\end{equation}
Comparing Equations~\ref{eq:Overlimiting conductance for no-anion-flux boundary condition at anode} and~\ref{eq:Overlimiting conductance for reservoir boundary condition at anode}, even though the boundary conditions for cases 1 and 2 differ, both cases have the same expression for overlimiting conductance. Equation~\ref{eq:I at large V for no-anion-flux boundary condition at anode} predicts that for a sufficiently large $V$, the $V$ term dominates the $\sqrt{V}$ term and $I$ varies linearly with $V$ and the overlimiting conductance $\sigma_\textn{OLC}$ is the gradient of this linear relationship. Like in case 1, there are no restrictions on how large $V$ can be for $\rho_\textn{s} \leq 0$ but there is a finite maximum value of $V$, which is denoted by $V_\maxn^\textn{NAF}$, for $\rho_\textn{s} > 0$ for which the steady state $I$-$V$ relation is valid. The current that corresponds to $V_\maxn^\textn{NAF}$ is denoted as $I_\maxn^\textn{NAF}$. $V_\maxn^\textn{NAF}$ is determined by setting $c_+ = 0$, or equivalently, $c_- = -\frac{\rho_\textn{s}}{z_-e}$, and $\phi = -V_\maxn^\textn{NAF}$ at $x = L$, which results in the following nonlinear algebraic equation that is solved using MATLAB's $\mathtt{fsolve}$ or $\mathtt{fzero}$ function:
\begin{align}
  \left(1+\tilde{\rho}_\textn{s}\right) y^2 \ln y &= \left(1-y\right)\left \{\gamma_1\tilde{\rho}_\textn{s}\left(1+y\right) - \left[\tilde{\rho}_\textn{s}+2\gamma_1\left(1+\tilde{\rho}_\textn{s}\right)\right]y\right \}, \quad \tilde{\rho}_\textn{s} > 0, \\
  y &= \exp\left(z_-\tilde{V}_\maxn^\textn{NAF}\right), \\
  \gamma_1 &= \frac{1}{2}\left(1-\frac{z_-}{z_+}\right).
\end{align}

Using the nondimensionalization defined in Section~\ref{sec:Reservoir boundary condition at anode}, we obtain
\begin{align}
  \tilde{I}_\limn^\textn{NAF} &= 1, \\
  \tilde{c}_- &= \tilde{\alpha}_3\exp\left(-z_-\tilde{\phi}\right), \\
  \tilde{I}\tilde{x} &= \frac{1}{2}\left[\tilde{\alpha}_3 - \tilde{c}_- + \frac{z_+\tilde{\rho}_\textn{s}}{z_+-z_-}\ln\left(\frac{\tilde{c}_-}{\tilde{\alpha}_3}\right)\right] = \frac{1}{2}\left \{\tilde{\alpha}_3\left[1-\exp\left(-z_-\tilde{\phi}\right)\right] - \frac{z_+z_-\tilde{\rho}_\textn{s}}{z_+-z_-}\tilde{\phi}\right \}, \\
  \tilde{I} &= \frac{1}{2}\left \{\tilde{\alpha}_3\left[1-\exp\left(z_-\tilde{V}\right)\right] + \frac{z_+z_-\tilde{\rho}_\textn{s}}{z_+-z_-}\tilde{V}\right \}. \label{eq:Dimensionless steady state I-V relation for no-anion-flux boundary condition at anode}
\end{align}
Like in case 1, it is possible to express $\tilde{c}_-$ and $\tilde{\phi}$ as explicit functions of $\tilde{x}$. We use the definitions for $\tilde{\alpha}_1$ and $\tilde{\alpha}_2$ in Section~\ref{sec:Reservoir boundary condition at anode}. If $\tilde{\rho}_\textn{s}=0$, then
\begin{align}
  \tilde{c}_- &= \tilde{\alpha}_3 - 2\tilde{I}\tilde{x}, \\
  \tilde{\phi} &= -\frac{1}{z_-}\ln\left(\frac{\tilde{c}_-}{\tilde{\alpha}_3}\right) = -\frac{1}{z_-}\ln\left(1-\frac{2\tilde{I}}{\tilde{\alpha}_3}\tilde{x}\right). \label{eq:Electrolyte electric potential as a function of x for rho_s = 0 for no-anion-flux boundary condition at anode}
\end{align}
If $\tilde{\rho}_\textn{s} \neq 0$, we obtain
\begin{align}
  \tilde{c}_- &= -\tilde{\alpha}_1 W\left[-\frac{\tilde{\alpha}_3}{\tilde{\alpha}_1}\exp\left(\frac{2\tilde{I}\tilde{x}-\tilde{\alpha}_3}{\tilde{\alpha}_1}\right)\right], \\
  \tilde{\phi} &= -\frac{\tilde{\alpha}_1}{\tilde{\alpha}_2}\ln\left(\frac{\tilde{c}_-}{\tilde{\alpha}_3}\right) = -\frac{\tilde{c}_- + 2\tilde{I}\tilde{x} - \tilde{\alpha}_3}{\tilde{\alpha}_2}. \label{eq:Electrolyte electric potential as a function of x for rho_s < 0 for no-anion-flux boundary condition at anode}
\end{align}
Noting that $\tilde{\phi}\left(\tilde{x}=1\right) = -\tilde{V}$, we can evaluate Equations~\ref{eq:Electrolyte electric potential as a function of x for rho_s = 0 for no-anion-flux boundary condition at anode} and~\ref{eq:Electrolyte electric potential as a function of x for rho_s < 0 for no-anion-flux boundary condition at anode} at $\tilde{x} = 1$ to express $\tilde{V}$ as an implicit function of $\tilde{I}$:
\begin{align}
  \tilde{V} &=
  \begin{cases}
    \frac{1}{z_-}\ln\left(1-\frac{2\tilde{I}}{\tilde{\alpha}_3}\right), & \tilde{\rho}_\textn{s} = 0 \\
    \frac{-\tilde{\alpha}_1 W\left[-\frac{\tilde{\alpha}_3}{\tilde{\alpha}_1}\exp\left(\frac{2\tilde{I}-\tilde{\alpha}_3}{\tilde{\alpha}_1}\right)\right] + 2\tilde{I} - \tilde{\alpha}_3}{\tilde{\alpha}_2}, & \tilde{\rho}_\textn{s} \neq 0 \label{eq:V as implicit function of I for no-anion-flux boundary condition at anode}
  \end{cases}
\end{align}
where we recall that $\tilde{\alpha}_3$ is a function of $\tilde{V}$. We solve Equation~\ref{eq:V as implicit function of I for no-anion-flux boundary condition at anode} using MATLAB's $\mathtt{fsolve}$ or $\mathtt{fzero}$ function.

\begin{figure}
  \centering
  \includegraphics[scale=0.6]{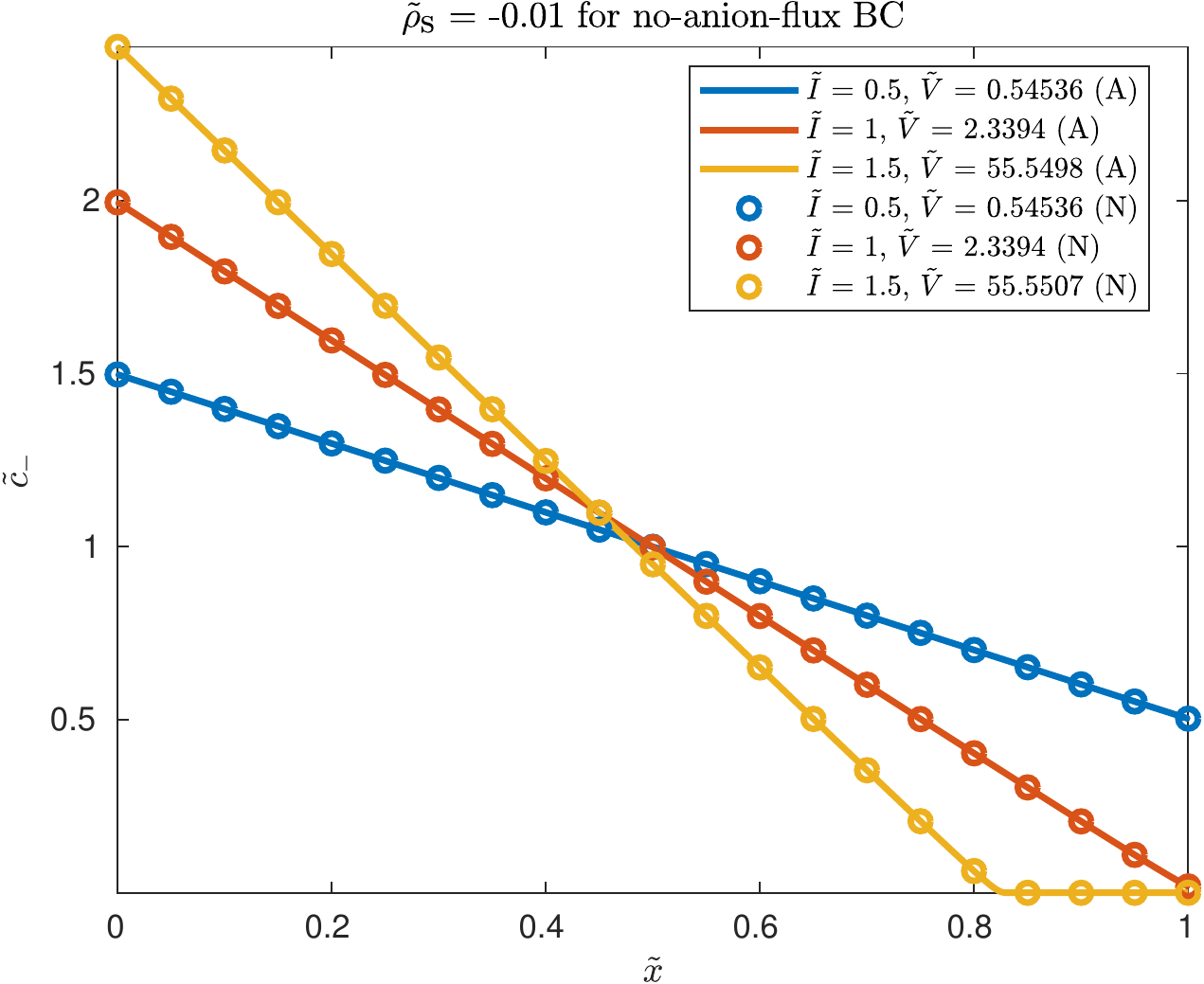}
  \includegraphics[scale=0.6]{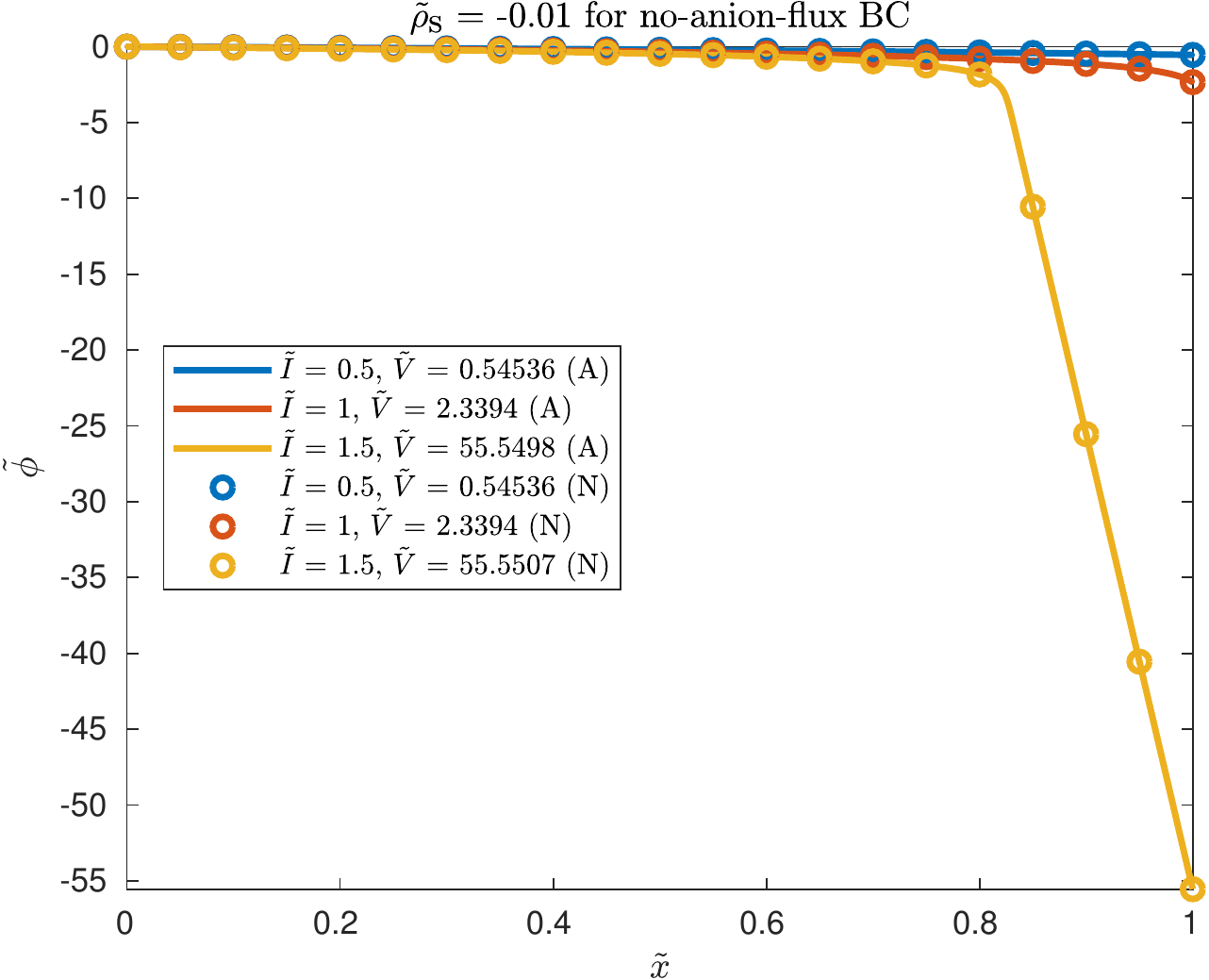}
  \includegraphics[scale=0.6]{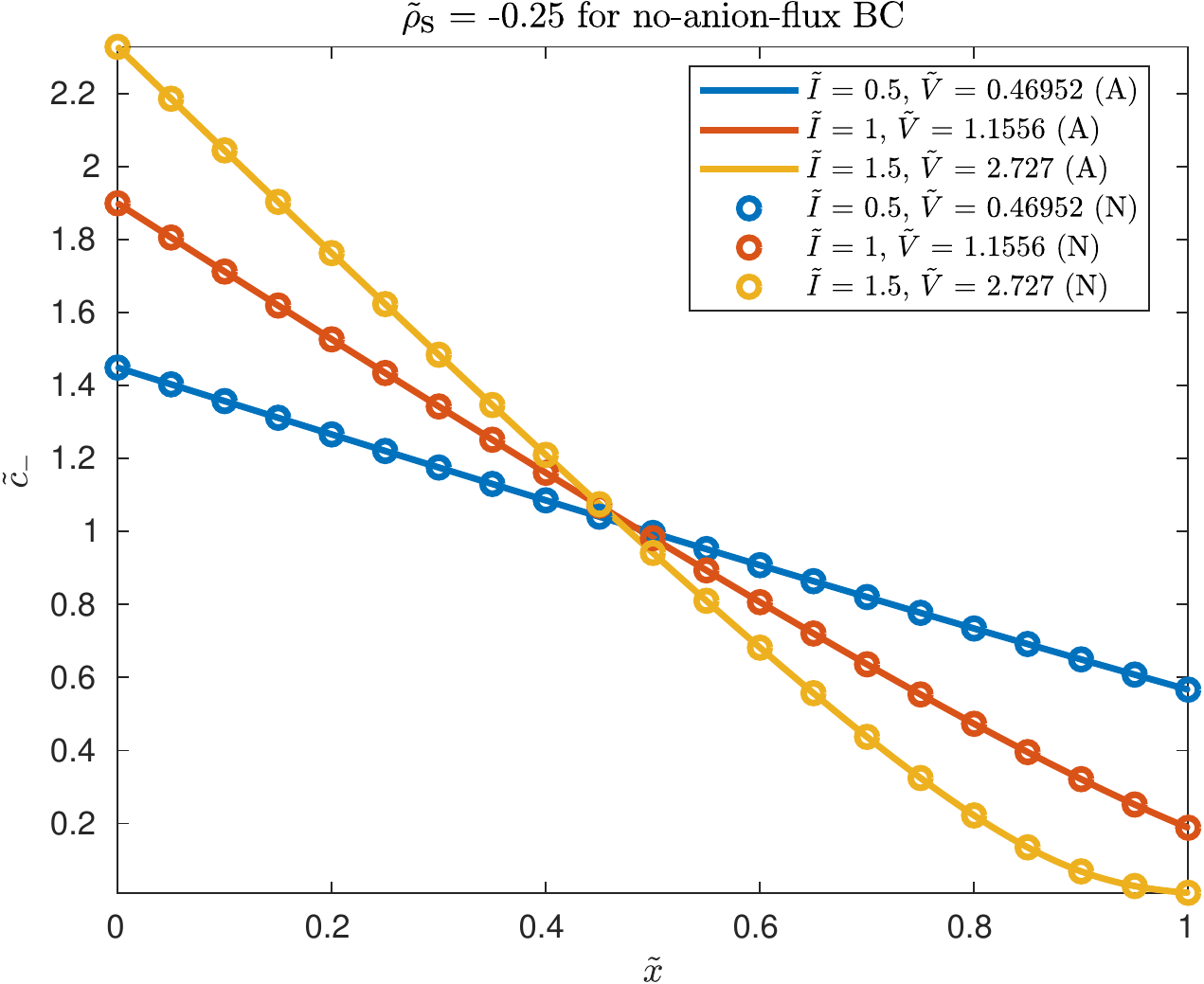}
  \includegraphics[scale=0.6]{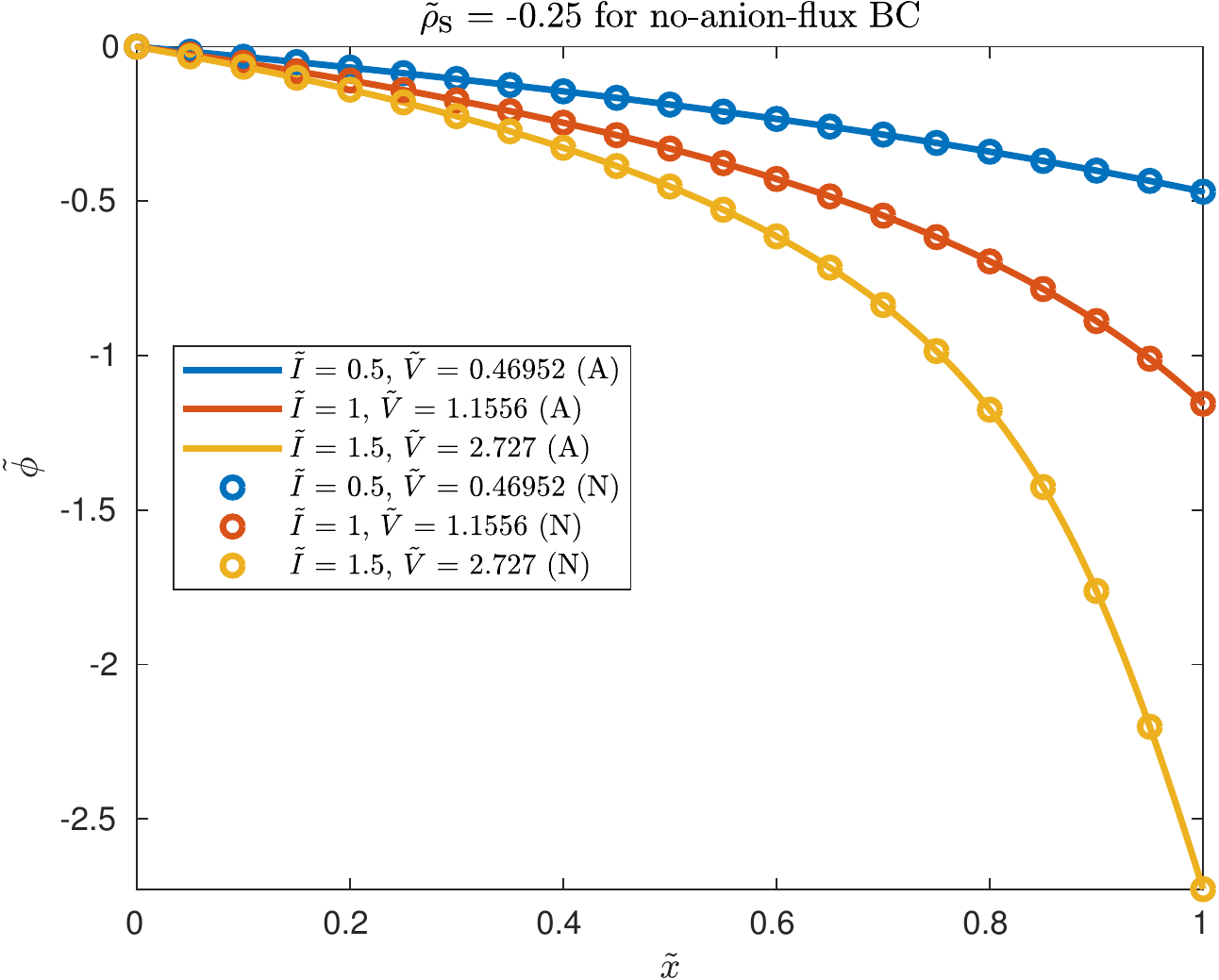}
  \caption{Plots of $\tilde{c}_-$ and $\tilde{\phi}$ against $\tilde{x}$ for $\tilde{\rho}_\textn{s} = -0.01, -0.25$ and $\tilde{I} = 0.5, 1, 1.5$ for no-anion-flux boundary condition at anode. (A) refers to analytical solutions and (N) refers to numerical solutions.}\label{fig:c_- and phi against x for rho_s < 0 for no-anion-flux boundary condition at anode}
\end{figure}

\begin{figure}
  \centering
  \includegraphics[scale=0.6]{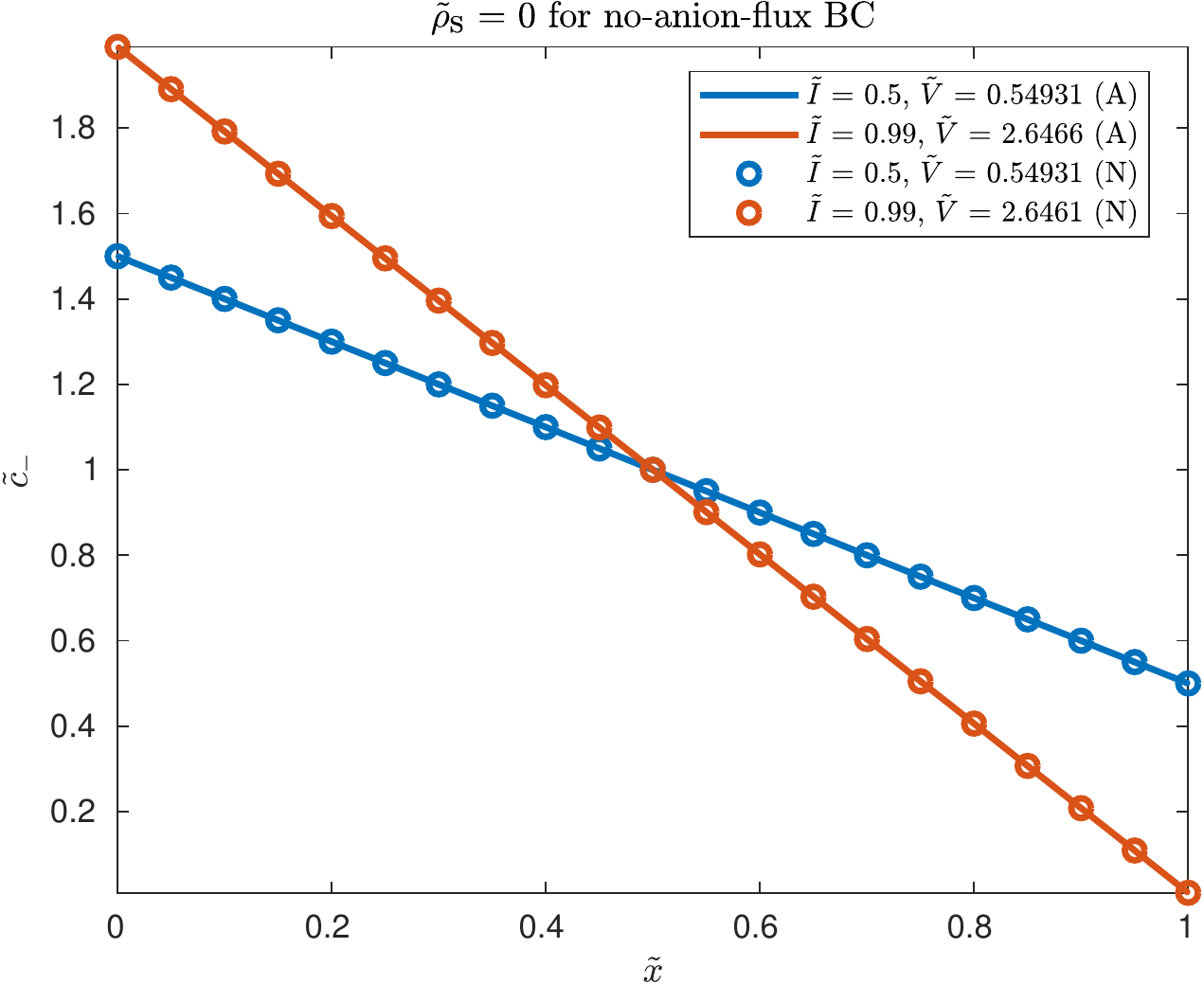}
  \includegraphics[scale=0.6]{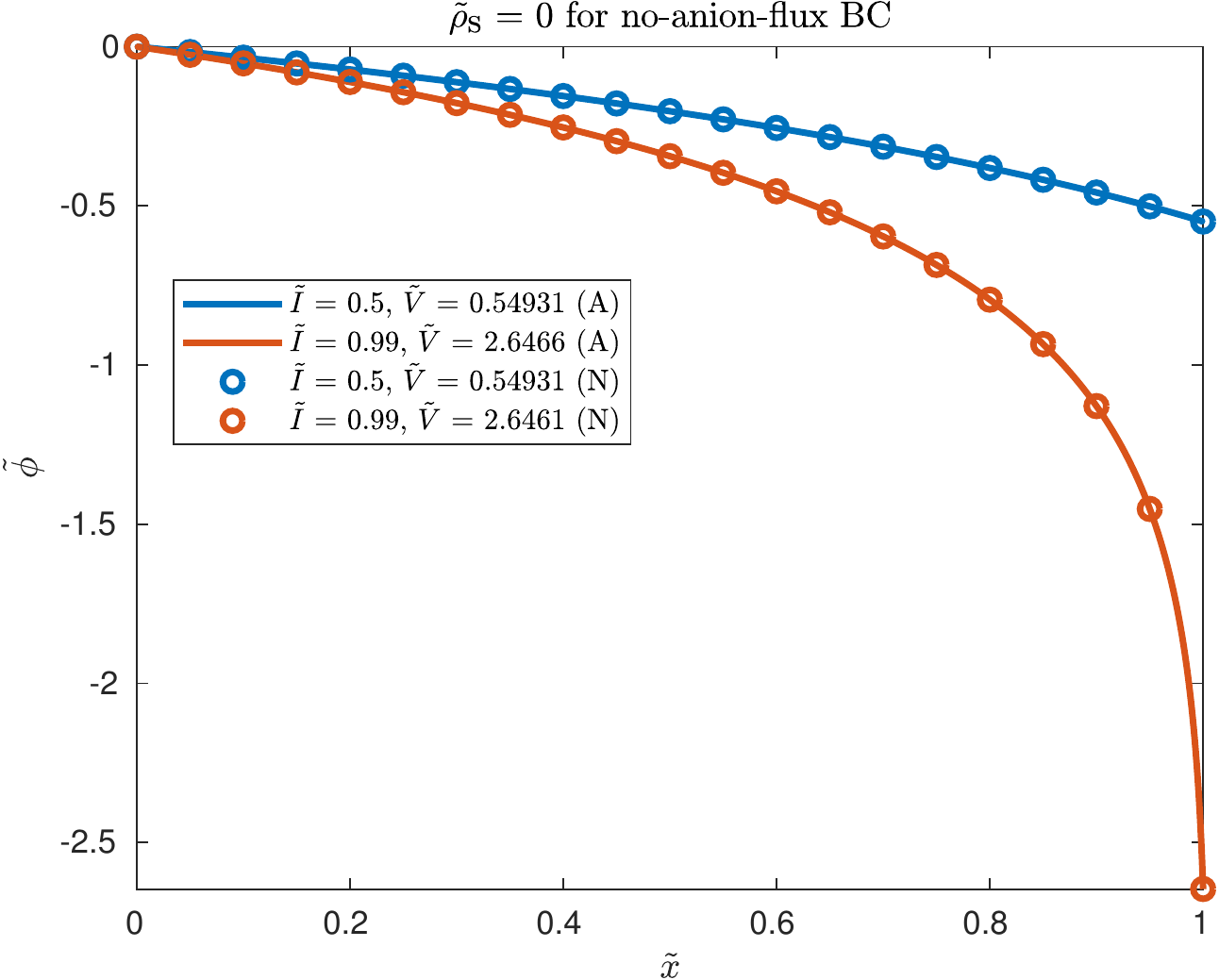}
  \includegraphics[scale=0.6]{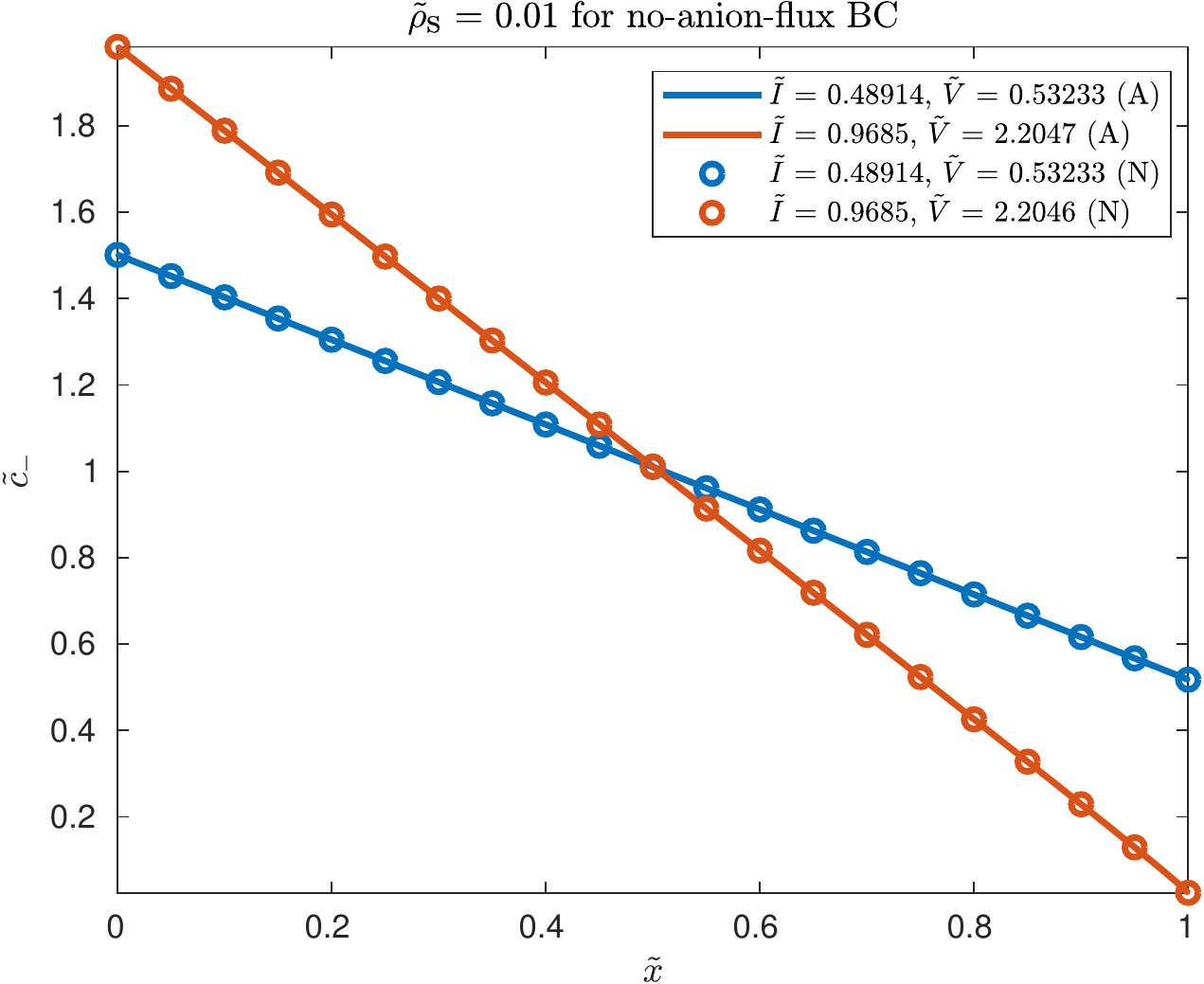}
  \includegraphics[scale=0.6]{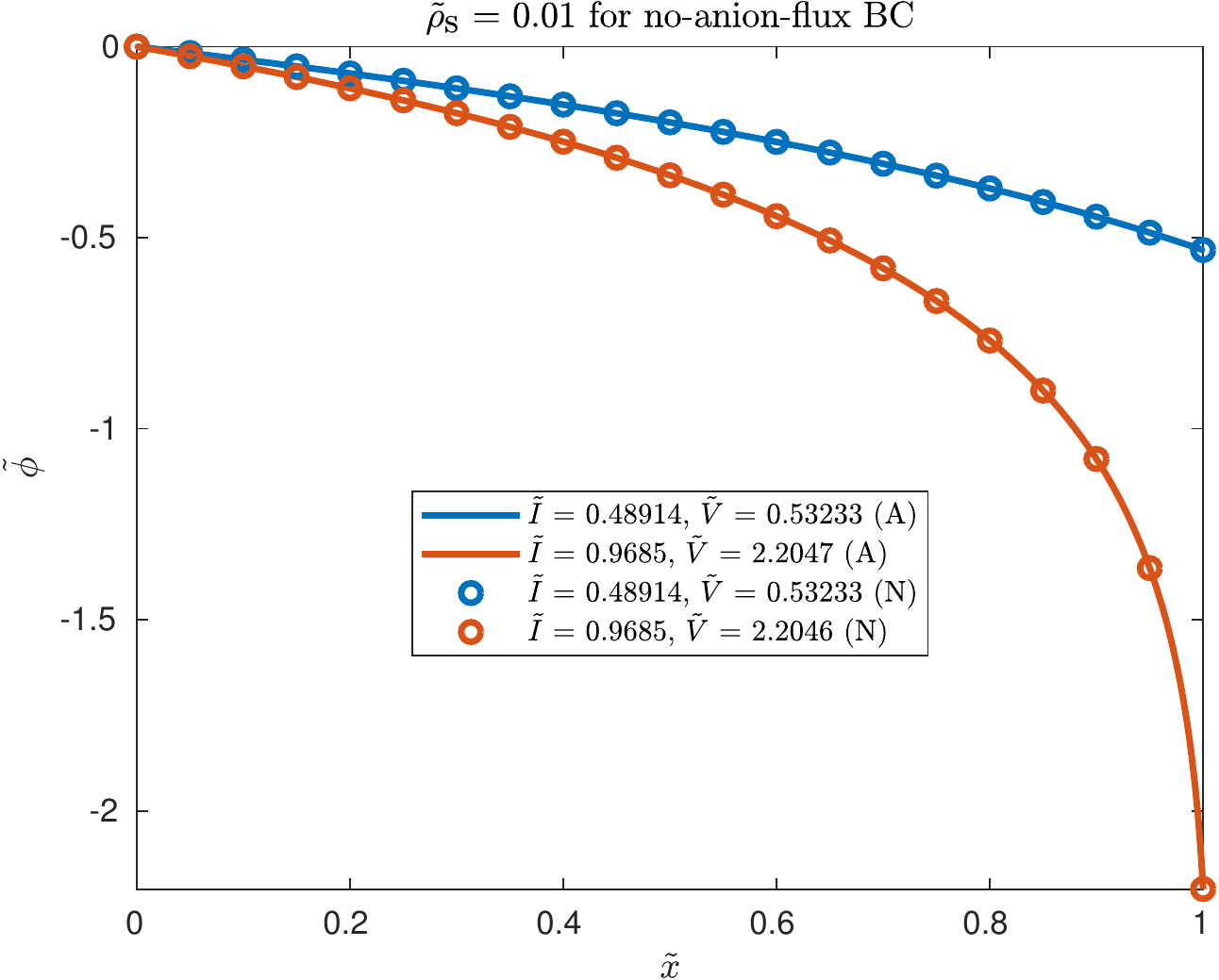}
  \includegraphics[scale=0.6]{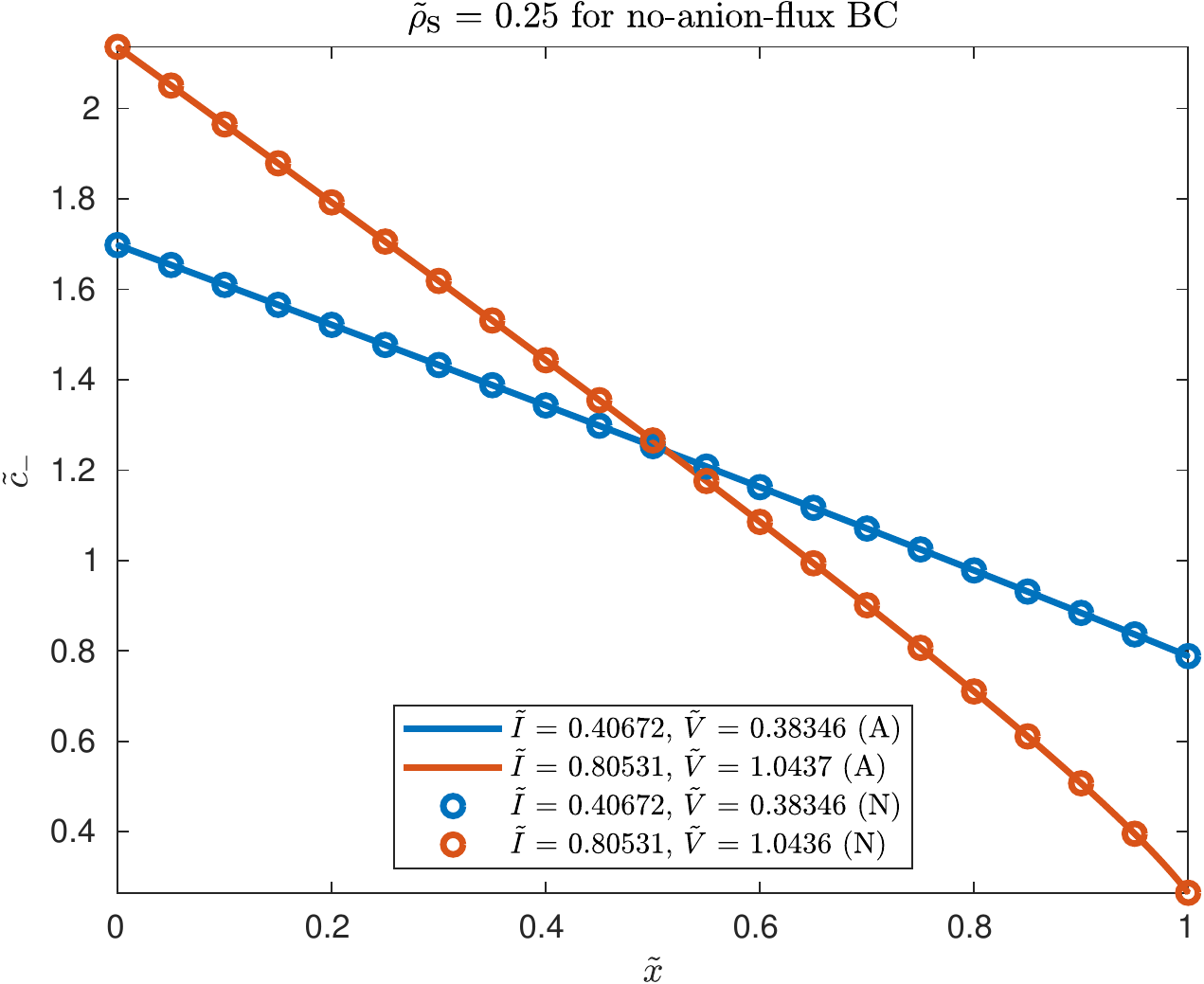}
  \includegraphics[scale=0.6]{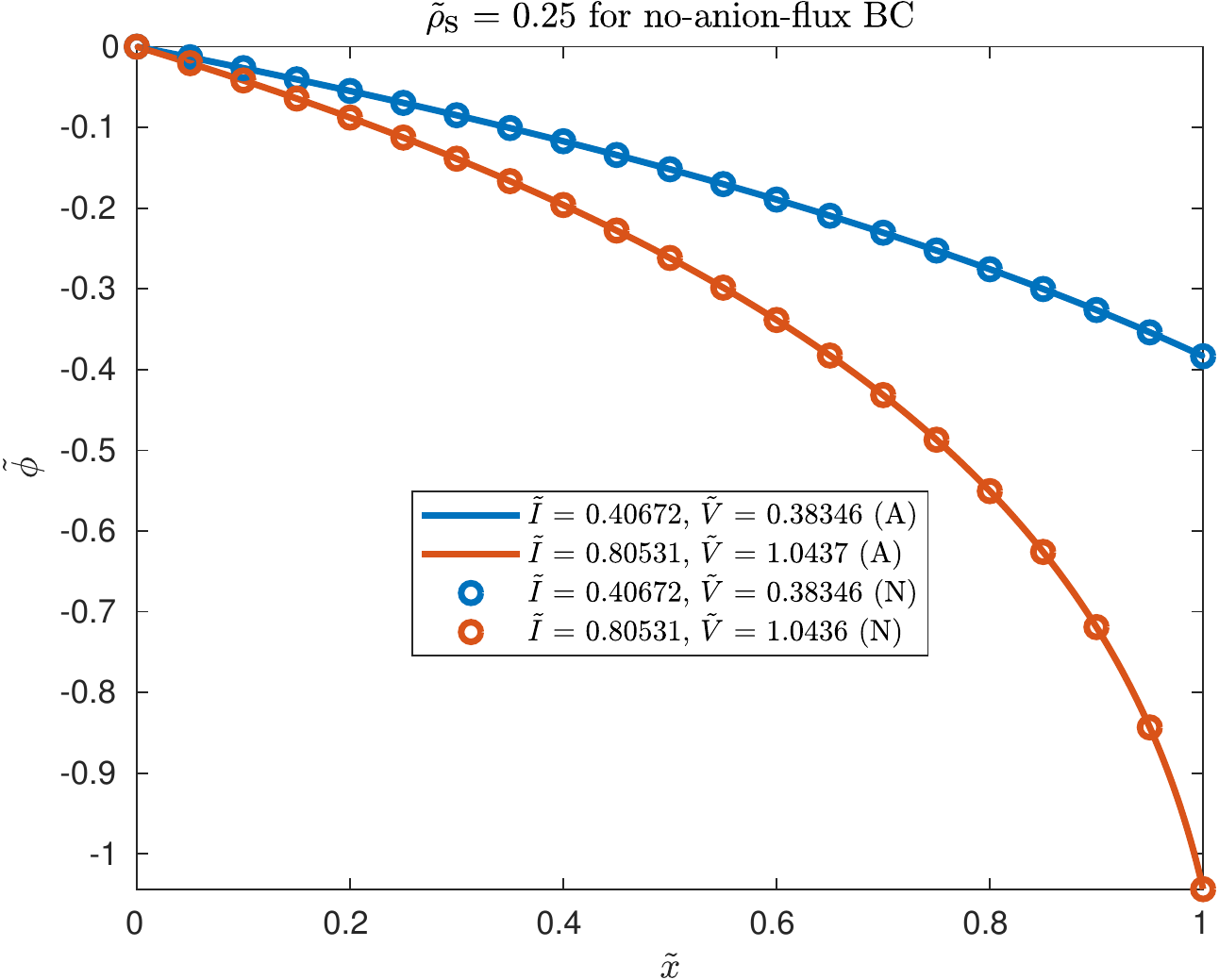}
  \caption{Plots of $\tilde{c}_-$ and $\tilde{\phi}$ against $\tilde{x}$ for 1) $\tilde{\rho}_\textn{s} = 0$ and $\tilde{I} = 0.5, 0.99$ (top row) and 2) $\tilde{\rho}_\textn{s} = 0.01, 0.25$ and $\tilde{I} = 0.5\tilde{I}_\maxn^\textn{NAF}, 0.99\tilde{I}_\maxn^\textn{NAF}$ (second and third rows) for no-anion-flux boundary condition at anode. (A) refers to analytical solutions and (N) refers to numerical solutions.}\label{fig:c_- and phi against x for rho_s >= 0 for no-anion-flux boundary condition at anode}  
\end{figure}

\begin{figure}
  \centering
  \includegraphics[scale=0.6]{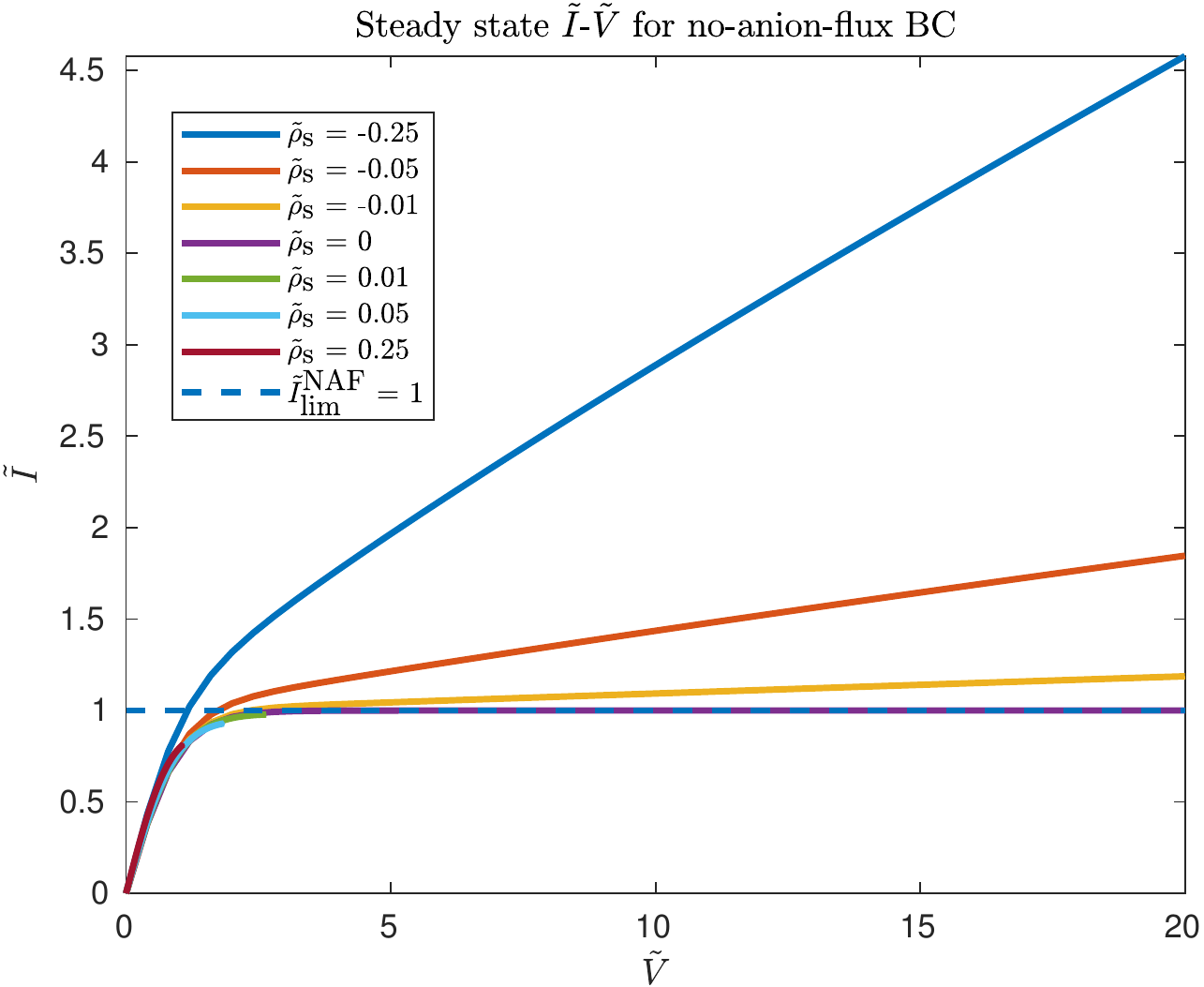}
  \includegraphics[scale=0.6]{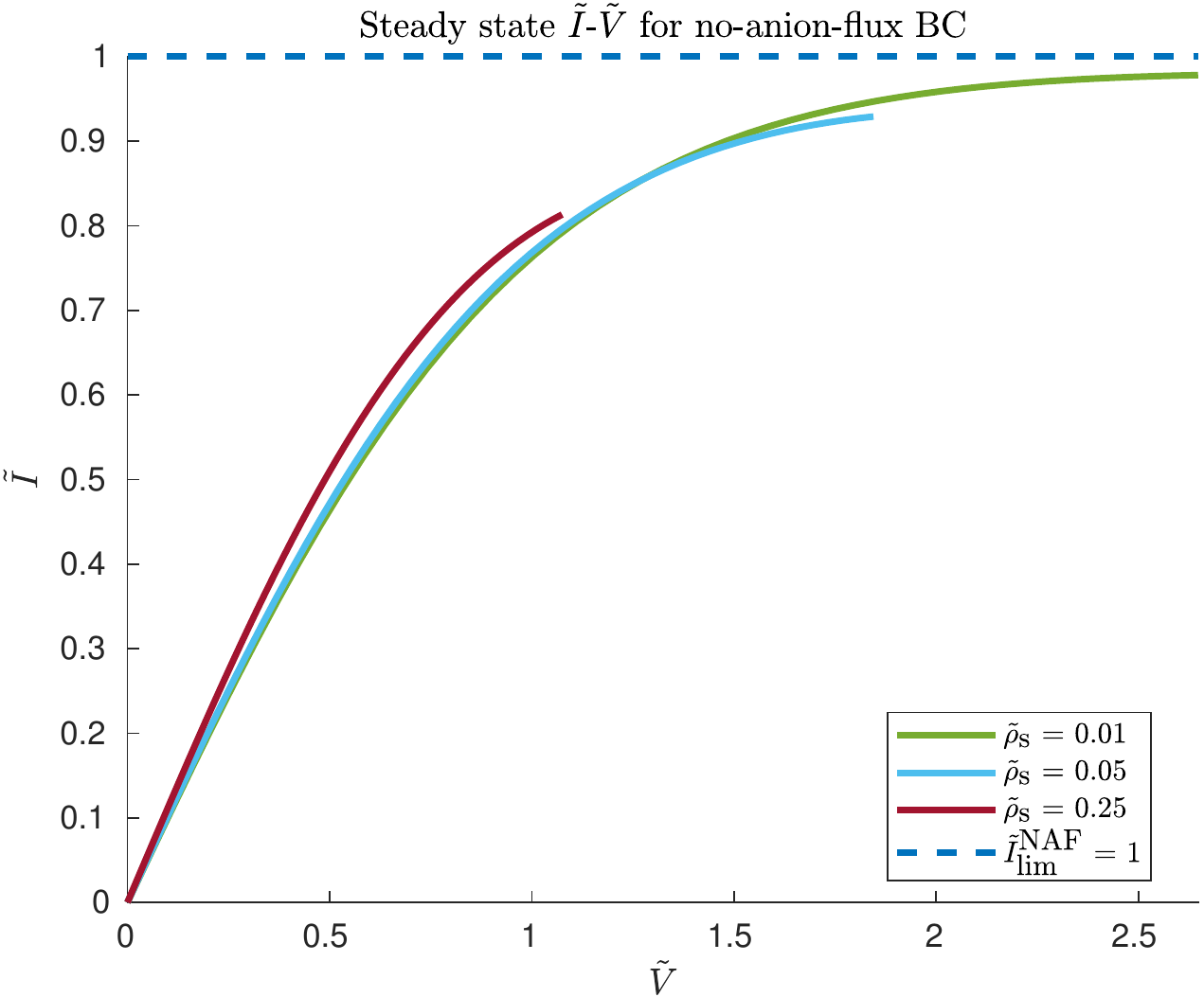}
  \caption{Left: steady state $\tilde{I}$-$\tilde{V}$ relations for $\tilde{\rho}_\textn{s} = 0, \pm 0.01, \pm 0.05, \pm 0.25$ for no-anion-flux boundary condition at anode. Right: zoom-in view of left plot for only $\tilde{\rho}_\textn{s} = 0.01, 0.05, 0.25$. The dashed line denotes $\tilde{I}_\limn^\textn{NAF} = 1$, which is the maximum $\tilde{I}$ that the system can reach when $\tilde{\rho}_\textn{s} = 0$.}\label{fig:Steady state I-V relations for no-anion-flux boundary condition at anode}
\end{figure}

For both analytical expressions and numerical solutions, we plot $\tilde{c}_-$ and $\tilde{\phi}$ as functions of $\tilde{x}$ for $\tilde{\rho}_\textn{s} = -0.01, -0.25$ in Figure~\ref{fig:c_- and phi against x for rho_s < 0 for no-anion-flux boundary condition at anode} and $\tilde{\rho}_\textn{s} = 0, 0.01, 0.25$ in Figure~\ref{fig:c_- and phi against x for rho_s >= 0 for no-anion-flux boundary condition at anode}. For $\tilde{\rho}_\textn{s} = 0$, we choose $\tilde{I} = 0.5, 0.99$ and avoid $\tilde{I} = \tilde{I}_\limn^\textn{NAF} = 1$ for the same reason discussed for case 1. For $\tilde{\rho}_\textn{s} = 0.01, 0.25$, we choose $\tilde{I} = 0.5\tilde{I}_\maxn^\textn{NAF}, 0.99\tilde{I}_\maxn^\textn{NAF}$ and for $\tilde{\rho}_\textn{s} = -0.01, -0.25$, we choose $\tilde{I} = 0.5, 1, 1.5$. We observe that the analytical expressions agree very well with the numerical solutions. The qualitative features of the $\tilde{c}_-$-$\tilde{x}$ and $\tilde{\phi}$-$\tilde{x}$ plots are very similar to that for case 1. Regardless of $\tilde{\rho}_\textn{s}$, when current is either underlimiting ($\tilde{I} = 0.5, 0.99, 0.5\tilde{I}_\maxn^\textn{NAF}, 0.99\tilde{I}_\maxn^\textn{NAF}$) or limiting ($\tilde{I} = \tilde{I}_\limn^\textn{NAF} = 1$), $\tilde{c}_-$ is approximately linear in $\tilde{x}$ because of predominant ambipolar diffusion. When current is overlimiting ($\tilde{I} = 1.5$), the depletion region extends for a finite distance from the cathode into the electrolyte. Because of the integral constraint on the anion concentration, anions can be exchanged across the anode to provide more conductivity to the electrolyte and it is possible for $\tilde{c}_-\left(\tilde{x}=0\right) > \tilde{\beta}_1$, in contrast to $\tilde{c}_-\left(\tilde{x}=0\right) = \tilde{\beta}_1$ for case 1. We also plot $\tilde{I}$ against $\tilde{V}$ for $\tilde{\rho}_\textn{s} = 0, \pm 0.01, \pm 0.05, \pm 0.25$ and $\tilde{V} \in \left[0,20\right]$ in Figure~\ref{fig:Steady state I-V relations for no-anion-flux boundary condition at anode}. For $\tilde{\rho}_\textn{s} = -0.01, -0.05, -0.25$, $\tilde{I}$ eventually becomes larger than $\tilde{I}_\limn^\textn{NAF}$ at a sufficiently large $\tilde{V}$ and becomes linear in $\tilde{V}$ with a gradient that is equal to the overlimiting conductance. In contrast, like in case 1, the presence of a positive background charge results in a finite maximum voltage, which corresponds to a finite maximum current that is smaller than the limiting current $\tilde{I}_\limn^\textn{NAF} = 1$.

\subsubsection{Case 3: Butler-Volmer boundary conditions at anode and cathode}\label{sec:Butler-Volmer boundary conditions at anode and cathode}

We repeat the analysis done in Section~\ref{sec:No-anion-flux boundary condition at anode} except that we replace the boundary conditions for $\phi$ given by Equations~\ref{eq:Reservoir boundary condition for electrolyte electric potential at anode} and~\ref{eq:Electrolyte electric potential at cathode} with Butler-Volmer boundary conditions at the anode and cathode
\begin{align}
  -J\left(x=0\right) &= \epsilon_\textn{p}J_\textn{F}^\textn{a}, \quad \phi_\textn{e}^\textn{a} = 0, \label{eq:Butler-Volmer boundary condition at anode} \\
  J\left(x=L\right) &= \epsilon_\textn{p}J_\textn{F}^\textn{c}, \quad \phi_\textn{e}^\textn{c} = -V, \quad V \geq 0, \label{eq:Butler-Volmer boundary condition at cathode}
\end{align}
where we use the expression for $J_\textn{F}$ given by Equation~\ref{eq:J_F} in Section~\ref{sec:Electrochemical reaction kinetics} for copper electrodeposition and electrodissolution.

We first compare the boundary conditions for $\phi$ given by the Butler-Volmer boundary conditions with the boundary conditions given by Equations~\ref{eq:Reservoir boundary condition for electrolyte electric potential at anode} and~\ref{eq:Electrolyte electric potential at cathode} that are used in cases 1 and 2. We define the electric potential difference across the electrolyte $\Delta\phi^\textn{electrolyte}\equiv\phi\left(x=L\right)-\phi\left(x=0\right)$ and the electric potential difference between the cathode and anode $\Delta\phi^\textn{electrode} \equiv \phi_\textn{e}^\textn{c}-\phi_\textn{e}^\textn{a} = \Delta\phi^\textn{c} + \Delta\phi^\textn{electrolyte} - \Delta\phi^\textn{a} = -V$ where the ``$\textn{a}$'' and ``$\textn{c}$'' superscripts denote the anode and cathode respectively. Experimentally, $\Delta\phi^\textn{electrode}$, not $\Delta\phi^\textn{electrolyte}$, is the quantity that we either impose under potentiostatic conditions or linear sweep voltammetry (LSV), or measure under galvanostatic conditions. The assumption we make in going from the Butler-Volmer boundary conditions to Equations~\ref{eq:Reservoir boundary condition for electrolyte electric potential at anode} and~\ref{eq:Electrolyte electric potential at cathode} is $\Delta\phi^\textn{electrolyte} \approx -V = \Delta\phi^\textn{electrode}$. We expect this approximation to become better as $\left\lvert\Delta\phi^\textn{electrode}\right\rvert$ increases. This is because a larger $\left\lvert\Delta\phi^\textn{electrode}\right\rvert$ results in lower cation and anion concentrations at the cathode that in turn result in a larger electric field at the cathode to sustain the current. This larger electric field at the cathode implies a larger $\left\lvert\Delta\phi^\textn{electrolyte}\right\rvert$, hence improving the approximation. Comparing cases 2 and 3, case 2 can be thought of as the limit of case 3 with $\textn{Da} \rightarrow \infty$ or $\frac{J_\textn{F}}{J_0} \rightarrow 0$, i.e., the reaction resistance tends to zero. Therefore, the expressions for limiting current, which is denoted as $I_\limn^\textn{BV}$, and overlimiting conductance are the same as that for case 2 given by Equations~\ref{eq:Limiting current for no-anion-flux boundary condition at anode} and~\ref{eq:Overlimiting conductance for no-anion-flux boundary condition at anode} respectively. The expression for overlimiting conductance is thus the same in all three cases regardless of boundary conditions. The main advantage of using such an approximation is that we can replace the nonlinear Butler-Volmer boundary conditions with linear Dirichlet boundary conditions for $\phi$, which have allowed us to derive an analytical expression for the steady state current-voltage relation in cases 1 and 2.

Unlike for cases 1 and 2, it is not possible to obtain an analytical expression for the steady state current-voltage relation for the nonlinear Butler-Volmer boundary conditions. However, at steady state, when compared to case 2, the governing ODEs (ordinary differential equations) in the domain remain unchanged. Moreover, the Butler-Volmer boundary conditions are functions of only concentrations and electric potentials but not functions of their higher order spatial derivatives. Therefore, when compared to case 2, for a given $\tilde{I}$ and $\tilde{\rho}_\textn{s}$, the $\tilde{c}_-$ profile remains unchanged while the $\tilde{\phi}$ profile is shifted downwards by a constant that allows the system to achieve the necessary overpotential for driving the appropriate amount of Faradaic current density at both electrodes. This constant is a function of $\tilde{I}$ and $\tilde{\rho}_\textn{s}$ and is computed using Equations~\ref{eq:Butler-Volmer boundary condition at anode} and~\ref{eq:Butler-Volmer boundary condition at cathode} with MATLAB's $\mathtt{fsolve}$ or $\mathtt{fzero}$ function. Hence, we can obtain an analytical expression for $\tilde{c}_-$ and semi-analytical expressions for $\tilde{\phi}$ and steady state current-voltage relation. Regarding the steady state current-voltage relation, like in cases 1 and 2, there are no restrictions on how large $V$ can be for $\rho_\textn{s} \leq 0$. For $\rho_\textn{s} > 0$, there are still no restrictions on the value of $V$. However, as $V \rightarrow \infty$, $I$ tends to a finite maximum value, which is denoted as $I_\maxn^\textn{BV}$. For a given $\rho_\textn{s}$, $I_\maxn^\textn{BV}$ is equal to $I_\maxn^\textn{NAF}$ at that $\rho_\textn{s}$ value because as $V \rightarrow \infty$, the overpotential diverges and the reaction resistance tends to zero.

\begin{figure}
  \centering
  \includegraphics[scale=0.6]{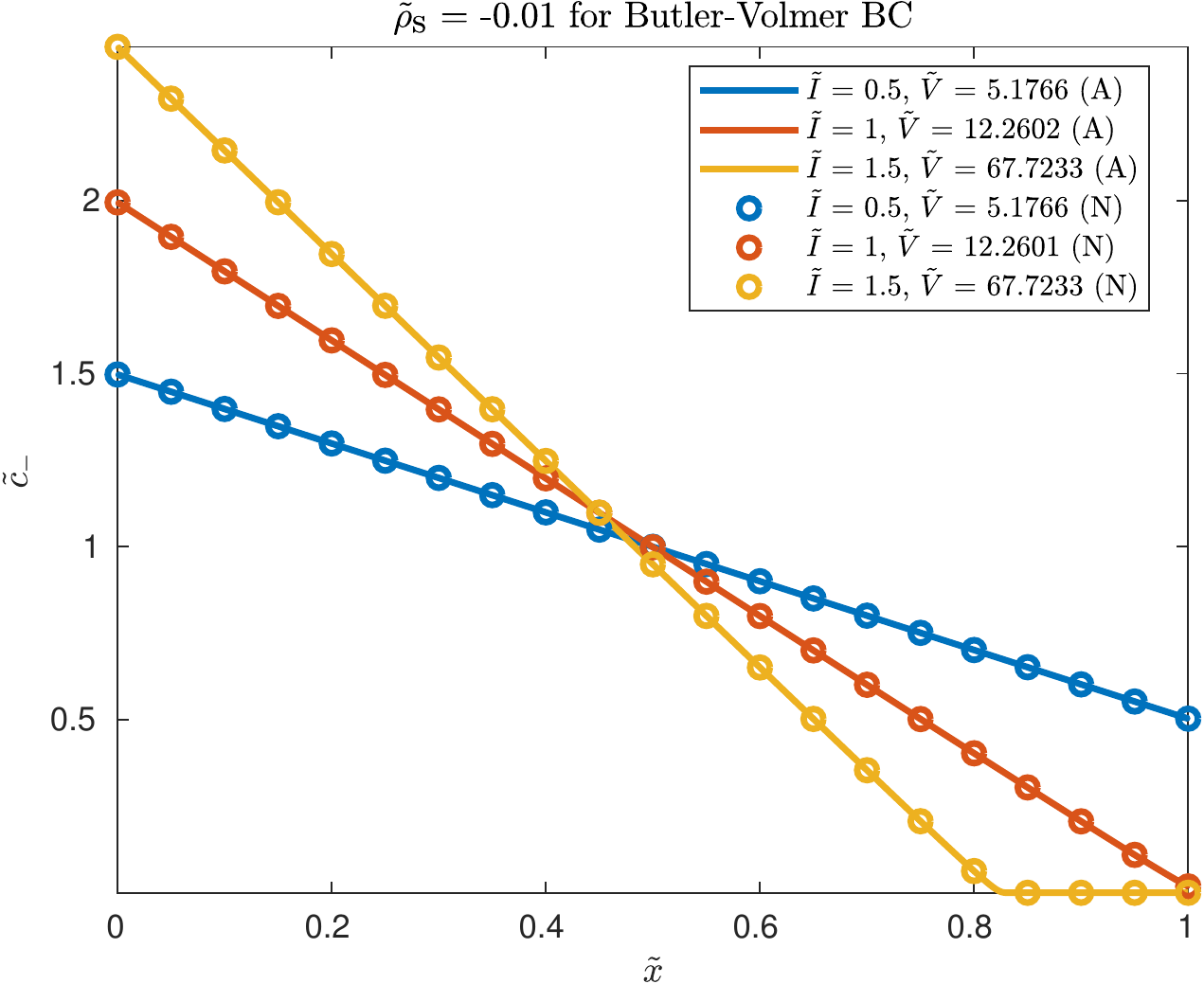}
  \includegraphics[scale=0.6]{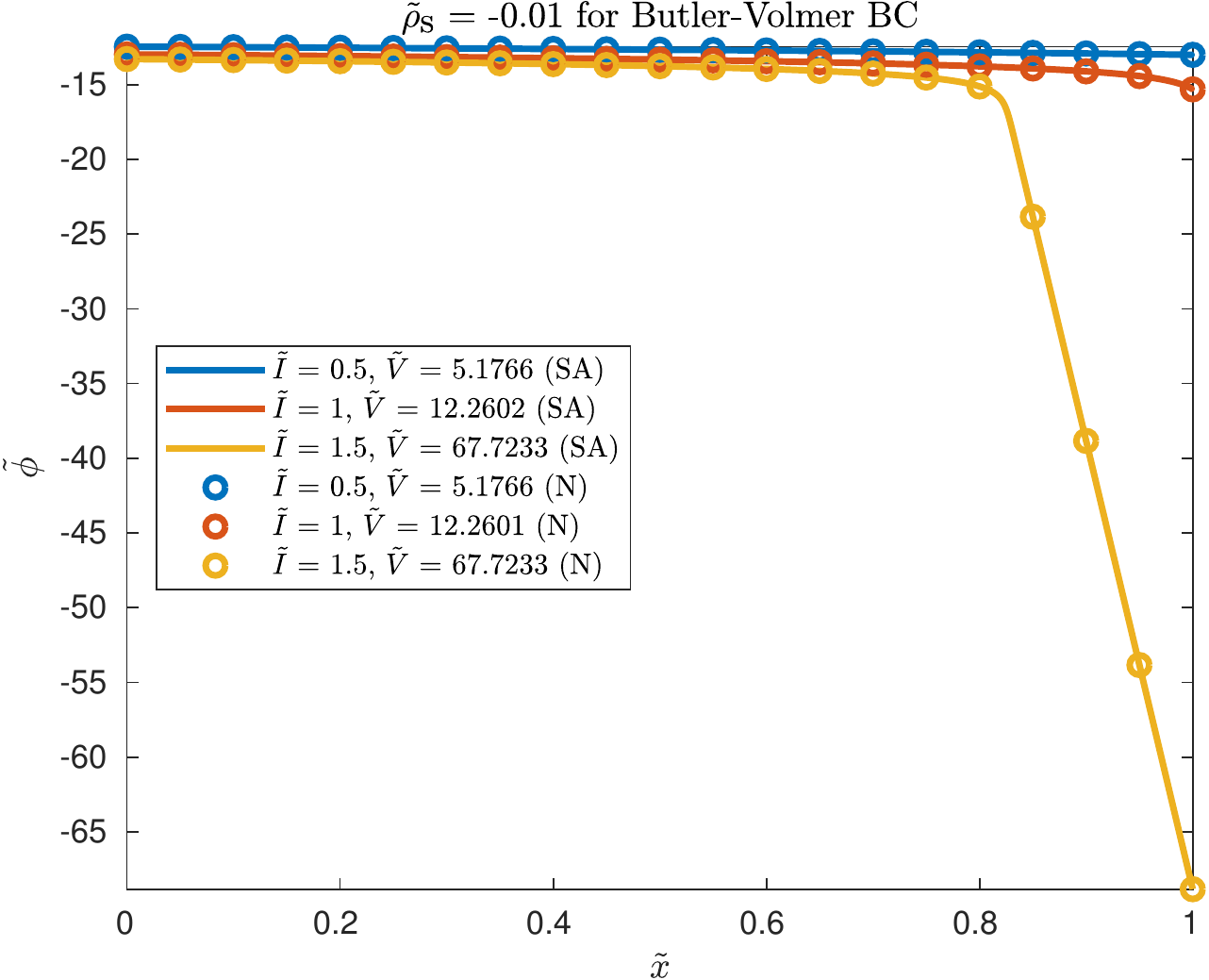}
  \includegraphics[scale=0.6]{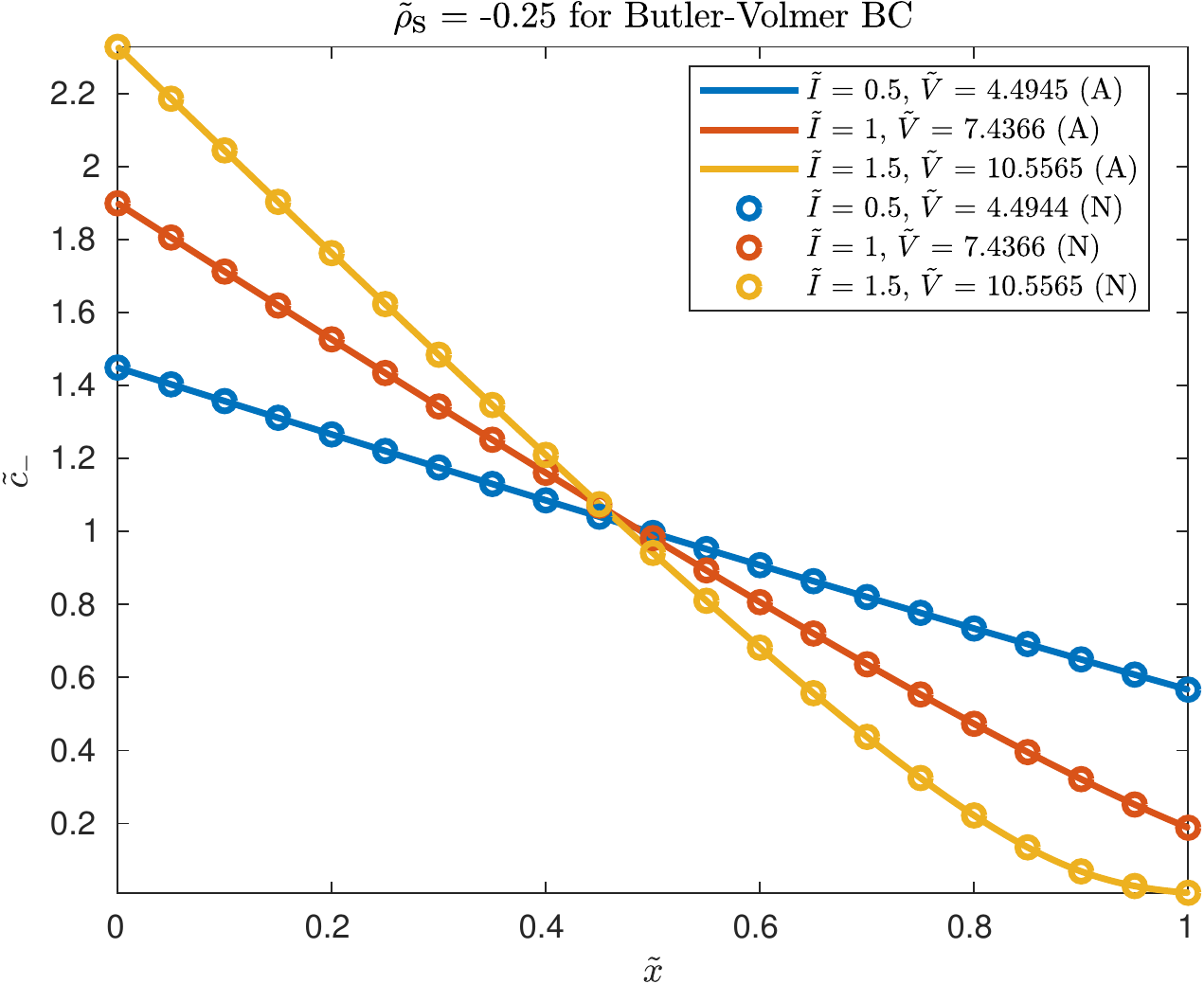}
  \includegraphics[scale=0.6]{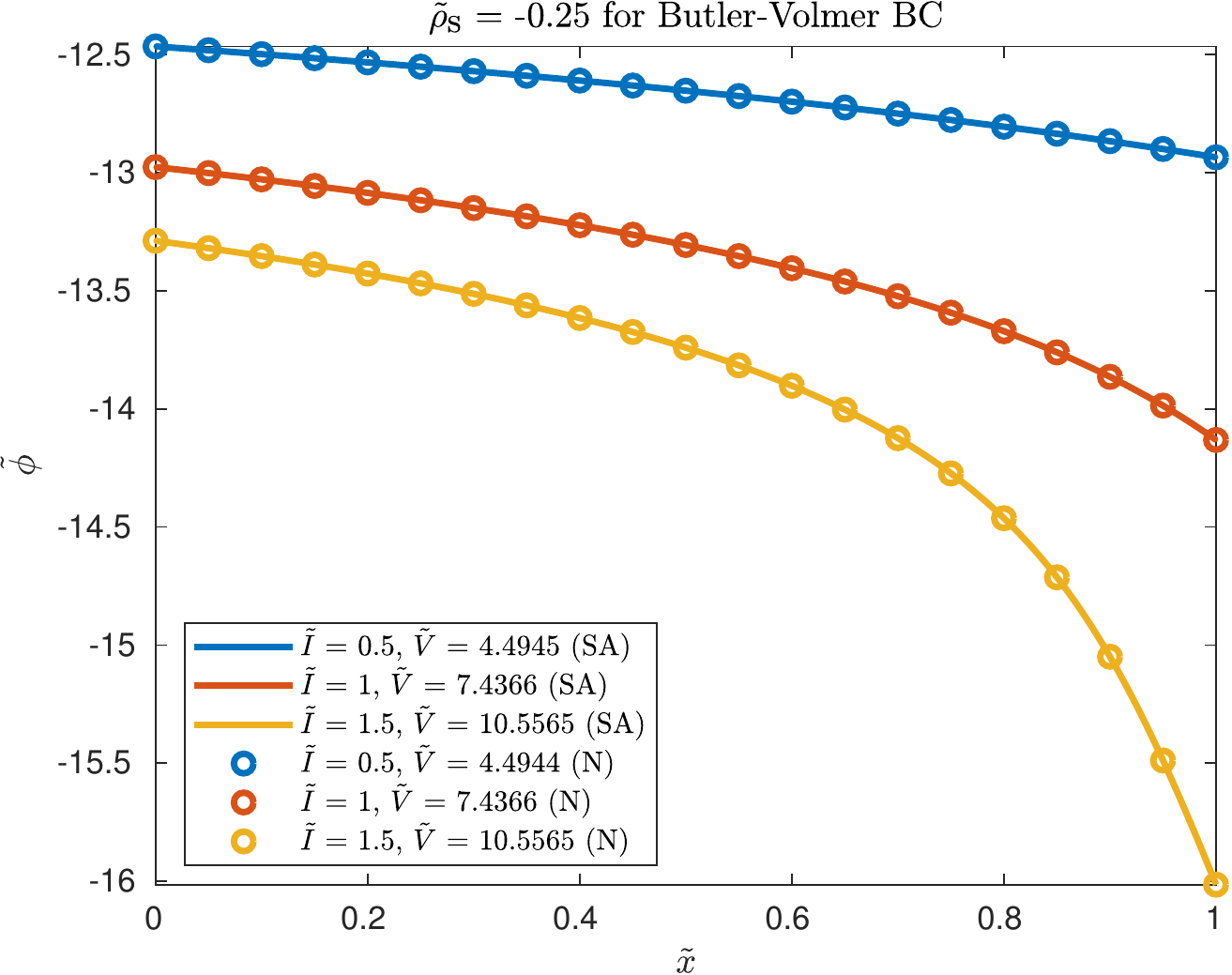}
  \caption{Plots of $\tilde{c}_-$ and $\tilde{\phi}$ against $\tilde{x}$ for $\tilde{\rho}_\textn{s} = -0.01, -0.25$ and $\tilde{I} = 0.5, 1, 1.5$ for Butler-Volmer boundary conditions at anode and cathode. (A) refers to analytical solutions, (SA) refers to semi-analytical solutions and (N) refers to numerical solutions.}\label{fig:c_- and phi against x for rho_s < 0 for Butler-Volmer boundary conditions at anode and cathode}
\end{figure}

\begin{figure}
  \centering
  \includegraphics[scale=0.6]{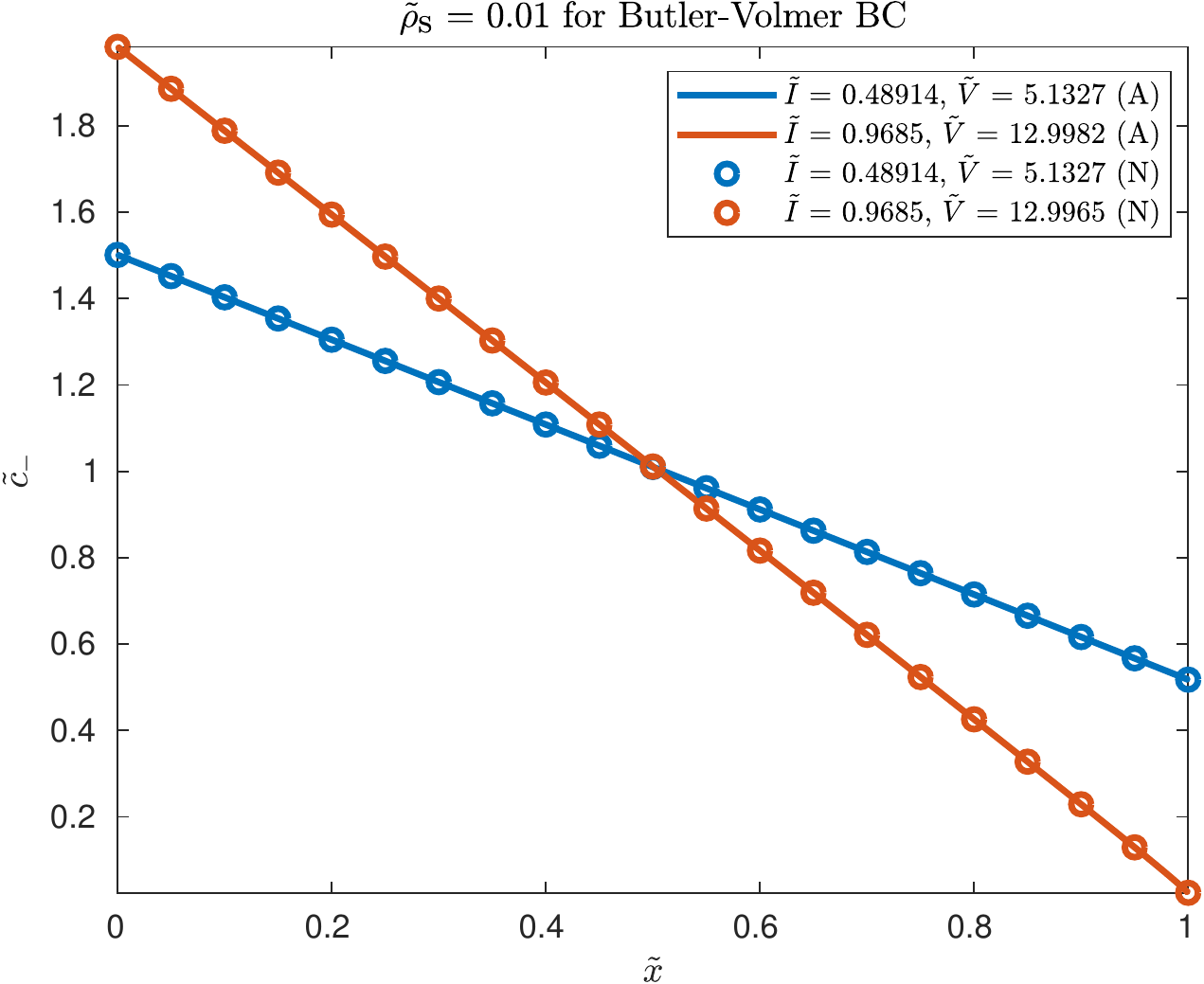}
  \includegraphics[scale=0.6]{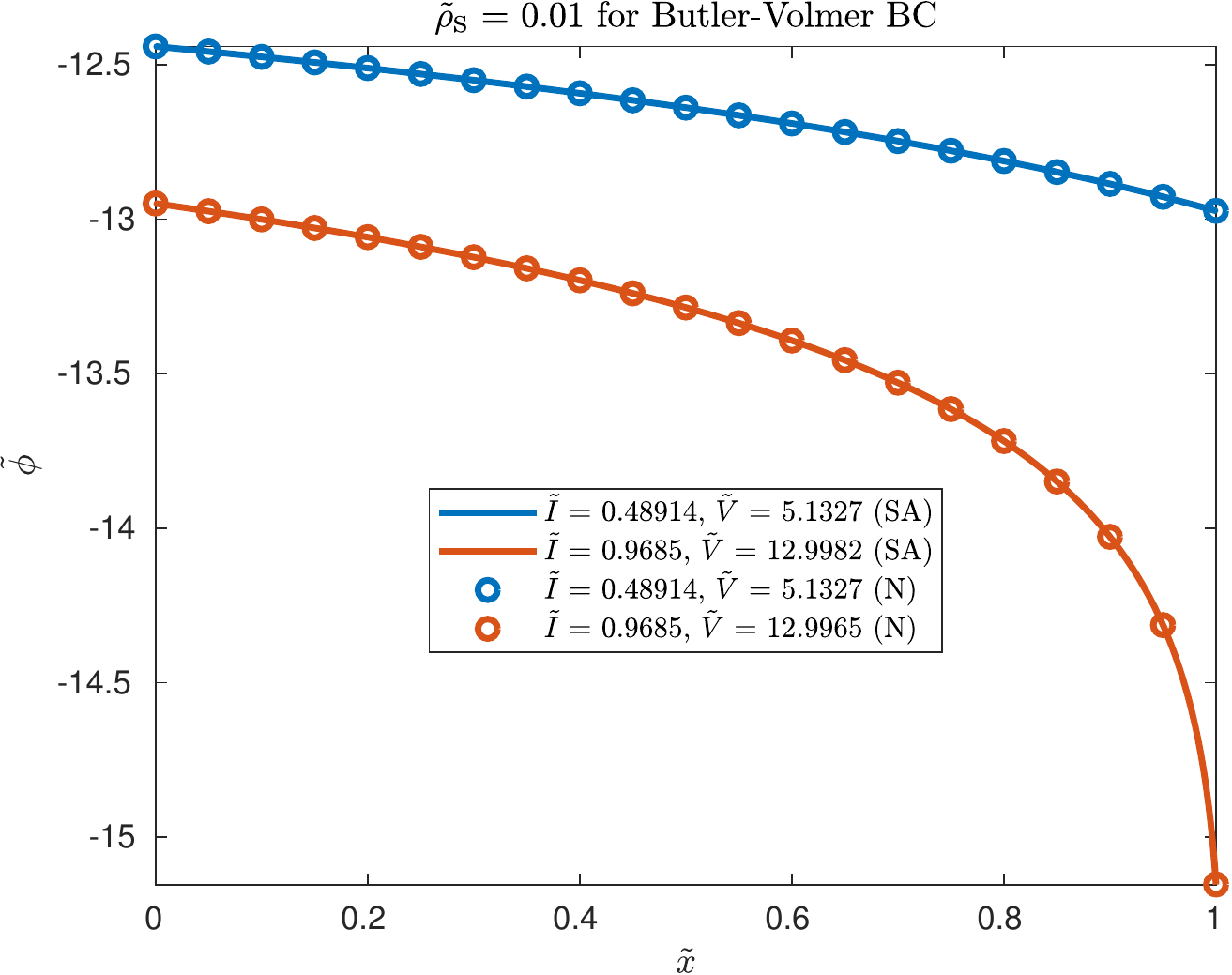}
  \includegraphics[scale=0.6]{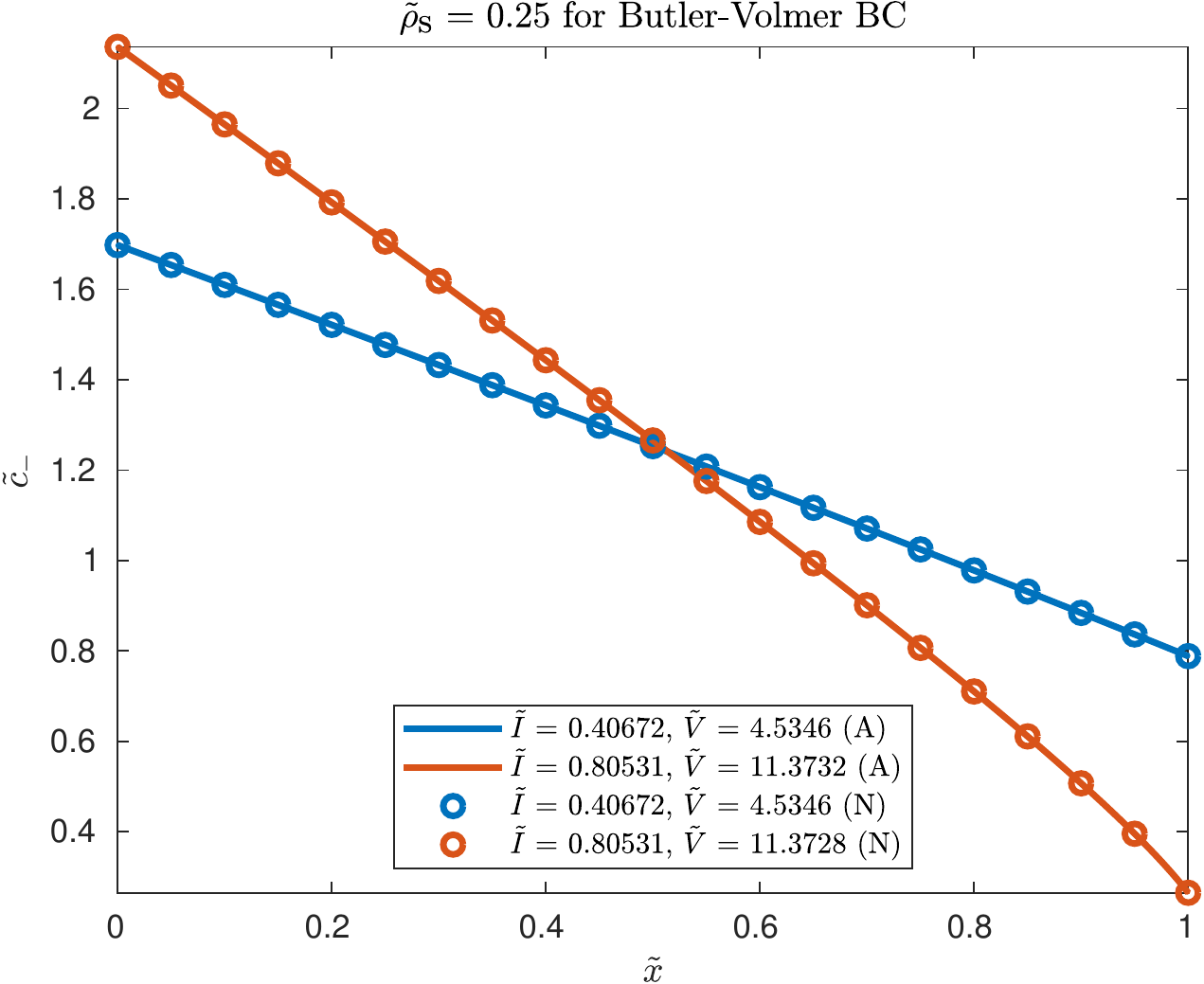}
  \includegraphics[scale=0.6]{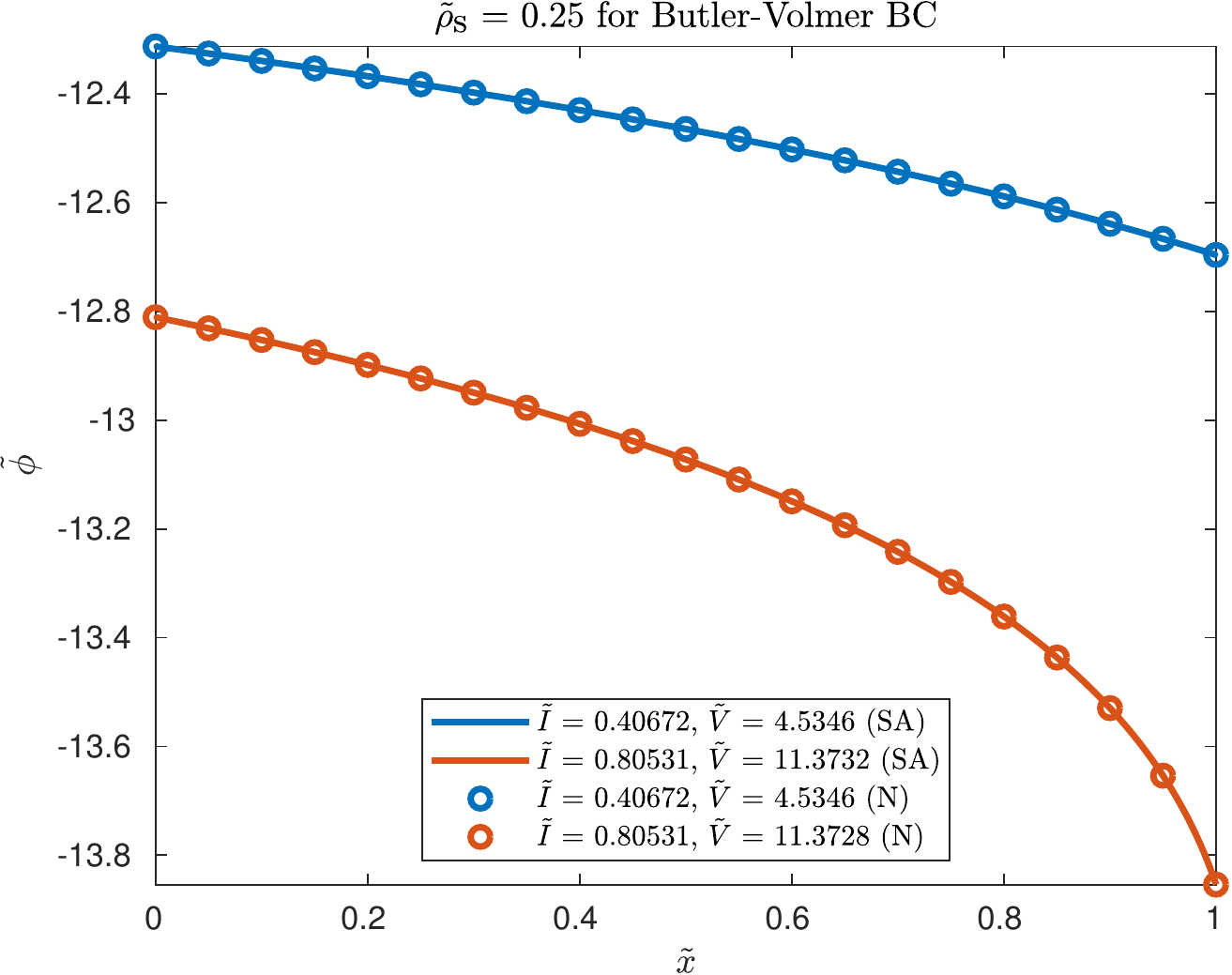}
  \caption{Plots of $\tilde{c}_-$ and $\tilde{\phi}$ against $\tilde{x}$ for $\tilde{\rho}_\textn{s} = 0.01, 0.25$ and $\tilde{I} = 0.5\tilde{I}_\maxn^\textn{BV}, 0.99\tilde{I}_\maxn^\textn{BV}$ for Butler-Volmer boundary conditions at anode and cathode. (A) refers to analytical solutions, (SA) refers to semi-analytical solutions and (N) refers to numerical solutions.}\label{fig:c_- and phi against x for rho_s > 0 for Butler-Volmer boundary conditions at anode and cathode}
\end{figure}

\begin{figure}
  \centering
  \includegraphics[scale=0.6]{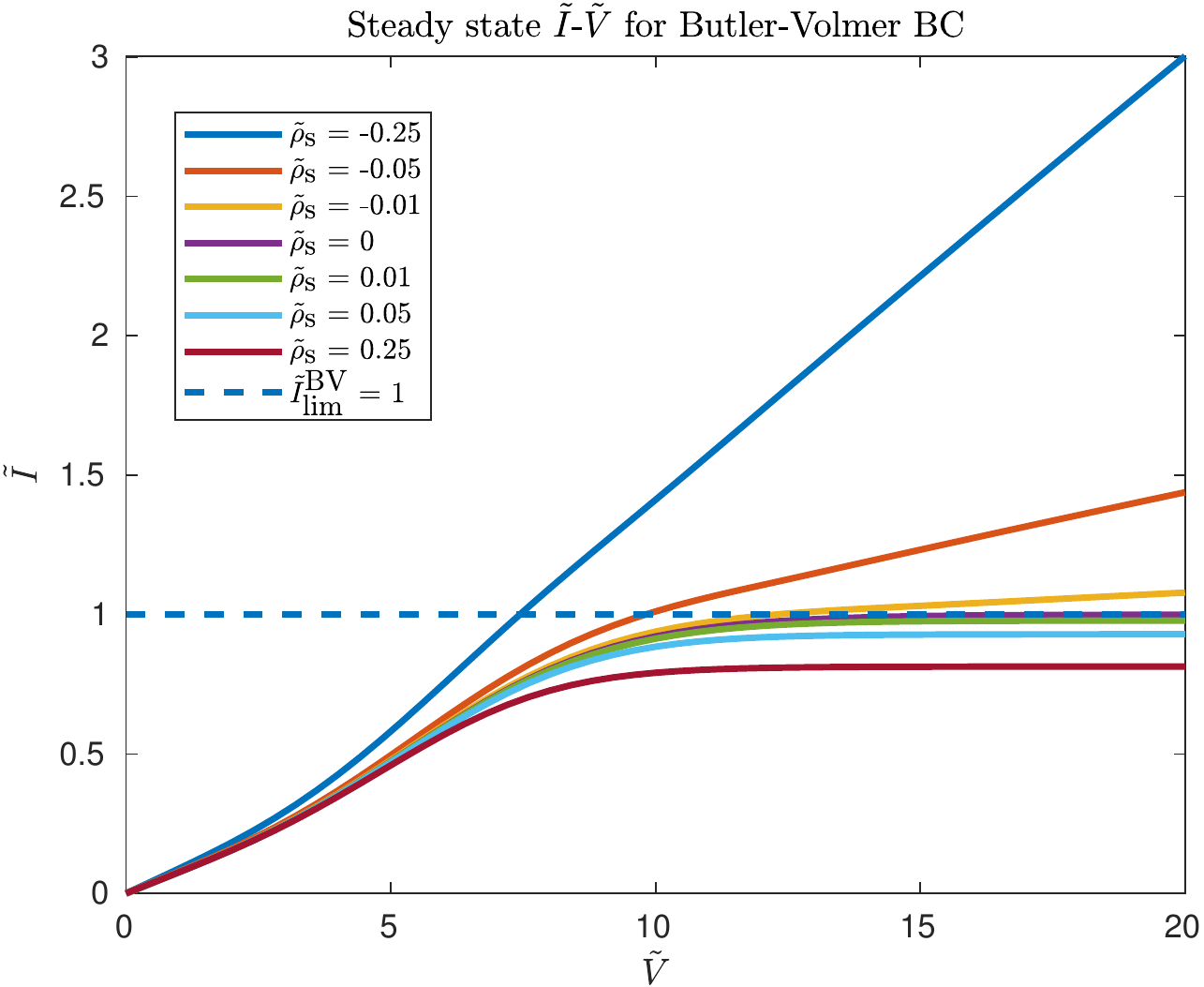}
  \includegraphics[scale=0.6]{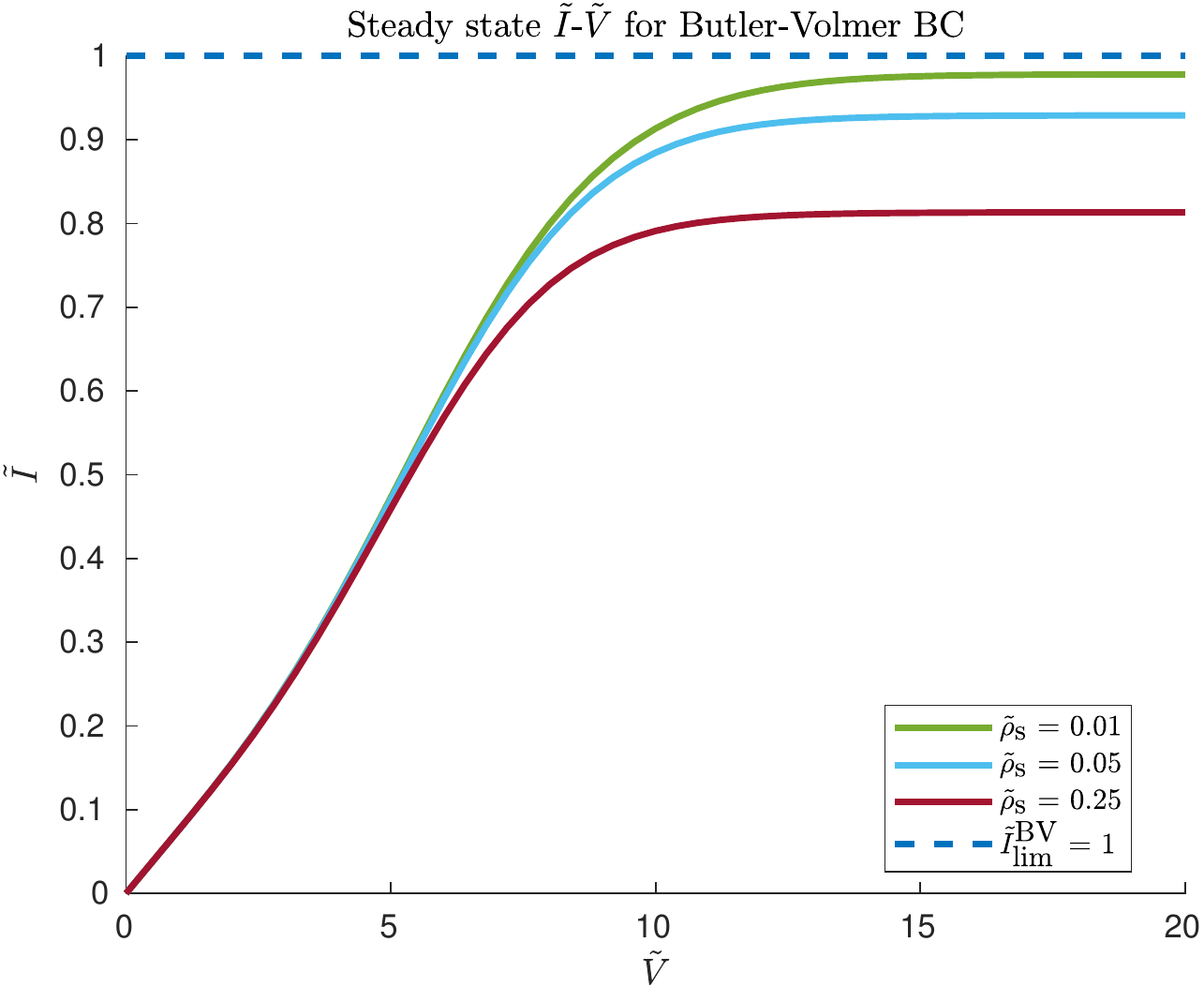}
  \caption{Left: steady state $\tilde{I}$-$\tilde{V}$ relations for $\tilde{\rho}_\textn{s} = 0, \pm 0.01, \pm 0.05, \pm 0.25$ for Butler-Volmer boundary conditions at anode and cathode. Right: zoom-in view of left plot for only $\tilde{\rho}_\textn{s} = 0.01, 0.05, 0.25$. The dashed line denotes $\tilde{I}_\limn^\textn{BV} = 1$, which is the maximum $\tilde{I}$ that the system can reach when $\tilde{\rho}_\textn{s} = 0$.}\label{fig:Steady state I-V relations for Butler-Volmer boundary conditions at anode and cathode}
\end{figure}

We plot $\tilde{c}_-$ and $\tilde{\phi}$ as functions of $\tilde{x}$ for $\tilde{\rho}_\textn{s} = -0.01, -0.25$ in Figure~\ref{fig:c_- and phi against x for rho_s < 0 for Butler-Volmer boundary conditions at anode and cathode} and $\tilde{\rho}_\textn{s} = 0.01, 0.25$ in Figure~\ref{fig:c_- and phi against x for rho_s > 0 for Butler-Volmer boundary conditions at anode and cathode}. For $\tilde{\rho}_\textn{s} = 0.01, 0.25$, we choose $\tilde{I} = 0.5\tilde{I}_\maxn^\textn{BV}, 0.99\tilde{I}_\maxn^\textn{BV}$ and for $\tilde{\rho}_\textn{s} = -0.01, -0.25$, we choose $\tilde{I} = 0.5, 1, 1.5$. As expected, the analytical and semi-analytical solutions agree very well with the numerical solutions. The features of the $\tilde{c}_-$-$\tilde{x}$ and $\tilde{\phi}$-$\tilde{x}$ plots are the same as that for case 2 except that for a particular $\tilde{I}$ and $\tilde{\rho}_\textn{s}$, $\tilde{V}$ is significantly larger than that for case 2 because additional electric potential differences and overpotentials are required to drive the Faradaic reactions at the electrodes. We also plot $\tilde{I}$ against $\tilde{V}$ for $\tilde{\rho}_\textn{s} = 0, \pm 0.01, \pm 0.05, \pm 0.25$ and $\tilde{V} \in \left[0,20\right]$ in Figure~\ref{fig:Steady state I-V relations for Butler-Volmer boundary conditions at anode and cathode}. Regardless of $\tilde{\rho}_\textn{s}$, we observe that the $\tilde{I}$-$\tilde{V}$ curve has a positive curvature at small $\tilde{V}$ because the system is reaction-limited and hence, Butler-Volmer reaction kinetics causes current to have an exponential dependence on voltage. At high $\tilde{V}$, the system becomes transport-limited in which surface conduction sustains OLC, therefore $\tilde{I}$ becomes linear in $\tilde{V}$ and the gradient of this linear relationship is equal to the overlimiting conductance.

\subsection{Copper electrodeposition and electrodissolution in AAO, CN and PE membranes}\label{sec:Copper electrodeposition and electrodissolution in AAO, CN and PE membranes}

Experimental steady state current-voltage relations are typically obtained using linear sweep voltammetry (LSV) with a sufficiently slow sweep rate. Using the steady state current-voltage relation for Butler-Volmer boundary conditions in Section~\ref{sec:Butler-Volmer boundary conditions at anode and cathode}, we perform nonlinear least squares fitting on the experimental datasets for copper electrodeposition and electrodissolution in charged nanoporous AAO~\cite{han_over-limiting_2014}, CN~\cite{han_dendrite_2016} and PE~\cite{han_dendrite_2016} membranes to demonstrate the usefulness of such a relation for extracting best-fit parameter values. We first use these best-fit parameter values to implement time-dependent LSV numerical simulations at various sweep rates to verify if the experimental sweep rate used is sufficiently slow for measuring quasisteady current-voltage relations. We then use these best-fit parameter values for computing steady state current-voltage relations and time-dependent LSV numerical simulations to see how well they compare with the experimental datasets. We also estimate the experimental overlimiting conductances for negatively charged membranes and compare them with the steady state overlimiting conductances that are computed using Equation~\ref{eq:Overlimiting conductance for no-anion-flux boundary condition at anode}.

The AAO membranes used in~\cite{han_over-limiting_2014} have parallel straight cylindrical pores with the same length and a constant pore radius, therefore the assumptions that $\epsilon_\textn{p}$, $\tau$, $a_\textn{p}$ and $h_\textn{p}$ are uniform and constant are reasonable. Denoting the pore radius as $r_\textn{p}$, we obtain $h_\textn{p} = \frac{\epsilon_\textn{p}}{a_\textn{p}} = \frac{r_\textn{p}}{2}$. In~\cite{han_over-limiting_2014}, boric acid ($\textn{H}_3\textn{BO}_3$) is added to reduce the rate of hydrogen evolution at high voltages by increasing the overpotential needed to do so~\cite{heidari_electrodeposition_2015}. We assume that boric acid is inert and does not dissociate at all, so the electrolyte consists of only $\textn{Cu}^{2+}$ and $\textn{SO}_4^{2-}$ ions. The CN and PE membranes used in~\cite{han_dendrite_2016} are random porous media with well connected pores, in contrast to the ordered AAO membranes that are a massively parallel network of non-intersecting straight cylindrical pores. In the absence of detailed geometrical information, we approximate $h_\textn{p} \approx \frac{r_\textn{p}}{2}$. In both~\cite{han_over-limiting_2014} and~\cite{han_dendrite_2016}, the electrolyte used is copper(II) sulfate ($\textn{CuSO}_4$) and the electrodes used are circular with a radius $r_\textn{e}$, therefore $A=\pi r_\textn{e}^2$ where $A$ is the total surface area of the anode or cathode. The geometrical parameters $r_\textn{p}$, $L$ and $r_\textn{e}$ for AAO, CN and PE membranes are given in Table~\ref{tab:Geometrical parameters for AAO, CN and PE membranes}. In Table~\ref{tab:Dataset labels and fitted parameter values for AAO, CN and PE membranes}, we label all the experimental datasets in~\cite{han_over-limiting_2014} and~\cite{han_dendrite_2016} based on the membrane identity (AAO, CN or PE), sign of membrane charge, sweep rate $\beta_\textn{LSV}$ and electrolyte concentration $c_0$.  

\begingroup
\begin{table} \caption{Geometrical parameters for AAO, CN and PE membranes. Values are based on product specifications.}\label{tab:Geometrical parameters for AAO, CN and PE membranes}
  \centering
  \begin{ruledtabular}
  \begin{tabular}{cccc}
    Parameter & AAO membranes & CN membranes & PE membranes \\ \hline  
    $r_\textn{p}\,/\,\textn{nm}$ & $175$ (mean of $150-200$) & $125$ (mean of $100-150$) & $25$ \\
    $L\,/\,\mu\textn{m}$ & $60$ & $130$ & $20$ \\
    $r_\textn{e}\,/\,\textn{mm}$ & $6$ & $6.5$ & $6.5$ \\
  \end{tabular}
  \end{ruledtabular}
\end{table}
\endgroup

\begingroup
\begin{table} \caption{Dataset labels (first three columns) and fitted parameter values (last five columns) for AAO, CN and PE membranes.}\label{tab:Dataset labels and fitted parameter values for AAO, CN and PE membranes}
  \centering
  \begin{ruledtabular}
  \begin{tabular}{ccc|ccccc}  
    Label & $\beta_\textn{LSV}\,/\,\textn{mV}\,\textn{s}^{-1}$ & $c_0\,/\,\textn{mM}$ & $\sigma_\textn{s}\,/\,\textn{e}\,\textn{nm}^{-2}$ & $\tau$ & $J_0^\textn{ref}\,/\,\textn{mA}\,\textn{cm}^{-2}$ & $\alpha_1$ & $\epsilon_\textn{p}$ \\ \hline  
     $\textn{AAO}_1(-)$ & $-1$ & $10$ & $-0.591$ & $1.00$ (fixed) & $1.65$ & $1.00$ & $0.500$ \\
     $\textn{AAO}_1(+)$ & $-1$ & $10$ & $1.63$ & $1.00$ (fixed) & $2.76$ & $0.750$ & $0.375$ \\
     $\textn{AAO}_2(-)$ & $-10$ & $100$ & $-0.517$ & $1.00$ (fixed) & $4.90$ & $1.00$ & $0.443$ \\
     $\textn{AAO}_2(+)$ & $-10$ & $100$ & $2.64$ & $1.00$ (fixed) & $4.59$ & $0.650$ & $0.400$ \\
     $\textn{CN}_1(-)$ & $-1$ & $10$ & $-0.0723$ & $1.83$ & $14.4$ & $0.287$ & $0.803$ \\
     $\textn{CN}_1(+)$ & $-1$ & $10$ & $0.617$ & $1.91$ & $13.0$ & $0.845$ & $0.664$ \\
     $\textn{CN}_2(-)$ & $-10$ & $100$ & $-0.0478$ & $2.11$ & $16.6$ & $0.612$ & $0.684$ \\
     $\textn{CN}_2(+)$ & $-10$ & $100$ & $1.78$ & $2.07$ & $5.20$ & $0.985$ & $0.761$ \\
     $\textn{PE}(-)$ & $-2$ & $10$ & $-0.0549$ & $5.44$ & $6.15$ & $0.995$ & $0.409$ \\
     $\textn{PE}(+)$ & $-2$ & $10$ & $0.0752$ & $7.84$ & $0.500$ & $0.918$ & $0.470$ \\
  \end{tabular}
  \end{ruledtabular}
\end{table}
\endgroup

The exchange current densities and charge transfer coefficients are generally sensitive to experimental conditions such as the method of electrode preparation and electrode surface roughness. There are also no estimates for the surface charge densities of the polyelectrolyte multilayers used in~\cite{han_over-limiting_2014,han_dendrite_2016}. Using tortuosities that deviate from the Bruggeman relation is not uncommon in porous membranes such as the porous separators used in batteries~\cite{thorat_quantifying_2009}. In our case, for the CN and PE membranes, which are polymeric porous separators commonly used in lithium-ion batteries~\cite{zhang_review_2007}, tortuosity can be used as a fitting parameter. The membrane porosities are also typically specified as a range and may not be known with certainty. Therefore, when using the steady state current-voltage relation for Butler-Volmer boundary conditions to perform nonlinear least squares fitting on the experimental datasets, we pick $\tilde{\rho}_\textn{s}$, $\tau$ (only for CN and PE membranes; fixed at $1$ for AAO membranes), $\tilde{J}_0^\textn{ref}$, $\alpha_1$ and $\epsilon_\textn{p}$ as fitting parameters. This nonlinear least squares fitting is carried out using MATLAB's $\mathtt{lsqnonlin}$ function and the initial guesses and lower and upper bounds for the fitting parameters are given in Table I in Section III of the Supplementary Material. All parameters that are not fitting parameters or given in Table~\ref{tab:Geometrical parameters for AAO, CN and PE membranes} are taken from Table~\ref{tab:Parameters for copper electrodeposition and electrodissolution for AAO membranes}. The fitted parameter values that are obtained for all the experimental datasets are given in Table~\ref{tab:Dataset labels and fitted parameter values for AAO, CN and PE membranes}; for $\tilde{\rho}_\textn{s}$ and $\tilde{J}_0^\textn{ref}$, we report their dimensional values $\sigma_\textn{s}$ and $J_0^\textn{ref}$ respectively.

\begin{figure}
  \centering
  \includegraphics[scale=0.6]{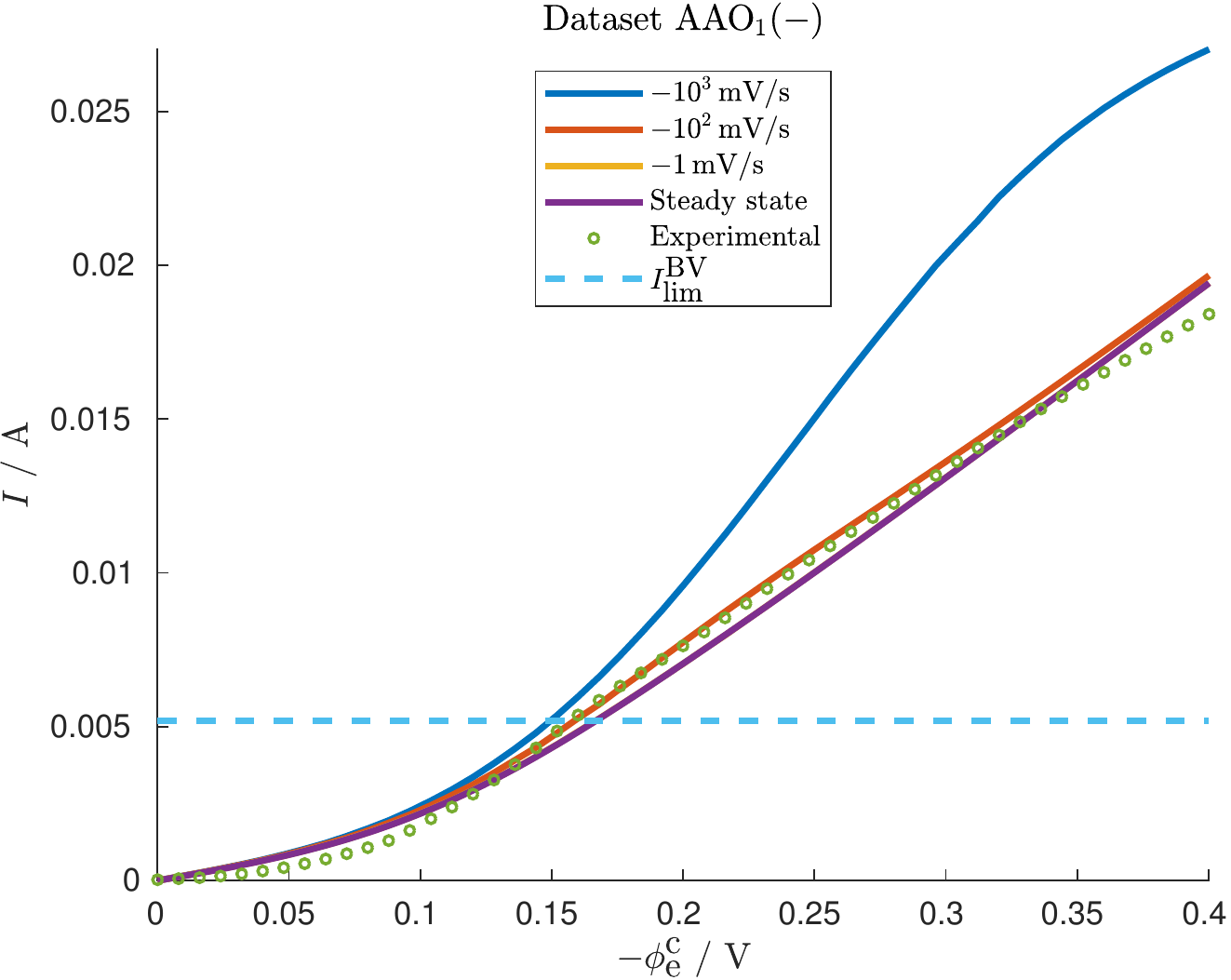}
  \includegraphics[scale=0.6]{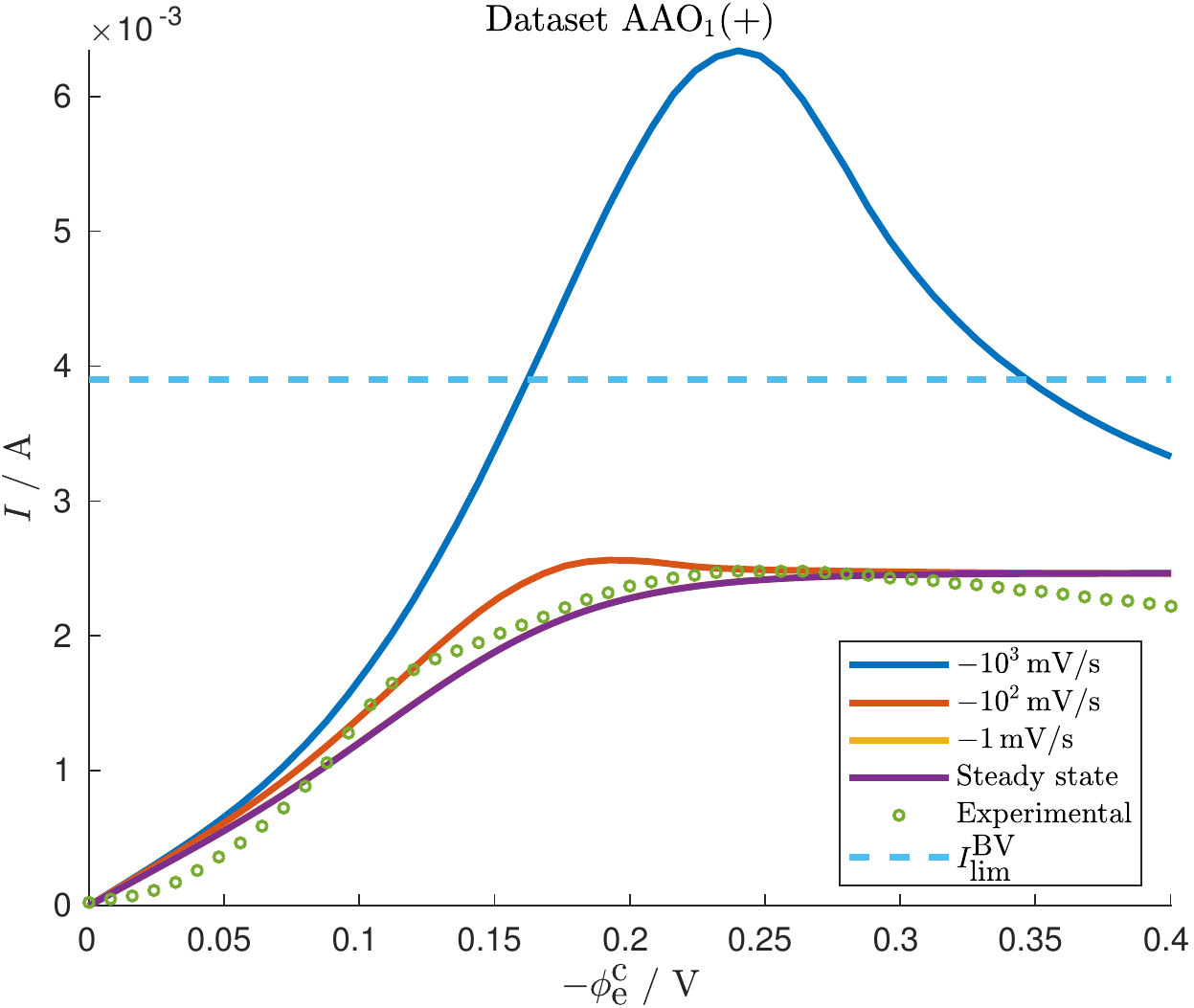}
  \caption{Numerical current-voltage relations obtained by linear sweep voltammetry with $\beta_\textn{LSV} = -10^3\,\textn{mV}/\textn{s}, -10^2\,\textn{mV}/\textn{s}, -1\,\textn{mV}/\textn{s}$ for copper electrodeposition and electrodissolution from copper(II) sulfate ($\textn{CuSO}_4$) for datasets $\textn{AAO}_1(+/-)$. Steady state and experimental current-voltage relations are also plotted. $-\phi_\textn{e}^\textn{c}$ is the negative of the cathode electric potential while $I$ is the current. Note that the lines for $\beta_\textn{LSV} = -1\,\textn{mV}/\textn{s}$ and steady state overlap.}\label{fig:Current-voltage relations for c_0 = 10 mM with different sweep rates}  
\end{figure}

Experimentally, to generate the steady state current-voltage relations in a reasonable amount of time, linear sweep voltammetry (LSV) with a sufficiently slow sweep rate is used. Therefore, we would like to first use numerical simulations to determine if the sweep rate used in experiments is slow enough for the experimental current-voltage relations to accurately approximate the true steady state ones. For datasets $\textn{AAO}_1(+/-)$, a sweep rate of $-1\,\textn{mV}/\textn{s}$ is used. To determine if this sweep rate is sufficiently slow, in our numerical simulations, we pick $\beta_\textn{LSV} = -10^3\,\textn{mV}/\textn{s}, -10^2\,\textn{mV}/\textn{s}, -1\,\textn{mV}/\textn{s}$. We also plot the semi-analytical steady state current-voltage relations for case 3 discussed in Section~\ref{sec:Butler-Volmer boundary conditions at anode and cathode} and the experimental current-voltage relations in Figure~\ref{fig:Current-voltage relations for c_0 = 10 mM with different sweep rates}. Figure~\ref{fig:Current-voltage relations for c_0 = 10 mM with different sweep rates} shows that the current-voltage relations for $\beta_\textn{LSV} = -1\,\textn{mV}/\textn{s}$ are almost identical with the steady state current-voltage relations, and both agree reasonably well with the experimental current-voltage relations. Therefore, we conclude that the sweep rate of $-1\,\textn{mV}/\textn{s}$ used experimentally is sufficiently slow. On the other hand, the sweep rates of $-10^3\,\textn{mV}/\textn{s}$ and $-10^2\,\textn{mV}/\textn{s}$ are too fast because at every value of $-\phi_\textn{e}^\textn{c}$, they result in currents that are significantly larger than their corresponding steady state and experimental values. In particular, for dataset $\textn{AAO}_1(+)$, the current significantly overshoots the limiting current $I_\limn^\textn{BV}$, which is caused by diffusion limitation as the time scale for the voltammetry is smaller than the diffusion time scale~\cite{moya_numerical_2015,yan_theory_2017}.

\begin{figure}
  \centering
  \includegraphics[scale=0.6]{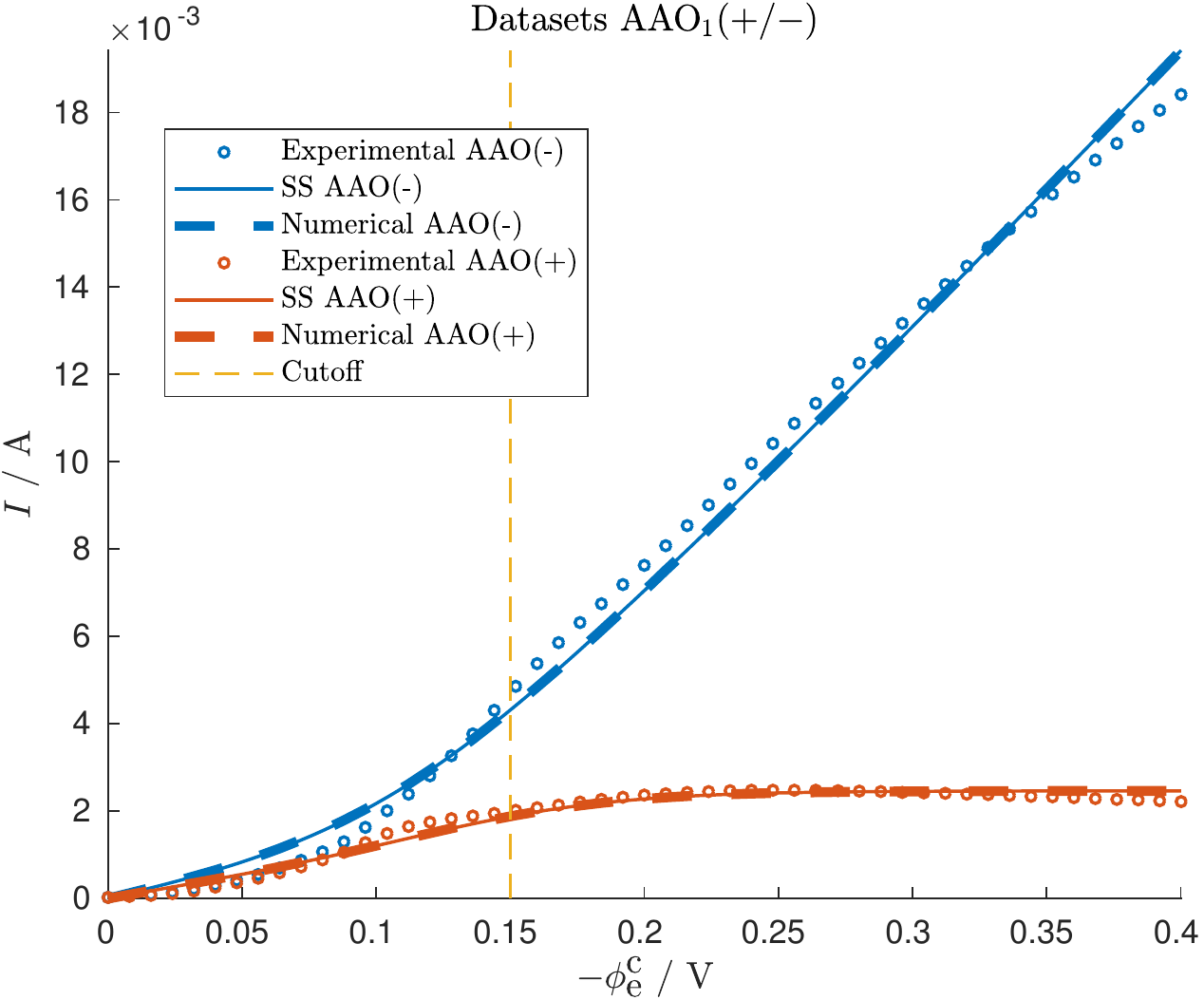}
  \includegraphics[scale=0.6]{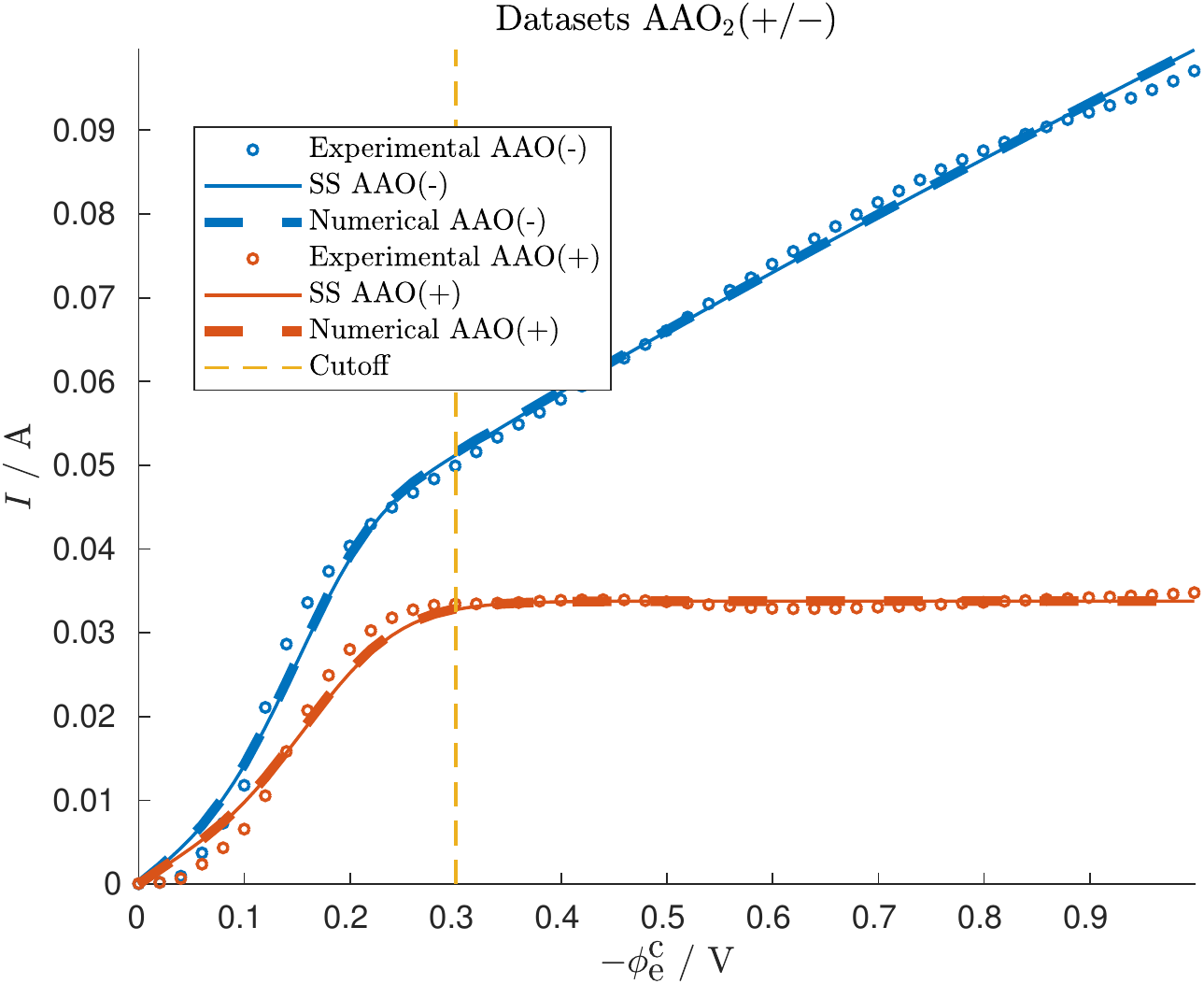}
  \includegraphics[scale=0.6]{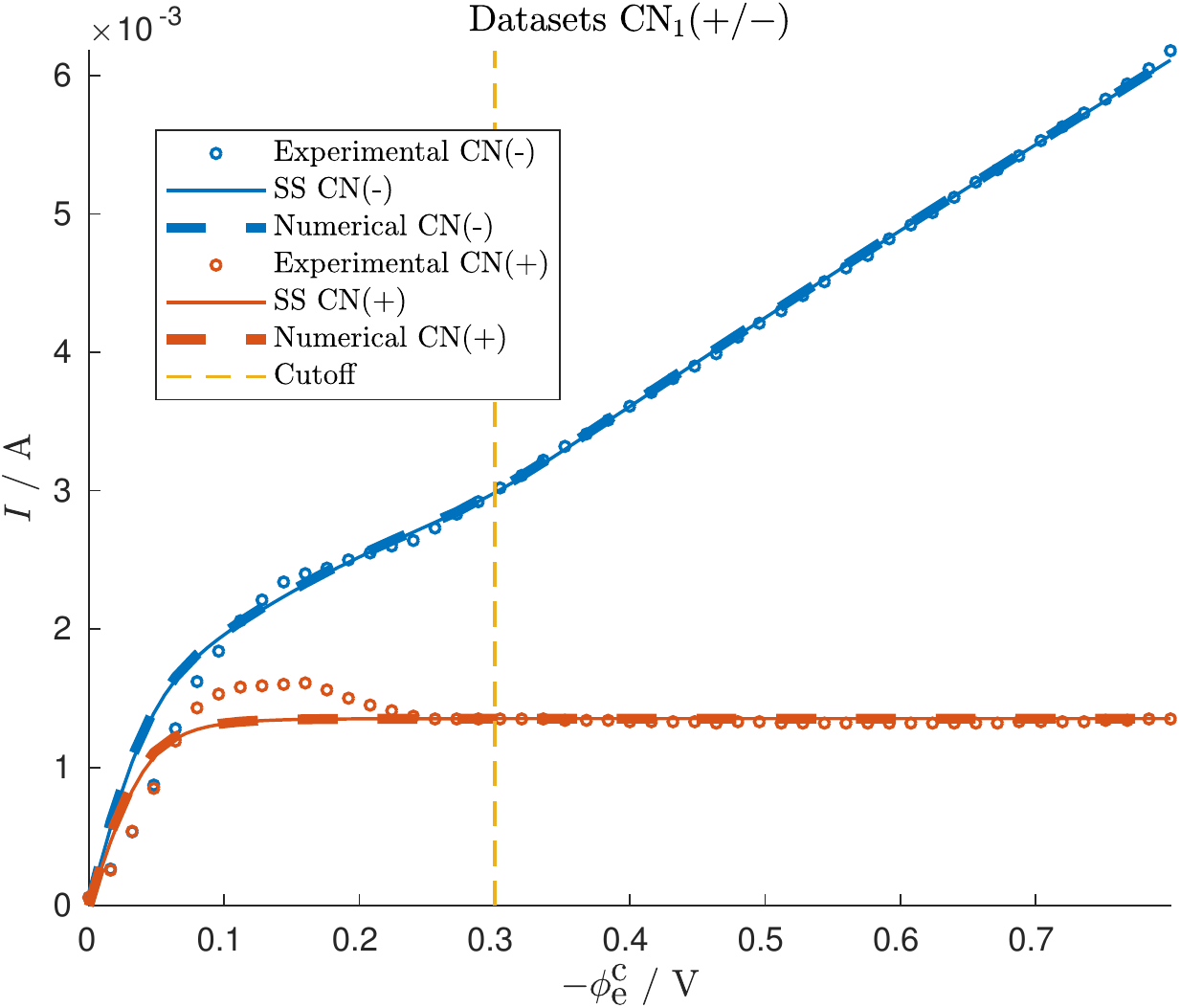}
  \includegraphics[scale=0.6]{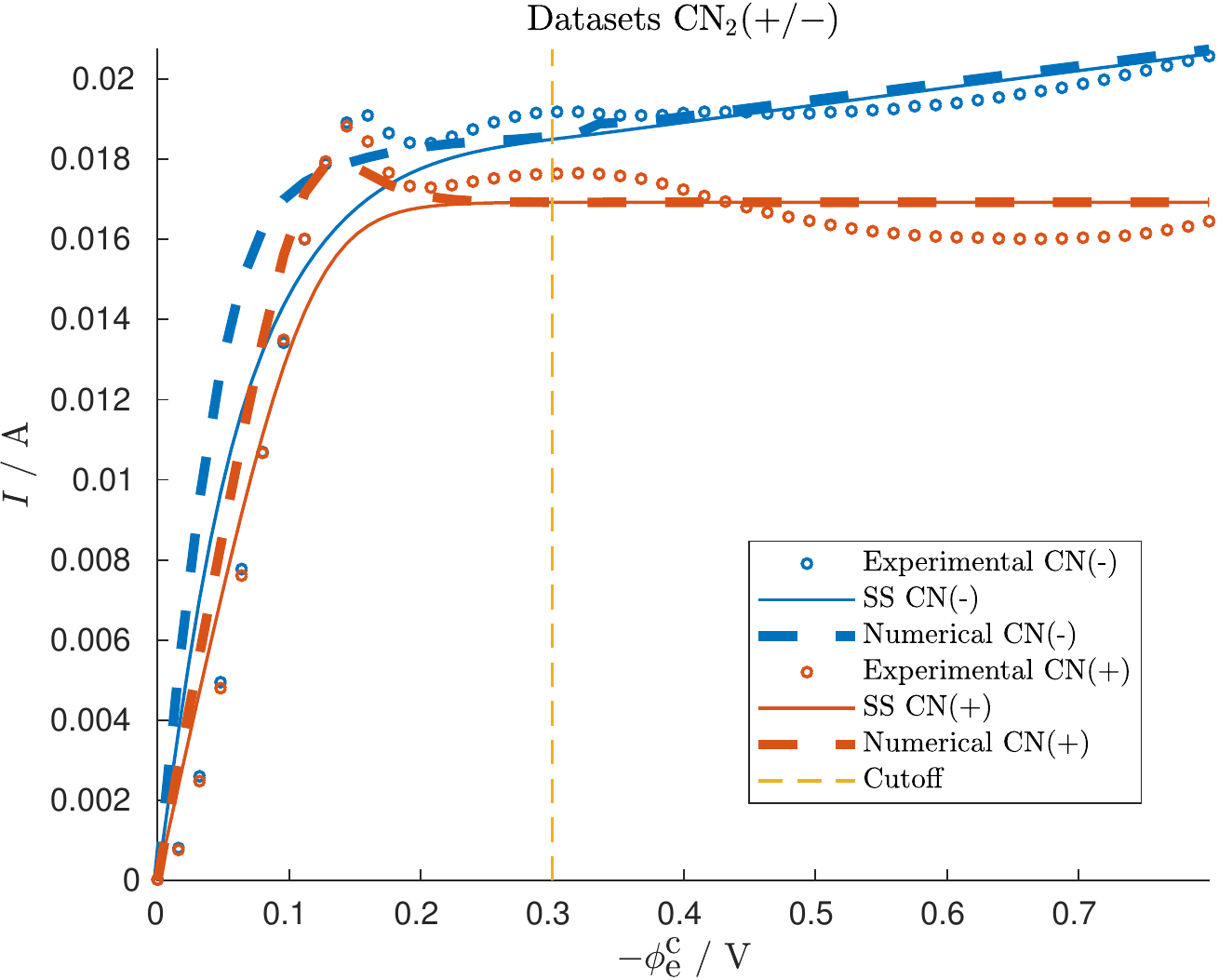}
  \includegraphics[scale=0.6]{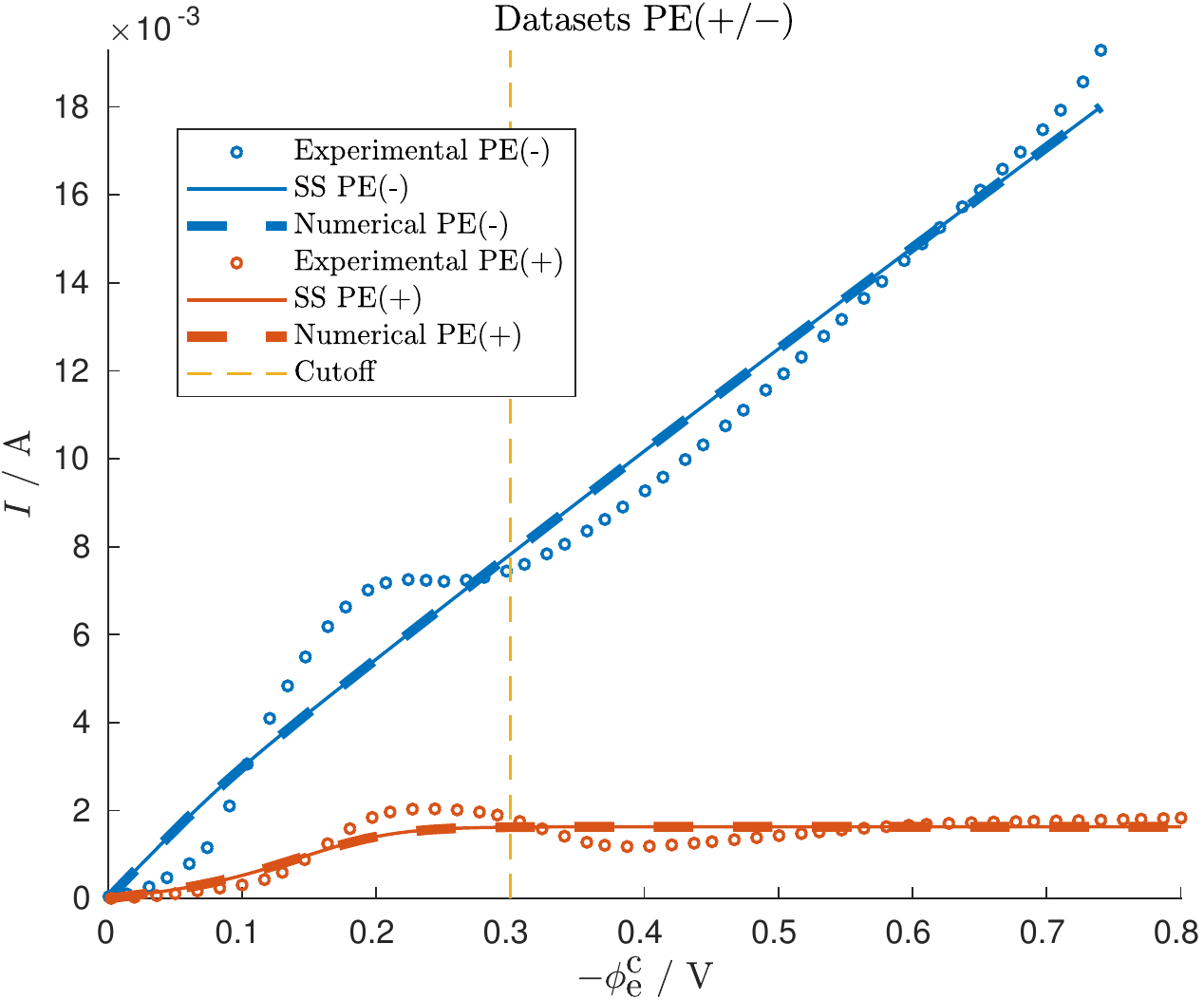}
  \caption{Current-voltage relations for copper electrodeposition and electrodissolution from copper(II) sulfate ($\textn{CuSO}_4$) in different charged nanoporous media. $-\phi_\textn{e}^\textn{c}$ is the negative of the cathode electric potential while $I$ is the current. (-) and (+) refer to negatively and positively charged membranes respectively. SS stands for steady state and the cutoff line indicates the cutoff potential for estimating overlimiting conductance.}\label{fig:Current-voltage relations for all experimental datasets}  
\end{figure}

Using the fitted parameter values in Table~\ref{tab:Dataset labels and fitted parameter values for AAO, CN and PE membranes}, we compute the steady state current-voltage relations for case 3 described in Section~\ref{sec:Butler-Volmer boundary conditions at anode and cathode} and also perform time-dependent LSV numerical simulations. We plot and compare both sets of current-voltage relations with the experimental datasets in Figure~\ref{fig:Current-voltage relations for all experimental datasets}. Note that although the nonlinear least squares fitting is performed on full experimental datasets, these datasets have too many data points to be plotted clearly. Therefore, we only plot $51$ points per dataset in Figure~\ref{fig:Current-voltage relations for all experimental datasets}. Generally, the steady state and numerical current-voltage relations agree well with the experimental ones, therefore demonstrating the usefulness of the steady state current-voltage relation for case 3 in Section~\ref{sec:Butler-Volmer boundary conditions at anode and cathode} for extracting important best-fit parameters such as $\sigma_\textn{s}$, which may be difficult to measure directly in experiments. In addition, the generally close agreement of the steady state current-voltage relations with the experimental and numerical ones indicates that the experimental sweep rates used are slow enough to generate quasisteady current-voltage relations. For datasets $\textn{PE}(+/-)$, the current bumps at around $-\phi_\textn{e}^\textn{c} = 0.2\,\textn{V}$ cannot be captured by the steady state and numerical current-voltage relations. In the context of our model, these current bumps are not caused by an overly fast sweep rate because the steady state current-voltage relation agrees very well with the numerical one. Instead, they are probably caused by unaccounted side reactions that contribute a current peak at $-\phi_\textn{e}^\textn{c} = 0.2\,\textn{V}$ that can for example be described by the ``modified'' Randles-Sevcik equation given by Equation 33 of~\cite{yan_theory_2017}.

For negatively charged membranes, the experimental current-voltage relations become approximately linear above a cutoff potential and the gradient of this line is the overlimiting conductance. Therefore, using MATLAB's $\mathtt{polyfit}$ function, we can estimate the experimental overlimiting conductance $\sigma_\textn{OLC}$ by performing a linear fit of the linear portion of the experimental current-voltage relation. We also compute the steady state overlimiting conductance given by Equation~\ref{eq:Overlimiting conductance for no-anion-flux boundary condition at anode}. We tabulate all the cutoff potentials and experimental and steady state overlimiting conductances in Table~\ref{tab:Cutoff potentials and overlimiting conductances for all negatively charged membranes}. Generally, the experimental overlimiting conductances agree well with the steady state ones, showing that the nonlinear least squares fitting procedure accurately fits the linear portions of the experimental current-voltage relations.

\begingroup
\begin{table} \caption{Cutoff potentials and overlimiting conductances for all negatively charged membranes.}\label{tab:Cutoff potentials and overlimiting conductances for all negatively charged membranes}
  \centering
  \begin{ruledtabular}
  \begin{tabular}{cccc}
    Dataset & Cutoff potential / $V$  & Experimental $\sigma_\textn{OLC}\,/\,\Omega^{-1}$ & Steady state $\sigma_\textn{OLC}\,/\,\Omega^{-1}$ \\ \hline  
    $\textn{AAO}_1(-)$ & $0.15$ & $0.0553$ & $0.0567$ \\
    $\textn{AAO}_2(-)$ & $0.3$ & $0.0685$ & $0.0439$ \\
    $\textn{CN}_1(-)$ & $0.3$ & $0.00630$ & $0.00461$ \\
    $\textn{CN}_2(-)$ & $0.3$ & $0.00261$ & $0.00225$ \\
    $\textn{PE}(-)$ & $0.3$ & $0.0268$ & $0.0195$ \\
  \end{tabular}
  \end{ruledtabular}
\end{table}
\endgroup

\section{Conclusion}

We have coupled transport described by the leaky membrane model, which is capable of predicting OLC, with Butler-Volmer boundary conditions and studied the resulting model at steady state in order to derive analytical and semi-analytical expressions for quantities of interest, namely concentration profiles, electric potential profiles, current-voltage relations and overlimiting conductances. These results generalize the ones in~\cite{dydek_overlimiting_2011,dydek_nonlinear_2013,tikekar_stability_2014} to a binary electrolyte that is asymmetric with unequal diffusivities and to Butler-Volmer boundary conditions. We have also analyzed linear sweep voltammetry with the model, building on the work of Yan et al~\cite{yan_theory_2017}, and validated its predictions against experimental data for copper electrodeposition in a variety of charged nanoporous media, with reasonable agreement for a simple, analytically tractable model.

Throughout the paper, we have assumed concentration-independent diffusivities, but this is generally not the case in concentrated solutions, where Stefan-Maxwell coupled fluxes and concentration-dependent activity coefficients contribute to the effective diffusion process~\cite{newman_electrochemical_2004}. Even the most basic concentration dependence of the Debye-Huckel theory for dilute-solution activity, or its generalization to concentrated solutions~\cite{schlumpberger_simple_2017}, can significantly affect the steady state concentration and electric potential profiles, as well as the current-voltage relation, in a leaky membrane~\cite{dydek_nonlinear_2013}. It would be interesting in future work to analyze how such effects couple with the highly nonlinear Butler-Volmer boundary conditions. In addition, copper(II) sulfate and boric acid, which is commonly added to suppress hydrogen evolution at high voltages, are slightly acidic, thus it is possible that charge regulation and pH changes provide additional conductivity~\cite{andersen_current-induced_2012}. We have also used the simplest reaction model for copper electro-deposition/dissolution, but more sophisticated reaction models do not assume any rate-determining step and take into account additional phenomena such as the adsorption of copper(I) ions on the electrode surface~\cite{huerta_garrido_eis_2006,lasia_remarks_2007,huerta_garrido_reply_2008}. Using these models may help with achieving better predictions for the current-voltage relation, especially at low voltages when the system is reaction-limited. We have assumed that macroscopic electroneutrality holds when the coion concentration is depleted at a current higher than its diffusion-limited value. In a free solution, above the diffusion-limited current, macroscopic electroneutrality does not hold and the electric double layers are no longer at equilibrium~\cite{bazant_current-voltage_2005,chu_electrochemical_2005}. A more detailed analysis of the structure of the electric double layers above the diffusion-limited current in charged porous media would be useful for determining if the assumption of macroscopic electroneutrality is valid at such a current.  

\begin{acknowledgments}
  E. Khoo acknowledges support from the National Science Scholarship (PhD) funded by Agency for Science, Technology and Research, Singapore (A*STAR). We acknowledge J.-H. Han and M. Wang for providing the raw experimental datasets, H. Zhao and K. M. Conforti for useful suggestions regarding data visualization, J. Song for discussion regarding the Lambert W function and P. M. Biesheuvel for helpful comments on the manuscript.
\end{acknowledgments}

\appendix

\section{Symbols for variables, parameters and constants.}\label{sec:Symbols for variables, parameters and constants}

Table~\ref{tab:Symbols for variables, parameters and constants} lists the symbols for variables, parameters and constants used throughout the paper.

\begingroup
\squeezetable  
\begin{table} \caption{Symbols for variables, parameters and constants. $\pm$ subscript refers to cation and anion and TS stands for transition state.}\label{tab:Symbols for variables, parameters and constants}
  \centering
  \begin{ruledtabular}
  \begin{tabular}{cc}
    Symbol & Variable / Parameter / Constant \\ \hline  
    $a_\pm$ & Ion activity \\
    $a_\textn{p}$ & Internal pore surface area/volume ratio \\
    $c$ & Neutral salt bulk concentration \\
    $c_\pm$ & Ion concentration \\
    $c_\pm^\Theta$ & Ion standard concentration \\
    $\hat{c}_\pm$ & Ion concentration normalized by standard concentration \\
    $D_\pm^\textn{m}$ & Ion molecular (free solution) tracer diffusivity \\
    $D_{\pm0}$ & Ion macroscopic tracer diffusivity in dilute limit \\
    $D_{\pm0}^\textn{m}$ & Ion molecular (free solution) tracer diffusivity in dilute limit \\
    $\textn{Da}$ & Damkohler number \\
    $E^\Theta$ & Standard electrode potential \\
    $F_\pm$ & Ion diffusional molar flux \\
    $h_\textn{p}$ & Effective pore size \\
    $I$ & Current \\
    $J$ & Current density \\
    $J_0$ & Exchange current density \\
    $J_\textn{F}$ & Faradaic current density \\
    $k_\textn{B}$ & Boltzmann constant \\
    $M_\textn{m}$ & Atomic mass of solid metal and electroactive cations \\
    $\hat{n}$ & Unit normal pointing outwards from electrolyte \\
    $r$ & Position vector \\
    $r_\textn{m}^\textn{a,c}$ & Position of anode/electrolyte or cathode/electrolyte interface \\
    $T$ & Temperature \\
    $z_\pm$ & Ion charge number \\
    $\alpha$ & Charge transfer coefficient \\
    $\gamma_\pm$ & Ion activity coefficient \\
    $\gamma_\ddagger^\textn{r}$ & Activity coefficient of TS for Faradaic reaction \\
    $\gamma_{\ddagger,\pm}^\textn{d}$ & Activity coefficient of TS for activated diffusion of ion \\
    $\epsilon_\textn{p}$ & Porosity \\
    $\eta$ & Overpotential \\
    $\mu_\pm$ & Ion electrochemical potential \\
    $\mu_\pm^\Theta$ & Ion standard electrochemical potential \\
    $\nu_\pm$ & Subscript of ion in chemical formula of neutral salt \\
    $\rho_\textn{m}$ & Mass density of solid metal \\
    $\rho_\textn{s}$ & Volume-averaged background charge density \\
    $\sigma_\textn{s}$ & Pore surface charge/area ratio \\
    $\tau$ & Tortuosity \\
    $\phi$ & Electrolyte electric potential \\
    $\phi_\textn{e}$ & Electrode electric potential \\
    $\Omega_\textn{m}$ & Atomic volume of solid metal \\
  \end{tabular}
  \end{ruledtabular}
\end{table}

\newpage

\bibliography{bibliography_bibtex}

\end{document}